\documentclass[reprint,aps,pra,notitlepage,nofootinbib,longbibliography,floatfix,superscriptaddress]{revtex4-2}

\usepackage[utf8]{inputenc}
\usepackage{float}
\usepackage{graphicx,nicefrac}  
\usepackage{physics}
\usepackage{esint}
\usepackage{amssymb}  
\usepackage{siunitx}
\usepackage{comment}
\usepackage{mathtools}
\usepackage{amsmath}
\usepackage{amsthm}
\usepackage{wrapfig}
\usepackage{mathrsfs}
\usepackage{enumitem}
\usepackage[colorlinks=true,linkcolor=blue,citecolor=magenta,urlcolor=magenta,plainpages=false,pdfpagelabels]{hyperref}
\usepackage[dvipsnames]{xcolor}
\usepackage{color,colortbl}

\usepackage{bm}
\usepackage[most]{tcolorbox}
\usepackage[normalem]{ulem}

\usepackage{multirow}
\makeatletter 
    
\renewcommand\onecolumngrid{
\do@columngrid{one}{\@ne}%
\def\set@footnotewidth{\onecolumngrid}
\def\footnoterule{\kern-6pt\hrule width 1.5in\kern6pt}%
}

\renewcommand\twocolumngrid{
        \def\footnoterule{
        \dimen@\skip\footins\divide\dimen@\thr@@
        \kern-\dimen@\hrule width.5in\kern\dimen@}
        \do@columngrid{mlt}{\tw@}
}%

\makeatother    

\newcommand{\beq}{\begin{equation}}
\newcommand{\eeq}{\end{equation}}

\newcommand{\bea}{\begin{eqnarray}}
\newcommand{\eea}{\end{eqnarray}}

\definecolor{coolblack}{rgb}{0.0, 0.18, 0.39}
\definecolor{darkred}{rgb}{0.5,0,0}
\definecolor{darkgreen}{rgb}{0,0.5,0}
\definecolor{darkblue}{rgb}{0,0,0.5}
\definecolor{lapislazuli}{rgb}{0.15, 0.38, 0.61}
\definecolor{venetianred}{rgb}{0.78, 0.03, 0.08}
\definecolor{bleudefrance}{rgb}{0.19, 0.55, 0.91}
\definecolor{dogwoodrose}{rgb}{0.84, 0.09, 0.41}

\newcommand{\mathcheck}[1]{\color{dogwoodrose}{#1}}
\newcommand{\tcp}[1]{\mathcheck{\tilde{c}_\tp}}

\newcommand{\ci}{\mathrm{i}}

\newcommand{\ee}{\mathrm{e}}
\newcommand{\tp}{\mathrm{p}}
\newcommand{\tv}{v}

\newcommand{\lr}[1]{\left(#1\right)}
\newcommand{\lrsq}[1]{\left[#1\right]}

\newcommand{\bs}[1]{\boldsymbol{#1}}
\newcommand{\bS}{\boldsymbol{S}}

\begin{document}

\title{Entanglement from superradiance and rotating quantum fluids of light}

\author{Adri\`a Delhom}
\email{adria.delhom@gmail.com}
\affiliation{Department of Physics and Astronomy, Louisiana State University, Baton Rouge, LA 70803, U.S.A.}
\author{Killian Guerrero}
\affiliation{Laboratoire Kastler Brossel, Sorbonne Universit\'e, CNRS, ENS-Universit\'e PSL, Coll\`ege de France, 4 Place Jussieu, Paris, 75005, France.}
\author{Paula Calizaya Cabrera}
\affiliation{Department of Physics and Astronomy, Louisiana State University, Baton Rouge, LA 70803, U.S.A.}
\author{K\'evin Falque}
\affiliation{Laboratoire Kastler Brossel, Sorbonne Universit\'e, CNRS, ENS-Universit\'e PSL, Coll\`ege de France, 4 Place Jussieu, Paris, 75005, France.}
\author{Alberto Bramati}
\affiliation{Laboratoire Kastler Brossel, Sorbonne Universit\'e, CNRS, ENS-Universit\'e PSL, Coll\`ege de France, 4 Place Jussieu, Paris, 75005, France.}
\author{Anthony J. Brady}
\affiliation{Department of Electrical and Computer Engineering, University of Arizona,  Tucson, Arizona 85721, USA}
\affiliation{Ming Hsieh Department of Electrical and Computer Engineering,
University of Southern California, Los Angeles, California 90089, USA}
\author{Maxime J. Jacquet}
\email{maxime.jacquet@lkb.upmc.fr}
\affiliation{Laboratoire Kastler Brossel, Sorbonne Universit\'e, CNRS, ENS-Universit\'e PSL, Coll\`ege de France, 4 Place Jussieu, Paris, 75005, France.}
\author{Ivan Agullo}
\affiliation{Department of Physics and Astronomy, Louisiana State University, Baton Rouge, LA 70803, U.S.A.}

\begin{abstract}
The amplification of radiation by superradiance is a universal phenomenon observed in numerous physical systems. We demonstrate that superradiant scattering generates entanglement for different input states, including coherent states, thereby establishing the inherently quantum nature of this phenomenon. To put these concepts to the test, we propose a novel approach to create horizonless ergoregions, which are nonetheless dynamically stable thanks to the dissipative dynamics of a polaritonic fluid of light. We numerically simulate the system to demonstrate the creation of a stable ergoregion. Subsequently, we investigate rotational superradiance within this system, with a primary focus on entanglement generation and the possibilities for its enhancement  using current techniques. Our methods permit the investigation of quantum emission by rotational superradiance in state-of-the-art experiments, in which the input state can be controlled at will.
\end{abstract}

\maketitle

\section{Introduction}

Superradiance describes the amplification of radiation by time-independent potentials, 
a phenomenon that occurs across a broad spectrum of systems, including optics~\cite{Dicke:1954zz}, fluid dynamics~\cite{Zeldovich1}, electrodynamics~\cite{Klein:1929zz,Sauter:1931zz,Schwinger:1951nm}, particle physics~\cite{Arvanitaki:2009fg, Arvanitaki:2010sy}, and fields on curved spacetimes~\cite{Zeldovich1,Penrose:1969pc, Penrose:1971uk,Press:1972zz, Teukolsky:1974yv, Blandford:1977ds}.

Rotational superradiance \cite{Zeldovich1, Zeldovich2} occurs in spinning systems featuring an ergoregion ---a region within which  physical probes are dragged around and forced to co-rotate with the system (see \cite{brito_superradiance_2020} for a review).\footnote{This is different from Dicke superradiance that describes the collective amplification of radiation by coherent emitters~\cite{Dicke:1954zz}.} 
While all probes have positive energy outside the ergoregion, frame dragging  permits the existence of waves with negative energies within. The mixing of positive- and negative-energies across the boundary of the ergoregion (the ergosurface) allows for incoming waves to be reflected with larger amplitudes.

The availability of negative energy states implies a quantum instability~\cite{Zeldovich1, Zeldovich2, Starobinskii:1973hgd, Unruh:1974bw}, by which the  vacuum decays into pairs of field excitations. This entails the unavoidable presence of a finite population of excitations inside the ergosurface. If this population is not dissipated away by some mechanism, it will be further amplified by superradiance, leading to an instability.
Ergoregions are thus intrinsically unstable~\cite{Friedman:1978wla}.

In known stable configurations, dissipation is provided by a horizon inside the ergosurface (as, for instance, in spinning black holes). A horizon is a one-way membrane, which effectively dissipates negative energy modes out of the system. However, horizons emit Hawking radiation, meaning that ergosurfaces are constantly exposed to an energy flux from the horizon. In other words, the input state of rotational superradiance in such configurations is never the vacuum, but  thermal Hawking radiation instead. This has prevented the isolation of quantum emission solely due to rotational superradiance.

In studies of superradiance, a coherent state is typically employed to externally illuminate the ergoregion (see, e.g., \cite{brito_superradiance_2020}). The ensuing emission statistics are dominated by classical amplitudes vastly surpassing the contributions from vacuum emission, including emissions from the horizon. Any potential quantum effects produced in the emission, such as entanglement, get lost among classical data (noise, classical correlations etc.). For this reason, superradiance is commonly understood as a classical amplification phenomenon.

This is the context within which rotational superradiance has been investigated experimentally~\cite{torres_rotational_2017,braidotti_measurement_2022}. In these experiments, the focus was on observing the reflection of the incoming coherent state. Energy conservation demands that the observed amplified wave is accompanied by a partner negative energy excitation in the ergoregion. However, experimental limitations were such that correlations across the ergosurface were not observed.

\begin{figure*}    \centering\includegraphics[width=0.82\textwidth]{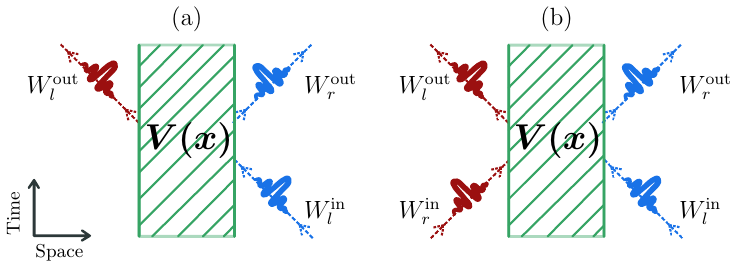}
    \caption{Illustration of wave packet scattering by a time-independent potential. The hatched region indicates the interaction region ---the region where the gradient of the total potential $V(x)$ is non-zero.
     (a) An incoming left-moving wave packet scatters into an outgoing left-moving (transmitted) wave packet and a  right-moving wave packet (reflected). (b) General scattering with incoming modes approaching the interacting region from both sides.}
    \label{fig:scattering}
\end{figure*}

This paper focuses on quantum emission by superradiance. On the one hand, we show how superradiance generates entanglement for a variety of input states, establishing it as an entanglement-generating phenomenon (see \cite{davies1993QuantumSuperradiance,Agullo:2023pgp} for previous discussions of entanglement in rotating black holes). 
On the other hand, we propose a new experimental setup to test our theoretical findings, involving the creation of horizonless configurations with an ergoregion that is nevertheless stable. We use a two-dimensional quantum fluid of light made of microcavity polaritons (hybrid particles composed of a photon and an electric dipole).  The dynamics of this system are intrinsically dissipative, a property we use in our advantage to control the ergoregion instability while minimizing data loss.
We numerically simulate an irrotational vortex featuring an ergosurface. We study the field equations describing acoustic perturbations and show 
the occurrence of rotational superradiance in that configuration.

The ability to create stable horizonless ergoregions is promising, as it permits the investigation of the quantum properties of superradiant emission by controlling the input state at will. Observing entanglement generated in pair-production processes has remained a challenge. We argue that the ability of ergoregions to amplify entanglement combined with the unique capabilities of fluids of light offer a promising avenue to achieve this benchmark. This article lays the theoretical groundwork toward the observation of entanglement from rotational superradiance. The numerical configuration found here can be implemented experimentally, paving the way for the experimental investigation of rotational superradiance from stable ergosurfaces.

\section{Superradiant wave scattering in a Nutshell}\label{ClassicalSR}

Superradiant scattering occurs when the transmission and/or reflection coefficients that describe a stationary scattering process exceed unity.  
In such cases, the amplitudes of the scattered waves are larger than that of the incident wave, at the expense of the potential, which supplies the necessary energy for the amplification process to occur.

Our objective in this section is to delineate the conditions under which superradiance universally occurs, and pinpoint the key mechanisms behind it. To that end, we focus on scalar waves and leave the potential unspecified. In the subsequent sections, we will restrict attention to a specific system demonstrating rotational superradiance.

 For the sake of clarity, we momentarily focus on one-dimensional wave propagation. Due to the time-independence of the potential, the frequency $\omega$ of the waves remains conserved. Therefore, it suffices to focus on a single frequency at a time.

Imagine a scenario akin to the one depicted in Fig.~\ref{fig:scattering}, where an ingoing wave packet centered around the frequency $\omega$  scatters off the potential, resulting in two outgoing wave packets, each centered at the same frequency $\omega$. The outgoing wave packets propagate away from the potential ---the reflected part towards the right and the transmitted wave packet towards the left. This scattering scenario is described by a wave $\Psi$ such that
\beq \Psi|_{t\to -\infty}=W^{\rm in}_{l}\  \xrightarrow{\text{time}} \  \Psi|_{t\to \infty}=t\, W^{\rm out}_{l}+r\, W^{\rm out}_{r}\, ,\eeq
where $t$ and $r$ denote the transmission and reflection amplitudes, respectively, and the subscripts $r$ and $l$ indicate rightward and leftward propagation, respectively. Valuable insights into these coefficients can be obtained by examining conserved quantities of the wave equation.  

To gain some intuition, we first consider a scattering process described by the standard Schr\"odinger equation, with $\int \dd x\, |\Psi(t,x)|^2$ being a conserved quantity. This is the familiar conservation of probability in non-relativistic quantum mechanics. This quantity is strictly positive, and equal to $1$ for normalized wave packets. Equating its value at early and late times, one finds
$1=|t|^2+|r|^2$, implying that neither $|t|$ nor $|r|$ can exceed unity. In other words, amplification cannot occur in the framework of Schr\"odinger theory, by virtue of the conservation of probability. 

Bosonic field theories behave differently. Consider a scalar field $\Phi(t,x)$ which satisfies the Klein-Gordon equation in Minkowski spacetime (the argument below  readily generalizes to curved spacetimes and to other bosonic theories, such as electromagnetic fields or gravitational perturbations)
\beq
\lrsq{\nabla^{\mu}\nabla_{\mu} +m^2+\nu (x)}\, \Phi(t,x)=0\, ,
\label{eq:KGE} 
\eeq
where $\nabla_{\mu}$ is a covariant derivative, accounting for possible interactions with external gauge fields, and $\nu(x)$ is an external potential. The total potential describing the scattering process, $V(x)$, can be thought of as the combination of the gauge potentials contained in $\nabla_{\mu}$ and the external potential $\nu(x)$. $m\geq 0$ is the mass of the field, and we are setting the speed of light to one. 
This constitutes a relativistic field theory, with $\Phi(t,x)$ representing the value of the Klein-Gordon field at a point in spacetime. The conserved quantity for this theory is 
 \beq\label{eq:KGNorm}Q_{\rm KG}(\Phi) \equiv \ci/\hbar \int (\Phi^*\, \Pi-\Pi^*\, \Phi)\,  dx\, ,\eeq
where $\Pi$ is the momentum canonically conjugate to $\Phi$. The conservation of this quantity  can be readily checked by computing its time derivative and utilizing the Klein-Gordon equation~\eqref{eq:KGE} to show that it vanishes. (This quantity is commonly referred to as ``symplectic norm'').
An important property of $Q_{\rm KG}$ is that it can take negative values. 
Consequently, some potentials admit scattering solutions in which the right- and left-moving scattered wave packets carry different signs of $Q_{\rm KG}$.
In such a scenario, the conservation of $Q_{\rm KG}$ yields
\beq {1= |r|^2-|t|^2}\label{eq:SupCondCoeff} \, , \eeq 
which implies that  $|r|$ must be larger than unity. Therefore, there is amplification, and the scattered outgoing waves have larger amplitudes than the incident one. This is superradiance.


We see that superradiance is a consequence of having modes with the same frequency but carrying opposite sign of $Q_{\rm KG}$,
{ a possibility attributed} 
to the bosonic character of the field.\footnote{For fermionic fields, the conserved charge is positive definite ---e.g., $Q_{\rm D}\equiv \int dx \, \Psi^{\dagger}\Psi$ for a Dirac field--- thus precluding the existence of superradiance. This positivity is directly related to the fermionic statistics and Pauli's exclusion principle.}
{ Superradiance typically occurs in the presence of external gauge fields, whose potential modifies the form of the canonical momentum on one side of the interacting region. A well-known simple example is a charged scalar field interacting with a static electric field that is compactly supported in space  \cite{brito_superradiance_2020}.} 

We finish this section with a simple characterization of superradiant scattering, which will be useful in the subsequent sections of this article.

Consider the scattering of  two incoming wave packets depicted in Fig.~\ref{fig:scattering}, both narrowly centered around the frequency $\omega$, approaching the potential from opposing directions, denoted by  $W^{\rm in}_{r}$ and $W^{\rm in}_{l}$. The scattered wave will result in a linear combination of the normalized outgoing wave packets $W^{\rm out}_{r}$ and $W^{\rm out}_{l}$. This scattering phenomenon can be succinctly written as
\beq 
\label{modescatt} (W^{\rm in}_{r}, W^{\rm in}_{l})|_{t \to \infty} = (W^{\rm out}_{r}, W^{\rm out}_{l}) \cdot \bs{B}\, ,  \eeq
where \beq \label{B} \bs{B}=\left(\begin{array}{cc} T & r\cr R & t\end{array}\right)\, ,
\eeq
and $T,R,r,t\in\mathbb{C}$. The matrix $\bs{B}$ describes the scattering of wave packets with central frequency $\omega$. If the incident wave is of the form $a_1\,W^{\rm in}_{r} +a_2  \, {W^{\rm in}_{l}}$, the scattered wave will have the form $b_1\,W^{\rm out}_{r} +b_2  \, {W^{\rm out}_{l}}$, where $a_i, b_i\in\mathbb{C}$ are amplitudes. The incident and scattered wave amplitudes are related via
\beq \left(\begin{array}{c} b_1 \cr b_2\end{array}\right) = \bs{B}\cdot \left(\begin{array}{c} a_1 \cr a_2\end{array}\right)\, .\eeq
The main outcome of this discussion is encapsulated in the following theorem, the proof of which can be found in Appendix~\ref{app:proof}.\\
{\bf Theorem 1:} The matrix $\bs{B}$ describing the scattering from IN to OUT modes off a time-independent potential corresponds to superradiant scattering if and only if it is not a unitary matrix.

\section{Quantum description of superradiance}\label{sec:QuantumSR}
A simple way to translate the classical scattering described by Eq.~\eqref{modescatt} to the quantum theory, is to recall that each normalized wave packet $W_i$, where $i$ denotes a collective label, defines a creation and annihilation operator as follows. Consider the operator defined as
\beq \ci/\hbar \int (W_i^*\, {\hat \Pi}-{{\Pi}_{W_i}^*}\, \hat \Phi)\,  dx\, , \label{eq:afromW} \eeq
where $\hat \Phi(t,x)$ is the bosonic field operator{, $\hat{\Pi}(t,x)$ its canonically conjugate momentum, and $\Pi_{W_i}$ the conjugate momentum associated with the wave packet $W_i$ (see Appendix \ref{sec:SmatrixObtention})}. We will call this operator $\hat a_i$ whenever $Q_{\rm KG}(W_i)=1$, and   $-\hat a_i^{\dagger}$ if  $Q_{\rm KG}(W_i)=-1$. 

Using the canonical commutation relations one can check that $[\hat a_i,\hat a_i^{\dagger}]=| Q_{\rm KG}(W_i)|=1$. Hence, these are creation and annihilation operators, whose associated quanta are described by the wave packet $W_i$. We can define in this way a pair of such operators for each of the four wave packets involved in the scattering process. The creation and annihilation operators defined for the left- and right-moving IN modes commute, and the same holds for the OUT modes.
The quantum scattering for a fixed central frequency $\omega$ is thus described by the vector equation
\beq \label{transS} \hat{\bs{A}}_{\rm out}=\bs{S}\cdot \hat{\bs{A}}_{\rm in}\, ,\eeq
where  
\beq \label{vecA} \hat{\bs{A}}_{\rm out}=\left(\begin{array}{cc} \hat a^{\rm out}_{r} \cr \hat a^{\rm out}_{l} \cr \hat a^{\rm out\, \dagger}_{r}\cr \hat a^{\rm out\, \dagger}_{l} \end{array}\right)\, , \ \  \hat{\bs{A}}_{\rm in}=\left(\begin{array}{cc} \hat a^{\rm in}_{r} \cr \hat a^{\rm in}_{l} \cr \hat a^{\rm in\, \dagger}_{r}\cr \hat a^{\rm in\, \dagger}_{l} \end{array}\right)\, ,\eeq
and $\bs{S}$ is a complex $4\times 4$ matrix. As described in some detail in Appendix \ref{app:Gaussian}, $\bs{S}$ is a symplectic matrix. 

Using Eq.~\eqref{eq:afromW}, which establishes the relationship between creation and annihilation operators and wave packets, it is possible to derive the matrix $\bs{S}$ entirely from the elements of the matrix $\bs{B}$ defined above. However, caution is required when connecting these two matrices. This is due to the fact that a wave packet $W_i$ defines either a creation or an annihilation operator depending on the sign of $Q_{\rm KG}(W_i)$, as indicated right below Eq.~\eqref{eq:afromW}. Thus, we must distinguish between non-superradiant (NSR) scattering, for which the sign of $Q_{\rm KG}$ is the same for all wave packets involved, and superradiant (SR) scattering, where that is not the case.  
Taking this into consideration, the matrix $\bs{S}$ takes the form\footnote{{Without loss of generality, we have assumed that $Q(W^{\rm in}_{\rm r})=Q(W^{\rm out}_{l})=-1$ in the superradiant case. Other choices will swap the components of $\bs{S}_{\rm SR}$, but will not alter our conclusions (see Appendix \ref{app:proof} for further details).}}, respectively 
\begin{align} \bS_{\rm SR}= \begin{pmatrix}  
      0 & r & T & 0\\
      R^* & 0 & 0 & t^* \\
      T^* & 0 & 0 & r^*  \\
      0 & t & R & 0   
\end{pmatrix}\,,\,
\bS_{\rm NSR}= \begin{pmatrix}
T & r & 0 & 0 \\
R & t & 0 &0 \\
0 & 0 &T^* & r^* \\
0 & 0 &R^*& t^*   
\end{pmatrix}
 \label{eq:GeneralSuperradiantMatrix}
 \end{align}
where the coefficients $T,R,t$ and $r$ are the elements of $\bs{B}$ as defined in the previous section. 
Upon a mere inspection of these matrices, it becomes evident that superradiant scattering mixes creation and annihilation operators. This mirrors the classical scattering process, which mixes modes with different signs of $Q_{\rm KG}(W_i)$. Conversely, this mixing does not occur in the case of non-superradiant scattering.

Using the constraints satisfied by the coefficients $T,R,t$ and $r$ for superradiant and non-superradiant scattering (written in Appendix \ref{app:proof}) it is easy to see that $\bS_{\rm NSR}$ is a unitary matrix while $\bs{S}_{\rm SR}$ is not. We find, therefore, an application of Theorem 1 to the quantum theory:  the transformations between creation and annihilation operators \eqref{transS} describes superradiant scattering if and only if $\bs{S}$ is not a unitary matrix.

From a physical perspective, the matrices $\bS_{\rm NSR}$ and $\bs{S}_{\rm SR}$ describe very different transformations. The unitary nature of $\bS_{\rm NSR}$ has physical implications: it preserves the vacuum state and also preserves the total number of quanta $\hat N^{\rm in}_{r}+\hat N^{\rm in}_{l}$, thus conserving energy.
$\bS_{\rm NSR}$ functions as a beam splitter.
In contrast, $\bs{S}_{\rm SR}$ does {\em not} preserve either the vacuum state or the total number of particles. It does, however, leave the number difference, $\hat N^{\rm in}_{r}-\hat N^{\rm in}_{l}$, unchanged, implying the creation of excitations in pairs.
$\bs{S}_{\rm SR}$ is a two-mode squeezer,  which converts the vacuum state into a two-mode squeezed vacuum.

While it is well-known that superradiant scattering amplifies incoming field excitations, recognizing that superradiance can be described as a two-mode squeezing process reveals that wave amplification is tied to the generation of entanglement \cite{davies1993QuantumSuperradiance,Agullo:2023pgp}. Hence, entanglement is expected to be generated at the output both for classical and quantum input states. This is a fundamental aspect of this quantum phenomenon which highlights its entanglement-generating capabilities.

\subsection{Generation of entanglement} 

 In the remainder of this section, we quantify the entanglement generated by superradiant scattering for three representative initial states, namely vacuum,  a coherent state, and a single-mode squeezed state. These three states are pure and separable, meaning there is no initial entanglement between the two IN modes.

We consider a coherent state centered at $\langle  \hat a_{r}\rangle =\gamma_r$ and $\langle  \hat a_{l}\rangle =\gamma_{l}$, with  $\gamma_r,\gamma_{l}\in \mathbb{C}$. 

A single-mode squeezed state is obtained by acting with a squeezing operator on one of the two IN modes, for instance, the left-mover $W_l^{\rm in}$
\beq 
 |z_l,\phi_l \rangle_{\rm sqz} = {\ee^{\frac{1}{2}(\xi_l^*\,  \hat a_{l}^2 - \xi_l\,  \hat a_{l}^{\dagger}{}^2)} } |0\rangle \,  ,
\eeq
where $\xi_l=z_l\ee^{i \phi_l}$, and $\phi_l,z_l$ are real numbers representing the squeezing angle and intensity, respectively.

We evolve these  Gaussian states using $\bs{S}_{\rm SR}$, and quantify entanglement between the two OUT modes on either side of the potential barrier via the entanglement entropy $S_{\rm ent}(\omega)$. For this calculation, we use the tools spelled out in Appendix \ref{app:Gaussian}.\footnote{In section \ref{sec:ProductionOfEntanglement}, we extend the analysis presented here by considering thermal noise at the input as well as losses and detector inefficiencies.}

First, we calculate the flux (number of quanta per unit time and unit bandwidth) in each of the outgoing modes in terms of the elements of $\bs{S}_{\rm SR}$.
For the two-mode vacuum, we find
\beq \langle \hat N^{\rm out}_r\rangle =\langle \hat N^{\rm out}_{l}\rangle =|T|^2
\label{eq:NquantVac}. \eeq 
For the two-mode coherent state, we obtain 

\begin{align}
    \langle \hat{N}_r^{\rm out}\rangle=&|\gamma_l|^2+|T|^2+ \label{eq:NquantrCoh}\\&+|T|^2\lr{|\gamma_r|^2+|\gamma_l|^2}+2\text{Re}\lrsq{rT^*\gamma_r\gamma_l}\,,\nonumber\\
    \langle \hat{N}_l^{\rm out}\rangle=&|\gamma_r|^2+|T|^2+\label{eq:NquantlCoh}\\
    &+|T|^2\lr{|\gamma_l|^2+|\gamma_r|^2}+2\text{Re}\lrsq{tR^*\gamma_l\gamma_r}\nonumber.
\end{align}
Finally, for the one-mode squeezed  state
\begin{align}
    &\langle \hat{N}_r^{\rm out}\rangle=\sinh^2{z_l}+|T|^2+|T|^2\sinh^2{z_l}\,,\label{eq:NquantrSq}\\
    &\langle \hat{N}_l^{\rm out}\rangle=|T|^2+|T|^2\sinh^2{z_l}\,\label{eq:NquantlSq}.
\end{align}
In Eqs.~\eqref{eq:NquantrCoh} and \eqref{eq:NquantlCoh}, the first term corresponds to the finite number of quanta already present at the input, while the second term, equal to  $|T|^2$, represents the amplification of vacuum fluctuations \eqref{eq:NquantVac}.
The third and fourth terms represent stimulated emission (amplification) due to superradiance. Likewise, in Eqs.~\eqref{eq:NquantrSq} and~\eqref{eq:NquantlSq}, $|T|^2$ corresponds to amplification of vacuum fluctuations while $|T|^2\sinh^2{z_l}$ manifests the amplification of the initial quanta in mode $W_l^{\rm in}$ present in the initial state. 

\begin{figure}
    \centering\includegraphics[width=0.47\textwidth]{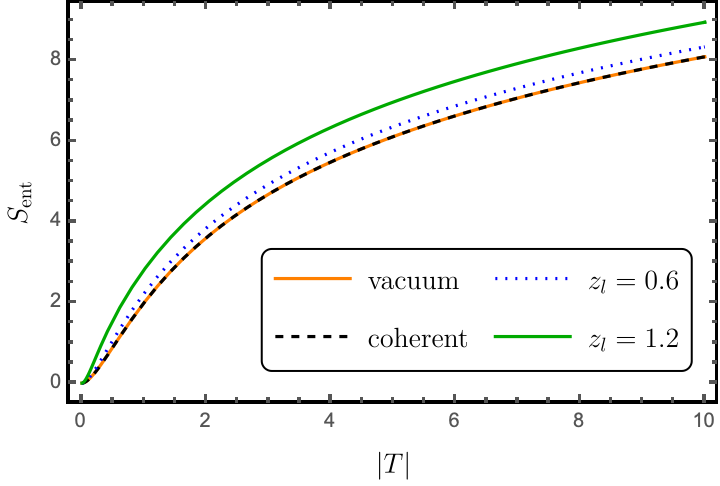}
    \caption{Entanglement entropy, $S_{\rm ent}(\omega)$, of the OUT modes as a function of the absolute value of the scattering coefficient $R$. We plot the result for different inputs: vacuum, an arbitrary coherent state (this plot is insensitive to $\gamma_{r}$ and $\gamma_{l}$), and states where the left IN mode is in vacuum but the right one is squeezed with squeezing intensity $z_{l}$ (the plot is insensitive to the squeezing angle $\phi_{l})$. To produce the plot, we have assumed, without loss of generality, that $Q(W^{\rm in}_{\rm r})=Q(W^{\rm out}_{l})$, so that $|R|^2\geq 1$.}
    \label{fig:EntropyGeneralSR}
\end{figure}

Importantly, not all created pairs at the output are entangled: superradiant scattering does not yield the same emission statistics for all input states. Fig.~\ref{fig:EntropyGeneralSR} shows the entanglement entropy of the output  $S_{\rm ent}(\omega)$ (i.e., the von Neumann entropy of either of the two output modes) as a function of $|T|$ for the three input states (and two values of initial squeezing in $W_l^{in}$).
On the one hand, we see that $S_{\rm ent}(\omega)$ is identical for vacuum and coherent input states.
This implies that, even though the amplitude of the coherent state gets enhanced after the scattering, entanglement at the output is due to the amplification of vacuum fluctuations only.
On the other hand, the amplification of single-mode squeezed states is different: $S_{\rm ent}(\omega)$ is higher than for the other two input states for any non-zero value of $z_l$.
So, in this case, entanglement at the output does not come exclusively from vacuum amplification  ---single-mode squeezing at the input enhances entanglement at the output.
Hence, we observe that superradiant scattering not only amplifies incoming radiation, but also its quantum properties.\footnote{A similar effect occurs in pair production from causal horizons, i.e., Hawking radiation.}

\section{Stable Ergoregions in Polariton Fluids}\label{sec:StableErgoregions}

Because of the spontaneous amplification of vacuum fluctuations by rotational superradiance, ergoregions are intrinsically dynamically unstable~\cite{Friedman:1978,comins_ergoregion_1978,Giacomelli:2020evu}.
In this section,  we propose that dissipative dynamics can quench this instability.
We evidence this numerically by engineering a stable ergosurface in a driven-dissipative fluid of light realized with microcavity exciton-polaritons (polaritons).

Polaritons are half-light half-matter bosonic quasi-particles resulting from the strong coupling of photons in a semiconductor cavity with an excitonic transition~\cite{carusotto_quantum_2013}.
Their dynamics in the cavity plane are intrinsically driven-dissipative, with steady-states reached only through constant pumping with a continuous wave laser.
All properties of the quantum fluid in the cavity are controlled by the pump laser and measured via photons exiting the cavity.

\subsection{Microcavity polaritons}\label{subsec:polaritons}
We consider a laser field inside a semiconductor microcavity made of quantum wells sandwiched between two Bragg mirrors.
Under the resonance-induced quantization of the photon wavevector perpendicular to the cavity, photons acquire an effective mass 5 orders of magnitude lower than the mass of the electron.

\begin{figure}
    \centering
    \includegraphics[width=.9\columnwidth]{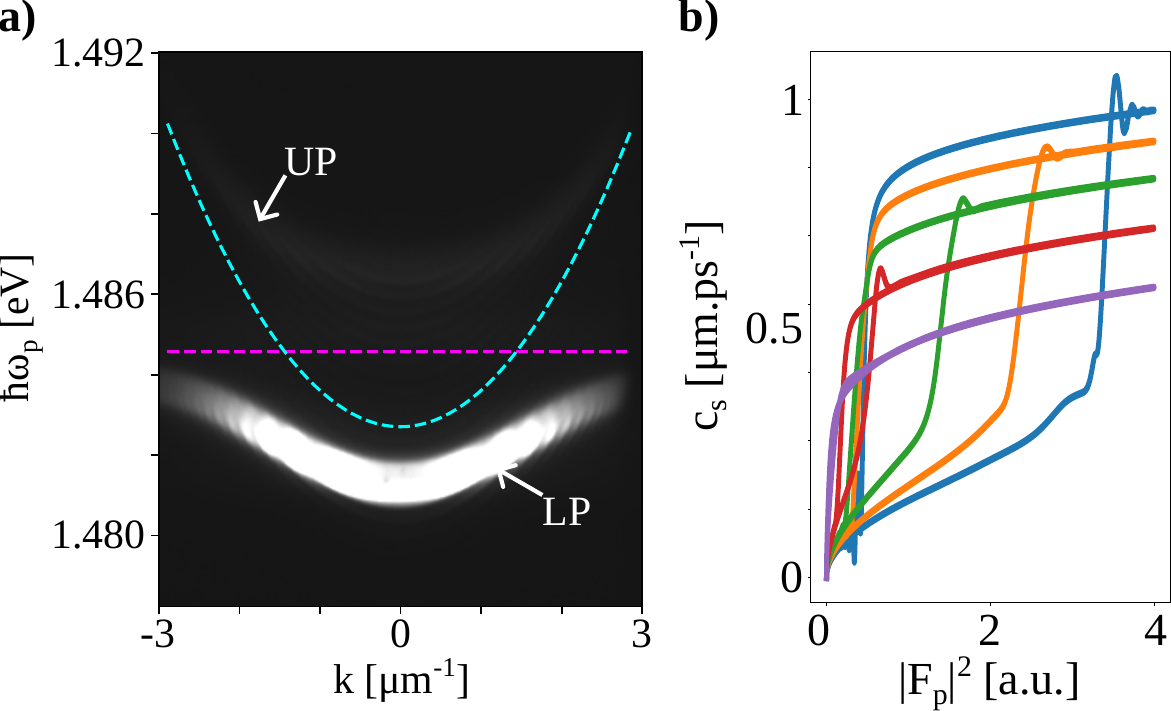}
    \caption{Microcavity polaritons.
    \textbf{a)} Dispersion of the light coming from the cavity (experimental measurement). Dashed turquoise: cavity photons; dashed purple: excitons; UP: upper polaritons; LP: lower polaritons. Exciton-photon energy difference $-\SI{1.88}{\milli\electronvolt}$.
    \textbf{b)} Optical bistability of the LP for a homogeneous fluid with $\delta_{k_p=0}=\SI{0.618}{\milli\electronvolt}$ incident on the cavity at different $k_p$ (numerical calculation): blue, $k_p=\SI{0}{\per\micro\meter}$; orange, $k_p=\SI{0.2}{\per\micro\meter}$; green, $k_p=\SI{0.3}{\per\micro\meter}$; red, $k_p=\SI{0.4}{\per\micro\meter}$; purple, $k_p=\SI{0.5}{\per\micro\meter}$.
    }
    \label{fig:polaritons}
\end{figure}

Cavity photons excite electric dipoles (bound electron-hole states) called excitons, which are massive self-interacting excitations.
The strong coupling of photons and excitons generates two new eigenstates of the system: two polariton branches with different dispersive properties, the upper and the lower polaritons (see Fig.~\ref{fig:polaritons}).
In practice, only the lower polaritons LP are excited and measured in experiments.\footnote{Since the Rabi energy $\hbar \Omega_R$, controlling the gap between the upper and lower polariton branches, is much larger than the characteristic energies of the system, in particular the interaction energy $\hbar g n$, the fluid contains only lower polaritons, so its wavefunction may be truncated to that of the lower polaritons $\psi(\bs{r}, t) = \psi_{LP} (\bs{r}, t)$.}

Polaritons inherit the properties of cavity photons and excitons, and thus behave collectively as an ensemble of self-interacting massive particles ---a quantum fluid of light~\cite{carusotto_quantum_2013,kasprzak_boseeinstein_2006,balili_bose-einstein_2007,lagoudakis_quantized_2008,utsunomiya_observation_2008,amo_superfluidity_2009}.

Because of the decay of excitons and cavity photons, polaritons are unstable and decay with a rate $\gamma$, yielding photons exiting the cavity at the same rate.
These photons carry all the information of the phase and intensity of the intra-cavity polariton field, allowing us to measure the quantum properties of the fluid with optical detectors.

The mean field $\psi(\bs{r}, t)$ of the polariton fluid is governed by the driven-dissipative Gross-Pitaevski equation~\cite{carusotto_quantum_2013}
\begin{eqnarray}
    \ci\hbar\frac{\partial\psi}{\partial t} &=& \left(\hbar\omega_{LP} -\frac{\hbar^2\nabla^2}{2m_{LP}}+ \hbar g|\psi|^2 -\ci\hbar\frac{\gamma}{2}\right)\psi \nonumber\\
    &+& \ci\hbar \mathcal{F} (\bs{r}, t),
    \label{eq:GPEpol}
 \end{eqnarray}
where $\omega_{LP}$ is the frequency of polaritons at wavevector $\bs{k}=0$, $-\hbar^2\nabla^2/2m_{LP}$ their kinetic energy in the cavity plane, $m_{LP}$ their effective mass, $\hbar g$ the strength of repulsive polariton self-interactions and $\mathcal{F} (\bs{r}, t) = F_p(\bs{r})\ee^{\ci\left(\phi_p(\bs{r})-\omega_p t\right)}$ represents the electromagnetic field of the laser pump.

Eq.~\eqref{eq:GPEpol} can be rewritten in terms of the density $n$ and phase $\phi_{LP}$ of the fluid by $\psi=\sqrt{n}\ee^{\ci\phi_{LP}}$, which leads to its hydrodynamical ---or Madelung--- representation. The resulting equations are the continuity and Euler equations for a fluid with density $n$ and velocity $\bs{\tv}(\bs{r})=\hbar\bs{\nabla}\phi_{LP}/m_{LP}$.

If the laser frequency $\omega_p$ is near resonance with the lower polariton branch, the steady state of the polariton fluid is controlled by the pump.
Concretely, its phase is inherited from the pump, $\psi(\bs{r}, t) = \sqrt{n(\bs{r})}\ee^{\ci\left(\phi_p(\bs{r})-\omega_p t\right)}$, and its density is related to the intensity of the pump and the detuning according to Eq.~\eqref{eq:GPEpol}.
In particular, for a homogeneous steady-state configuration excited by a plane wave at wavenumber $k_p$, in the rest frame of the fluid we find the equation of state
\begin{equation}
n \left[ \left(\frac{\gamma}{2} \right)^{2}+\left( gn - \delta \right) ^{2}\right] = |F_p|^{2},
\label{eq:bistab}
\end{equation}
with $\delta(\bs{r})=\omega-\omega_{LP}-\hbar k_p^2(\bs{r})/2m_{LP}$ the effective detuning. 
The number of solutions to this equation crucially depends on the value of this detuning: for $\delta>\sqrt{3}\gamma/2$, there can be either one or three solutions depending on the pump intensity.
This leads to a hysteresis loop, shown in  Fig.~\ref{fig:polaritons} \textbf{b)}, where the fluid can either be in a high or low density regime for some values of the pump intensity, known as optical bistability~\cite{baas_bista_2004}.

Collective excitations of the polariton fluid are studied by linearizing Eq.~\eqref{eq:GPEpol} via $\psi(\bs{r}, t) = \left(\psi^0 + \delta \psi\right)\ee^{\ci\left(\phi_{p}(\bs{r})-\omega_{p}t\right)}$.
Their spectrum in a spatially homogeneous region is given by the Bogoliubov dispersion relation, which is a quadratic form admitting two $\{ k,\omega\}$ solutions at positive laboratory-frame frequencies $\omega$~\cite{carusotto_quantum_2013,claude_2022}.
While these roots can be complex, when they are real there are two frequency solutions at each wavenumber~\cite{carusotto_quantum_2013,claude_2022}: positive (negative) rest-frame frequency modes have a positive (negative) norm~\eqref{eq:afromW}~\cite{macher_blackwhite_2009}.

The inhomogeneous mean field $\psi^0\ee^{\ci\left(\phi_{p}(\bs{r})-\omega_{p}t\right)}$ sets the kinematics of the collective excitations $\delta\psi$ (which effectively behaves as a Klein-Gordon field)~\cite{visser_acoustic_1998,jacquet_analogue_2022}.
Depending on the fluid velocity profile $\bs{\tv}(\bs{r})$ and the speed of sound $c_s(\bs{r})=\sqrt{\hbar (2gn^0(\bs{r})-\delta(\bs{r}))/m}$ in the inhomogeneous flow ($n^0$ the mean-field density associated to $\psi^0$), we can find configurations in which $\delta\psi$ experiences a scattering process as the one described in Eqn~\eqref{eq:GeneralSuperradiantMatrix}.
Specifically, there exist positive laboratory-frame frequencies at which there are two positive-norm ($Q_{\rm KG}>0$) modes in some region of space and two negative-norm ($Q_{\rm KG}<0$) modes in some other region of space.
In addition, in each spatial region, the two modes always have opposite sign of group velocity.
This configuration entails supperadiant scattering.

In the following subsection, we  show a simulation of a stable rotating polariton fluid with a horizonless ergoregion. Entanglement between collective excitations generated by superradiance is then quantified in section~\ref{sec:RSPolaritons}.

\begin{figure}
    \centering
    \includegraphics[width=.9\linewidth]{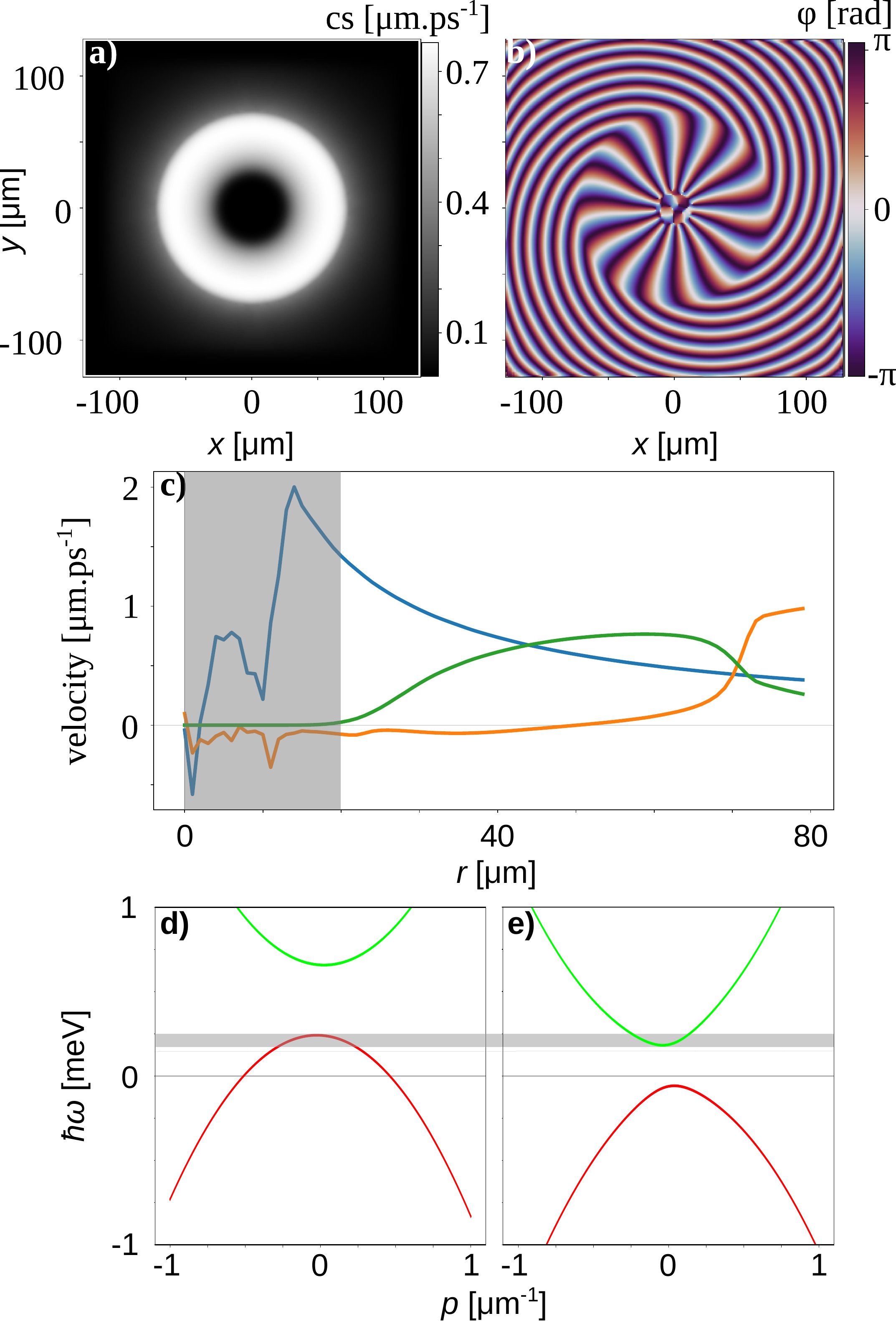}
    \caption{Vortex flow in a polariton fluid with $D=0$ and $C=15$. \textbf{a)}, \textbf{b)} and \textbf{c)} numerical simulation of Eq.~\eqref{eq:GPEpol}. \textbf{a)} speed of sound $c_s$; \textbf{b)}, phase $\phi_{LP}$; \textbf{c)} Velocities versus a radius. Blue: $v_\theta$; orange: $v_r$; green: $c_s$.
    Gray area: weak nonlinear interactions.
    Corresponding $\omega-p$ dispersion~\eqref{eq:spectrum} for $\ell=10$ at \textbf{d)} $r=\SI{25}{\per\micro\meter}$ and  \textbf{e)} $r=\SI{67}{\per\micro\meter}$. Green: positive-norm modes; red: negative-norm modes.
    Gray area: frequency interval for rotational superradiance in this configuration.}
\label{fig:fig_num_vort_pol}
\end{figure}

\subsection{Numerical simulations}\label{subsec:polexp}
Control of the fluid phase and density via the properties of the pump laser allows us to engineer polariton vortices by shaping the pump field into an optical vortex with $\phi_p(r,\theta)=D(r)+C\theta$, with $(r,\theta)$ being polar coordinates in the plane and where we have imposed rotational symmetry. The function $D(r)$ and the parameter $C$ control the flow velocity, whose target velocity profile is
\begin{equation}
    \label{eq:velocityfield}
        \bs{\tv}(r)=\frac{\hbar}{m_{LP}}\left(\frac{d D}{dr}\hat{\bs{r}}+\frac{C}{r}\hat{\bs{\theta}}\right),
\end{equation}
where hats denote unit vectors.

We numerically simulate the polariton dynamics described by Eq.~\eqref{eq:GPEpol}  {with the choice $D=0$ (this can be implemented by tuning the pump profile),} with  $m_{LP}=5.7\times 10^{-35}$kg and a polariton linewidth $\hbar\gamma=\SI{0.08}{\milli\electronvolt}$.
Fig.~\ref{fig:fig_num_vort_pol} shows the numerical data. 

The fluid density (we show $c_s(\bs{r})$ as a proxy to $n$) assumes the annular shape characteristic of polariton vortices.
The density is high enough to enable the nonlinear regime of interactions over $r\in\left[20,67\right]\SI{}{\micro\meter}$.
{ Note that, due to  crossing of the radial and sound velocities at $r=\SI{71}{\micro\meter}$, one could worry about the presence of an acoustic horizon emitting Hawking radiation, which could interfere with the superradiant effect we are interested in. However, in the region where this horizon could form, the density of the fluid is extremely low, and consequently there is no significant Hawking radiation.} 

The phase $\phi_{p}(r,\theta)$, shown in Fig.~\ref{fig:fig_num_vort_pol} \textbf{b)}, features 15 jumps from 0 to $2\pi$, indicating finite circulation of the phase along any given radius. 
The negligible curvature of phase jumps along $r$ in the high-density region manifests the near zero radial velocity $\tv_r$ of the fluid, as confirmed by the corresponding orange line nearing zero in Fig.~\ref{fig:fig_num_vort_pol} \textbf{f)}.
This implies that, in this configuration, $\bs{\tv}=v_\theta\hat{\bs\theta}$, so that in the relativistic analogy~\cite{visser_acoustic_1998}, the acoustic ergosurface in the simulation lies at $r_e=r_{v_\theta=c_s}=\SI{43}{\micro\meter}$.

The spectrum of collective excitations in this fluid is approximately described (locally) by a dispersion relation of the form~\cite{carusotto_quantum_2013}
\begin{multline}
        \omega-\bs{\tv}\cdot\bs{k}=-\frac{\ci\gamma}{2}
        \\
        \pm \sqrt{\frac{\hbar^2}{4 m^2_{LP}}k^4+c_s^2k^2+\lr{g n^0 - \delta}\lr{3 g n^0 - \delta}},
\label{eq:spectrum}
\end{multline}
¡where $k=|\bs{k}|$ and, for radial waves, the wavevector is of the form $\bs{k}=(p,\ell/r)$, where $\ell$ and $p$ are the azimuthal and radial wavenumbers --- see appendix \ref{app:dispbogo} for a detailed derivation of \eqref{eq:spectrum}.

 We use the Truncated Wigner approximation to simulate the full quantum dynamics of the polariton fluid. In this approximation, noise is added at every simulation step~\cite{carusotto_quantum_2013}. Under these circumstances, a steady state can only be reached if the system is dynamically stable against perturbations. Our numerical simulations show that the fluid reaches a steady state.
Furthermore, the existence of a steady state allows us to numerically extract the spectrum of collective excitations, which agrees well with Eq.~\eqref{eq:spectrum} (see appendix~\ref{app:dispbogo} for further details). The spectrum has two branches $\omega_\pm$, which correspond to modes with $Q_{\rm KG}(\omega_+)=1$ and $Q_{\rm KG}(\omega_-)=-1$, shown in green and red, respectively, in Fig.~\ref{fig:fig_num_vort_pol} \textbf{d)}-\textbf{e)}.

In our driven-dissipative fluid, the frequency detuning $\delta$ contributes to a gap between positive- and negative-norm modes (see Appendix~\ref{app:dispbogo} for an extended discussion on the spectrum and the gap). Due to this gap, propagating/oscillating modes with negative norm and positive lab frequencies exist only in the region $r\leq\SI{40}{\micro\meter}<r_{e}$.
The dispersion relation shown in Fig.~\ref{fig:fig_num_vort_pol} \textbf{d)}-\textbf{e)}  indicates that superradiant scattering between positive- and negative-norm modes is possible across the ergosurface inside the frequency interval $\omega\in\left[\mathrm{min}(\omega_+),\mathrm{max}(\omega_-)\right]$ (gray area in Fig.~\ref{fig:fig_num_vort_pol} \textbf{d)}-\textbf{e)}). 

It may be surprising at first that a steady state is reached despite the presence of superradiant scattering in this horizonless ergoregion configuration. We conclude this subsection by providing an explanation for the physical origin of the observed stability.

Within the superradiant frequency window, negative-norm modes in the ergoregion always have a finite amplitude, even in the idealized case of zero temperature, due to vacuum fluctuations. These excitations propagate with a velocity comparable to $c_s\leq\SI{0.7}{\micro\meter\per\pico\second}$. Given that the value of $\gamma$ corresponds to a lifetime of $\approx\SI{8}{\pico\second}$, the mean free path of these excitations is $\approx\SI{6}{\micro\meter}$. This implies that, for the same reason that the polariton density drops towards short radii, negative-norm excitations will not be able to propagate across the vortex and reach the ergosurface on the other side. The finite lifetime of polaritons quenches the characteristic instability of horizonless ergoregions.\footnote{In our configuration, the phase is set by the resonant pump, so the stabilization dynamics are different from configurations in which the phase is left free, e.g., in the formation of multiply charged vortices upon Bose-Einstein condensation~\cite{alperin_multiply_2021}.}

\section{Rotational superradiance in a quantum fluid of light}\label{sec:RSPolaritons}

In this section, we will compute the matrix $\bs{B}$ describing wave scattering by numerically solving the evolution equation of collective excitations in a model with a simple velocity profile modeling the shape shown in Fig.~\ref{fig:fig_num_vort_pol}. These calculations will demonstrate the presence of superradiant scattering in the corresponding frequency window, confirming the insights gained from the dispersion relation analysis in the previous section. 

In the hydrodynamic limit, the scalar field, denoted as $\delta\psi$, describing these excitations as they propagate on a polariton fluid  follows the equation~\cite{jacquet_polariton_2020}
\beq \label{eqpert}
\begin{split}
  &\Bigg[(\partial_t+\bs{\tv} \cdot \bs \nabla)^2-c_s^2 \bs{\nabla}^2 \\
  &+c_s^2 \left(\bs{\nabla}\cdot \frac{\bs \tv}{c_s^2}\right) (\partial_t+\bs{\tv} \cdot \bs{\nabla})\Bigg]\delta\psi(t,r,\theta)=0\,
\end{split}
\eeq 
{which in the WKB approximation leads to dispersion \eqref{eq:spectrum} with $c_s=\sqrt{g n/m}$, and we have taken $\delta=gn$.
Assuming a constant speed of sound $c_s$ and a velocity profile $\bs{v}(r)=v_{\theta}(r)\, \hat \theta$}, Eq.~\eqref{eqpert} simplifies to
\beq \label{simpeqpert}
\left [(\partial_t+\bs v \cdot \bs \nabla)^2-c_s^2 \Delta \right ]\delta\psi(t,r,\theta)=0\, .
\eeq 
These equations suggest defining the time coordinate $\tilde t$ associated with a frame co-rotating with the fluid, such that $\partial_{\tilde t}=\partial_{t}+\bs v \cdot \bs \nabla=\partial_{t}+\frac{v_{\theta}}{r} \partial_{\theta}$.

Solutions to \eqref{simpeqpert} are combinations of cylindrical waves  of the form 
\begin{equation}
\delta\psi_{\omega,\ell}(t,r,\theta)=\ee^{-\ci \omega t}\ee^{\ci \ell\theta}\, \varphi_{\omega\ell}(r),    
\end{equation} 
with $\varphi_{\omega\ell}(r)$ a solution of the radial differential equation 
\begin{equation}
\lrsq{\frac{1}{r}\frac{d}{dr}\lr{r\frac{d}{dr}}+V_{\omega\ell}(r)} \, \varphi_{\omega\ell}(r)=0\, ,
\label{eq:SpatialODE}
\end{equation}
where
\begin{equation}
V_{\omega\ell}(r)=\frac{1}{c_s^2}\lr{\omega-\frac{v_\theta(r)\ell}{r}}^2-\frac{\ell^2}{r^2}.
\end{equation}
The symmetries of the problem ensure the conservation of both $\omega$ and $\ell$. Therefore, modes characterized by different $\omega$ or $\ell$ values remain decoupled throughout their evolution, and it suffices to describe each of them individually. We drop these labels from now on to simplify the notation.

From Eq.~\eqref{eq:SpatialODE}, we see that wave scattering in this configuration reduces to a  one-dimensional problem in the radial direction described by the effective potential $V_{\omega\ell}(r)${, which can be written as}
\beq   (W^{\rm in}_{r}, W^{\rm in}_{l}) \xrightarrow{\text{time}} (W^{\rm out}_{r}, W^{\rm out}_{l})\cdot\bs{B}\, .  \eeq
with ``$r$'' and ``$l$'' corresponding to propagation in the radial direction, denoting local out- and inward propagating modes respectively.

\begin{figure}
    \centering
    \includegraphics[width=0.43\textwidth]{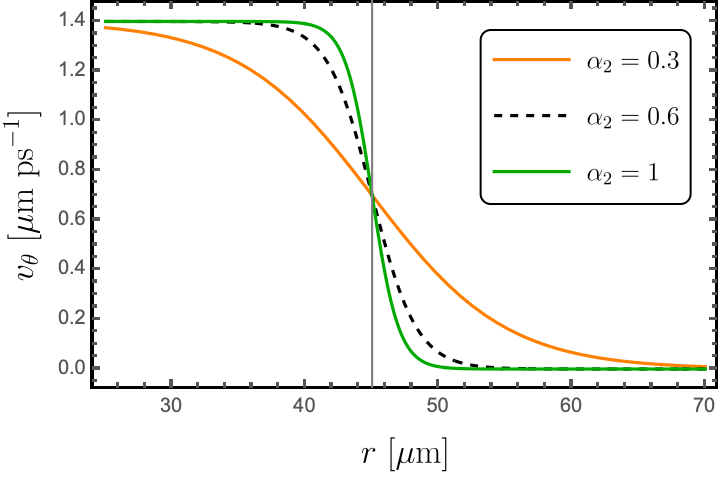}
    \caption{Velocity profile in \eqref{vtheta} with $\alpha_1=2c_s$, with $c_s=0.7\mu$m ps$^{-1}$, and $\alpha_3=45\mu$m. The grey vertical line signals the position of the ergosurface.}
    \label{fig:VthetaProfile}
\end{figure}

{To facilitate the computation of the coefficients of matrix $\bs{B}$,} we model the velocity profile $v_\theta/c_s$ across the ergosurface with
\beq \label{vtheta}
v_{\theta}(r)=\frac{\alpha_1}{2}\left[1+ \tanh{\alpha_2 (\alpha_3-r)}\right]\, ,
\eeq
which is parameterized by three constants, $\alpha_i$, $i=1,2,3$. 
As shown in Fig.~\ref{fig:VthetaProfile}, the variation from $v_{\theta}=\alpha_1$ for $r\ll\alpha_3$ to $v_{\theta}\to 0$ as $r\gg \alpha_3$ is monotonous, while  $\alpha_2$ governs the sharpness of the transition between the two asymptotic values of $v_{\theta}$.
Choosing $\alpha_1$ exceeding $c_s$  ensures the existence of an ergoregion at small radii, with its boundary situated at $r_{e}$, defined by $v_{\theta}(r_{e})=c_s$. 
We choose $\alpha_1=2c_s$ with $c_s=0.7\mu$m ps$^{-1}$ and $\alpha_3=45\mu$m, as suggested by the values in Fig.~\ref{fig:fig_num_vort_pol}, and $\alpha_2=0.3\mu$m$^{-1}$. 
This velocity profile is relevant only for radii greater than $r_{\rm min}$, below which the fluid density is insufficient to support the propagation of acoustic waves. 

The conserved quantity~\eqref{eq:KGE} associated with the equation of motion~\eqref{simpeqpert} reads
\beq 
Q_{\rm KG}(\phi)=\ci/\hbar\, \int^{\infty}_0dr\int^{2\pi}_0d\theta\, \frac{r}{c_s^2} \big[\phi^*\, \partial_{\tilde t}\phi-\partial_{\tilde t}\phi^*\, \phi\Big]\, ,
\eeq 
implying that modes with $\tilde \omega>0$ ($\tilde \omega<0$)  have positive (negative) norm $Q_{\rm KG}$. {Here, $\tilde\omega$ is defined as the frequency associated with $\partial_{\tilde t}$. Concretely, in our setup with cilyndrical symmetry and vanishing radial velocity $v_r=0$, we have $\tilde\omega=\omega-\frac{v_\theta(r)}{r}\ell$.}

Given that $v_{\theta}$ diminishes for large radii, in this asymptotic region ($r\gg r_{e}$) we have  $\partial_{\tilde t} =\partial_t$, and $\tilde \omega=\omega$. Consequently, the sign of $Q_{\rm KG}$ for wave packets initially localized at large radii is determined by the sign of $\omega$. This scenario applies to the wave packets $W^{\rm in}_{l}$ and $W^{\rm out}_{r}$, since they are localized at large radii at early and late times, respectively. Since $Q_{\rm KG}$  remains conserved throughout their evolution, these wave packets bear positive $Q_{\rm KG}$ at all times when $\omega>0$. 

Conversely, for wave packets localized in the vicinity of $r\approx r_{\rm min}$, { the relation between $\tilde \omega$ and $\omega$ is instead} $\tilde \omega=\omega -\frac{v_{\theta}(r_{\rm min})}{r_{\rm min}} \, \ell$. Consequently, these modes possess negative $Q_{\rm KG}$ values when  $\left(\omega -\frac{v_{\theta}(r_{\rm min})}{r_{\rm min}} \, \ell\right)<0$.  This applies to the modes $W^{\rm in}_{r}$ and $W^{\rm out}_{l}$, which  are localized near $r_{\rm min}$ at early and late times, respectively. 

In brief, whenever $\omega$ and $\left(\omega -\frac{v_{\theta}(r_{\rm min})}{r_{\rm min}} \, \ell\right)$ have different signs, we expect superradiant scattering to occur {near} the ergosurface.

Next, we proceed to solve the scattering process between the IN and OUT wave packets. Exact solutions to the radial Eq.~\eqref{eq:SpatialODE} exist when the velocity $v_{\theta}$ remains constant along the radial direction. In such a scenario, a basis of solutions is provided by Whittaker functions
\begin{equation}
    \begin{split}
&\frac{1}{\sqrt{r}}\text{WhittakerM}\lr{\ci\ell\frac{v_\theta}{c},\ell\sqrt{1-\frac{v_\theta^2}{c^2}},\ci\frac{2\omega}{c} r}\,,\\
&\frac{1}{\sqrt{r}}\text{WhittakerW}\lr{\ci\ell\frac{v_\theta}{c},\ell\sqrt{1-\frac{v_\theta^2}{c^2}},\ci\frac{2\omega}{c} r}\,,
\end{split}
\label{eq:Whittakers}
\end{equation}
which reduce to Bessel functions in the homogeneous limit $v_\theta\to0$. (In the definition of these functions, we adhere to the same conventions as those used in the software \texttt{Mathematica}.)

At a given radius $r$, these solutions can have either locally oscillatory or evanescent behavior, depending on the value of $\omega -\frac{v_{\theta}(r)}{r} \, \ell$. The behavior of a mode $(\omega,\ell)$ is determined by the WKB dispersion relation, which has a position-dependent gap equal to $2c_s \ell/r$ (see Appendix~\ref{app:dispbogo}). To a good approximation, if the frequency of a mode falls within this gap, it will be an evanescent mode, and wave packets constructed from it do not propagate energy. Evanescent modes are not part of the IN and OUT bases. 

On the other hand, for propagating modes,  negative-norm modes at inner radii have frequency  $\omega -\frac{v_{\theta}(r_{\rm min})}{r_{\rm min}} \, \ell<-\frac{c_s}{r_{\rm min}}\ell$, while positive-norm modes at $r_{\rm max}\gg\alpha_3$ have frequency $\omega -\frac{v_{\theta}(r_{\rm max})}{r_{\rm max}} \, \ell<-\frac{c_s}{r_{\rm max}}\ell$. In our system, the laboratory-frame frequency window of superradiant scattering is thus $6.54\, \ell<\hbar\omega<18.31\, \ell$ (in \SI{}{\micro\electronvolt}).

The velocity profile \eqref{vtheta} remains { almost uniform in}
$r$, except in the vicinity of $r=\alpha_3$. Consequently, we can employ exact solutions \eqref{eq:Whittakers} to define the IN and OUT modes within the asymptotic regions  $r\sim r_{\rm min}$ and $r\gg \alpha_3$, respectively. Purely incoming or outgoing waves are represented by specific linear combinations of these exact solutions, and in each of the asymptotic regions take the form
\begin{equation}\label{assymp}
\begin{split}
    &\exp\left\{\pm\ci\lr{\frac{\omega r}{c}-\frac{\ell v_\theta(r_{\rm min})}{c_s}\log{\frac{r}{r_{\rm min}}}}\right\}\qquad {\rm for} \ r\to r_{\rm min}\,, \\
&\exp\left\{\pm\ci\lr{\frac{\omega r}{c}-\frac{\ell v_\theta(r_{\rm max})}{c_s}\log{\frac{r}{r_{\rm max}}}}\right\}\qquad {\rm for} \ r\to r_{\rm max}\, ,
\end{split}
\end{equation}

We construct the wave packets $W^{\rm in}_{r}$, $W^{\rm in}_{l}$, $W^{\rm out}_{r}$, and $W^{\rm out}_{l}$, each associated with an azimutal number $\ell$ and narrowly centered at  frequency $\omega$, from these radial functions multiplied by  $\ee^{-\ci \omega t}\ee^{\ci \ell\theta}$. We then proceed to numerically solve the evolution of the IN wave packets with the velocity profile~\eqref{vtheta} (see Appendix~\ref{sec:SmatrixObtention} for details). 

Numerical calculations are needed because of the absence of closed analytical solutions for the radial differential equations corresponding to the velocity profile~\eqref{vtheta}.
We numerically solve the scattering problem across a parameter space grid encompassing  $(\omega, \ell$). As explained above, we consider frequencies in the range $6.54\, \ell<\omega<18.31\, \ell$ in $\mu$eV for each $\ell$, which remain well within the hydrodynamical regime, and for which the modes scatter superradiantly. For illustrative purposes, we consider integer values of $\ell$ within the range $[1,4]$.

The outcomes of these calculations are the elements of the scattering matrix
\beq \label{S} \bs{B}_{\omega\ell}=\left(\begin{array}{cc} T_{\omega\ell} & r_{\omega\ell}\cr R_{\omega\ell} & t_{\omega\ell}\end{array}\right)\, .\eeq
The resulting values of these coefficients corresponding to the selected values of $\omega$ and $\ell$ are displayed in  Table \ref{tab:TableData} in Appendix \ref{sec:SmatrixObtention}. All figures in the subsequent section are based on these calculated values. As an illustrative example, we present here the scattering coefficients corresponding to $\omega=17.00\mu$eV and $\ell=1$:
\beq
\bs{B}_{\omega\ell}
=
\begin{pmatrix}
    1.377+0.990\ci & 1.296+0.444\ci\\
    -1.296-0.444\ci & -1.695-0.069\ci
\end{pmatrix}
\, .
\eeq
This is {\em not} a unitary matrix, which confirms, in accordance with Theorem 1 in Section~\ref{ClassicalSR}, that the scattering process exhibits superradiance.\footnote{As an additional verification, we have confirmed that, if we extend our calculations to higher frequencies, the numerically obtained scattering matrix indeed becomes unitary outside of the superradiant regime, serving as a  test of our theoretical framework.} Furthermore, one can check that these numerically obtained coefficients satisfy the constraints in~\eqref{eq:SupConstraints}.

\section{Superradiant Production of Entanglement}\label{sec:ProductionOfEntanglement}

The calculations in the previous section yield the scattering coefficients $T_{\omega\ell}, R_{\omega\ell}, t_{\omega\ell},$ and $r_{\omega\ell}$ for the relevant range of $\omega$ and $\ell$ ---the superradiant modes within the hydrodynamical approximation. By substituting these coefficients into expression \eqref{eq:GeneralSuperradiantMatrix}, we derive the matrix $\boldsymbol{S}^{\omega\ell}_{SR}$, which describes the quantum scattering process for each mode $(\omega,\ell)$. This matrix provides all the information necessary to analyze the scattering, once the initial quantum state is specified. The primary objective of this section is to compute the entanglement between the two OUT modes following the scattering, considering a family of physically interesting initial states.

The content of this section shares certain similarities with the calculations performed in section \ref{sec:QuantumSR}, with a significant distinction being that, while the description in section \ref{sec:QuantumSR} was generically applicable to any superradiant process,  
we focus here on the specific rotating quantum fluid examined in the preceding sections. We also make use of the velocity profile depicted in Fig.~\ref{fig:VthetaProfile}.

For the initial state of acoustic perturbations, we first consider a thermal state at temperature $T_{\rm env}$, in order to incorporate thermal noise, a common element in real experiments. It will become evident that the temperature $T_{\rm env}$  influences the entanglement generated during the scattering in an important manner.

\begin{figure*}
  \centering
  \includegraphics[width=.45\linewidth]{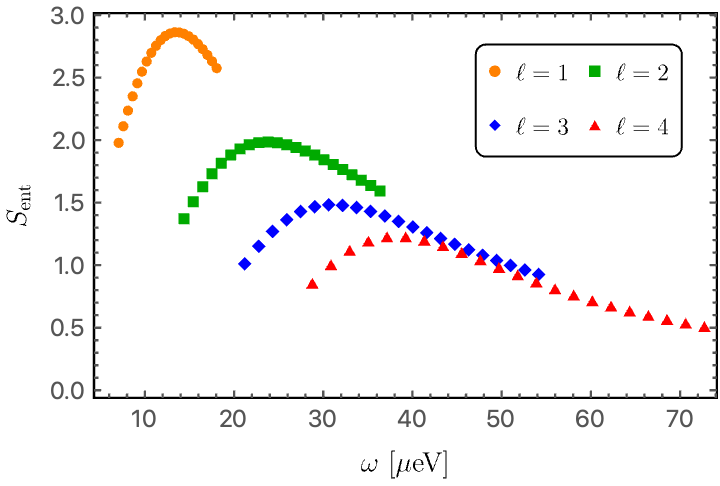}
  \hspace{.5cm}
  \includegraphics[width=.45\linewidth]{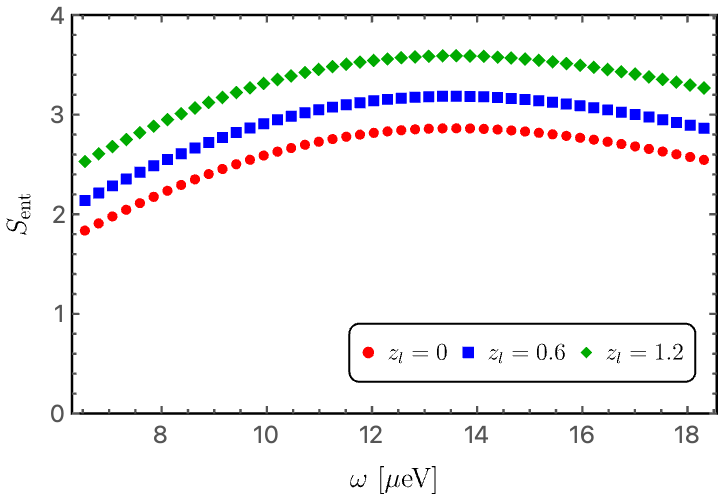}
  \caption{Entanglement entropy for OUT modes for vacuum input (left) and for a one-mode squeezed input with squeezing intensity $z_l$ for modes with $\ell=1$ (right)  as a function of the frequency. These plots are obtained for a polariton fluid with  velocity depicted in Fig.~\ref{fig:VthetaProfile}. Only frequencies describing propagating superradiant modes are shown.}
  \label{fig:EntropySqueezed}
\end{figure*}

However, given that thermal states are mixed quantum states,  entanglement entropy is no longer applicable to quantify entanglement, as entropy measures entanglement only when the total state is pure. An easily computable entanglement metric suitable for both mixed and pure states is the Logarithmic Negativity (LN) \cite{peres96,vidal02,plenio05}. It is defined as
\beq {\rm LN}(\hat \rho)=\log_2 ||\hat \rho^{\top_A}||_1\, \label{eq:defLN},\eeq
where $\hat \rho$ represents the density matrix of the system, $\hat \rho^{\top_A}$ is its partial transpose with respect to one of the two subsystems, which we denote as $A$, and $||\cdot||_1$ is the trace norm. A non-zero value of LN indicates a violation of the Positivity of Partial Transpose criterion for quantum states \cite{peres96}. For Gaussian states, and when one subsystem contains a single mode, regardless of the size of the other subsystem, LN is non-zero if and only if the state is entangled. Furthermore, LN is a faithful quantifier of entanglement, meaning that a higher LN value corresponds to a greater degree of entanglement. 
In our calculations, we will restrict to Gaussian states, which encompass the vacuum, coherent, squeezed, and thermal states. This family is sufficiently comprehensive to describe many interesting states in our setup. Additionally, evolving Gaussian states through a scattering process described by $\boldsymbol{S}^{\omega\ell}_{SR}$ or, more generally, by any Hamiltonian quadratic in the fields, can be efficiently accomplished using the Gaussian formalism (see, \cite{weedbrook2012,serafini17QCV} for reviews). While not needed for an understanding of the results in this section, we have succinctly compiled  the necessary tools to reproduce our calculations  in Appendix \ref{app:Gaussian}.

To facilitate comparison, we first show in
Figs.~\ref{fig:EntropySqueezed} our results for vacuum input. As this is a pure state, entanglement can be quantified through entanglement entropy. This plot manifestly shows that pair-creation at the ergoregion involves generation of quantum entanglement.

\begin{figure}
  \centering
  \includegraphics[width=.9\linewidth]{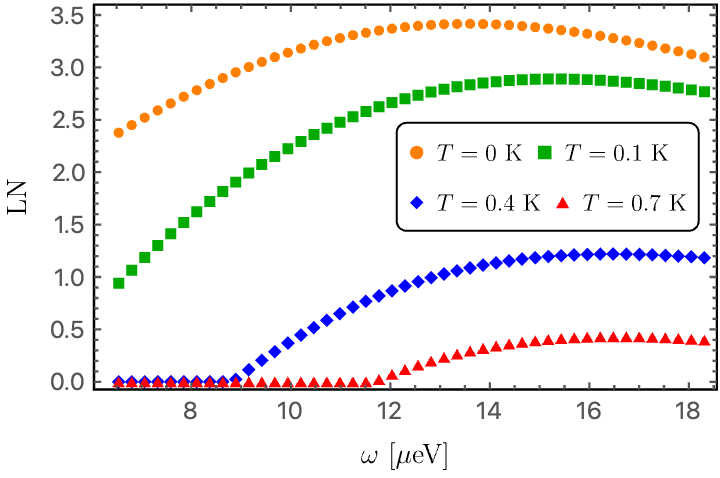}
  \caption{The LN for OUT modes with $\ell=1$ under four different thermal input states with varying temperatures. This plot illustrates how the presence of thermal noise degrades entanglement. The frequencies at which LN vanishes are explained by $T_c(\omega,\ell)$ and are also displayed in  Fig.~\ref{fig:TCrit}.}
  \label{thermalinput}
\end{figure}

\begin{figure}
  \centering
  \includegraphics[width=.9\linewidth]{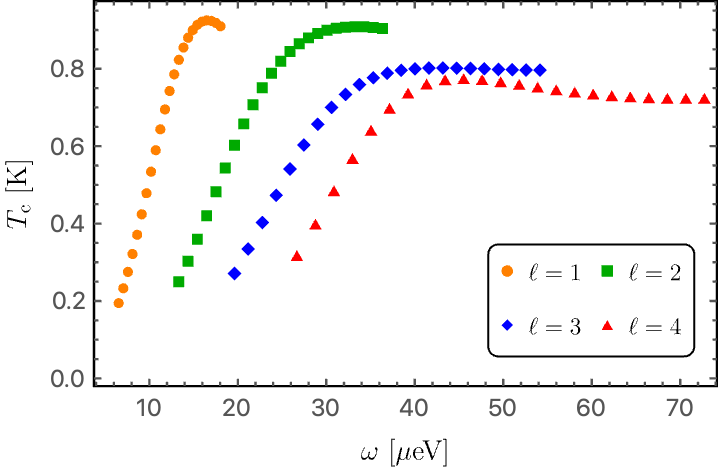}
  \caption{Critical  temperature, $T_c$, above which entanglement at the output vanishes, assuming {a thermal input state}. Only frequencies describing propagating superradiant modes are shown.}
  \label{fig:TCrit}
\end{figure}

Conversely, Fig.~\ref{thermalinput} shows the LN for a thermal input with various environmental temperatures, and for the $\ell=1$ modes (the result for other values of $\ell$ is qualitatively similar). This figure reveals that the presence of thermal quanta in the initial state hinders the generation of entanglement. This outcome aligns with the intuitive notion that thermal fluctuations act as an effective source of noise, contributing to the decoherence of the system and consequently reducing entanglement. In fact, for each mode $(\omega,\ell)$, there exists a threshold temperature beyond which the two out modes are not entangled. This ``critical'' temperature is given by\footnote{{The logarithmic negativity vanishes when the minimum symplectic eigenvalue of the partially transposed covariance matrix is equal to or larger than one $\tilde{\nu}_{\rm min}\geq1$ (see Appendix \ref{app:Gaussian}). These eigenvalues are a function of $|T_{\omega\ell}|$ and the initial state. For a thermal input state at temperature $T_{\rm env}$, they are proportional to $1+2n_{\omega}(T_{\rm env})$, where $n_{\omega}(T_{\rm env})$ is given by a Bose-Einstein distribution at temperature $T_{\rm env}$. Hence, $\tilde\nu_{\rm min}$ grows monotonically with temperature. $T_c(\omega,\ell)$ is the temperature at which $\tilde\nu_{\rm min}=1$ and, thus, at which entanglement vanishes.}}
\beq T_c(\omega,\ell)=\frac{\hbar\, \omega/k_{\rm B}}{\ln\left[1+\Big(|T_{\omega\ell}| (|T_{\omega\ell}|+\sqrt{1+|T_{\omega\ell}|^2})\Big)^{-1}\right]}\,  ,  \eeq
where $T_{\omega\ell}$ are the reflection coefficients, which we have obtained numerically, and $k_B$ is Boltzman's constant.  Fig.~\ref{fig:TCrit} shows $T_c(\omega,\ell)$ for $\ell=1,2,3$ and $4$, and for the relevant range of frequencies. This threshold temperature is important for experimental efforts seeking to detect entanglement. The range of temperatures at which entanglement exists is within current experimental capabilities with state-of-the-art cryostats.

All the findings discussed thus far remain unaffected if we replace the initial state with a coherent state. This holds true whether we consider a pure coherent state at zero temperature or add thermal noise to it. The rationale behind this is that a coherent state differs from the vacuum solely through a displacement of the mean value of the field operator, which entanglement remains insensitive to. Consequently, while the number of quanta in each of the OUT modes is amplified when the input is a coherent state (according to formulas \eqref{eq:NquantrCoh} and \eqref{eq:NquantlCoh}), entanglement remains unaffected.

Next, we discuss superradiant amplification of entanglement. As described in Section \ref{sec:QuantumSR}, entanglement in the final state can be  amplified by illuminating the ergoregion with a thermal-single-mode squeezed state. This state stems from the application of a single-mode squeezer operator, ${\ee^{\frac{1}{2}(\xi_l^*\,  \hat a_{l}^2 - \xi_l\,  \hat a_{l}^{\dagger}{}^2)} }$, to a thermal state ($\xi_l=z_l\, e^{i\phi_l}$). In keeping with Section \ref{sec:QuantumSR}, we focus on squeezing the { (radially inward propagating)} IN mode {$W_l^{\rm in}$}. It is worth noting that this initial state does not contain any entanglement between the two IN modes. However, initial squeezing represents a quantum resource~\cite{braunstein2005squeezing,asboth2005EntanglementPotential}, and the ergoregion has the capability to transform it into entanglement. Consequently, the resulting OUT state exhibits a higher degree of entanglement compared to what would have been achieved with thermal or coherent-thermal inputs at the same temperature. This amplification serves to counterbalance the detrimental effects of thermal noise.

\begin{figure}
    \centering
    \includegraphics[width=0.45\textwidth]{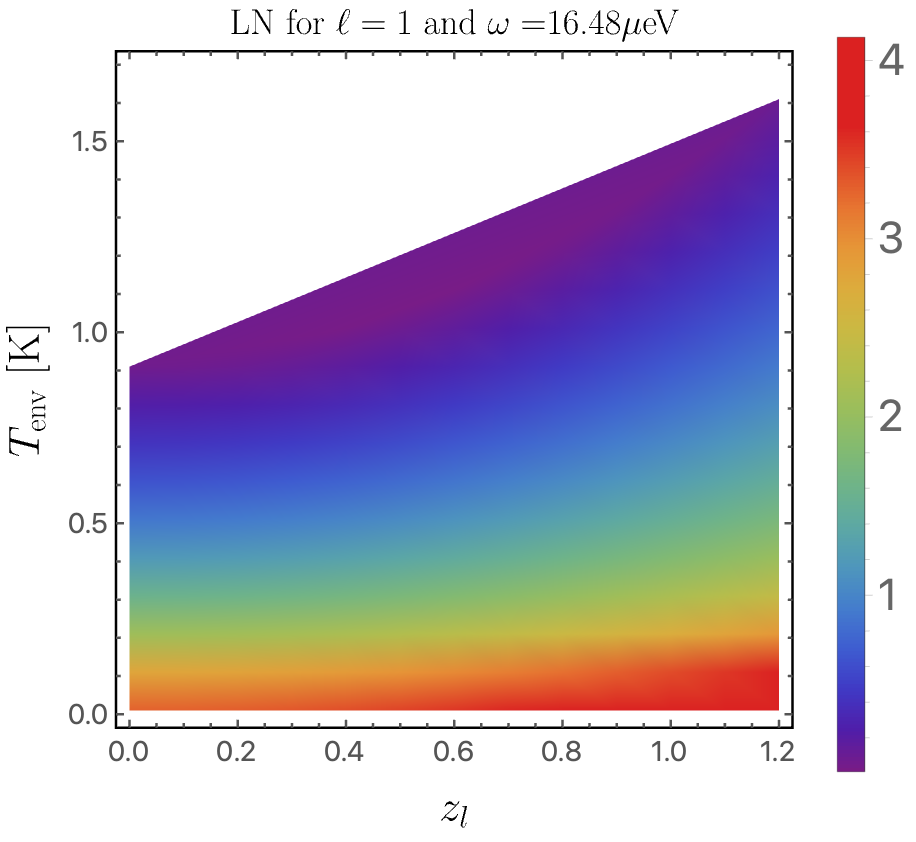}
    \caption{The LN of the OUT modes as a function of the squeezing intensity of the initial state $z_l$ and the temperature $T_{\rm env}$. White points have LN$=0$. The value of the LN is independent of the squeezing angle $\phi_l$. The results for other superradiant modes exhibit qualitatively similar behaviors. This figure  illustrates the  interplay between the detrimental effects of temperature and the amplifying influence of initial squeezing.}
    \label{fig:ColorLogNegTvsSq}
\end{figure}

These effects are shown in the right  panel of Fig.~\ref{fig:EntropySqueezed} and in Fig.~\ref{fig:ColorLogNegTvsSq}. In particular, the latter figure illustrates how initial squeezing can maintain the OUT state entangled, even for temperatures exceeding the threshold value that would render the state separable in the absence of initial squeezing.

To finish, we discuss the effects of {detection} losses and their impact on the entanglement in the OUT state. {These losses are not related to the cavity lifetime $\gamma$, which controls the amount of photons escaping the cavity, but rather to the efficiency in detecting these photons. The escaping photons provide exhaustive information on the state of the cavity fluid, and failure to detect all of them will hinder our ability to measure entanglement.} A suitable estimation of these effects can be achieved using a  pure loss channel (see, for example, \cite{serafini17QCV}). In this model, quanta have a probability $\eta$ of being observed and a probability (1-$\eta$) of being replaced by the vacuum {---or by environment modes.} The parameter $\eta$ represents the detection efficiency, {typically around $98\%$ in polariton experiments (see e.g. \cite{claude_2022}). This model can also be used to effectively parameterize sources of decoherence in a simple way. However, a precise model for decoherence will depend on the particular couplings between our system and the environment, and is beyond the scope of this work. As a final remark, we note that} this loss model maintains the Gaussian nature of quantum states. Further details are provided in Appendix \ref{app:Gaussian}.

Fig.~\ref{fig:LNetaz} shows the behavior of entanglement in the OUT state, quantified by the LN, as it varies with $\eta$.
The LN decreases rapidly when $\eta$ significantly deviates from unity.
Entanglement enhances with the squeezing intensity, but this is only notable for high detection efficiency $\eta>90\%$.

\begin{figure}
  \centering
  \includegraphics[width=.5\textwidth]{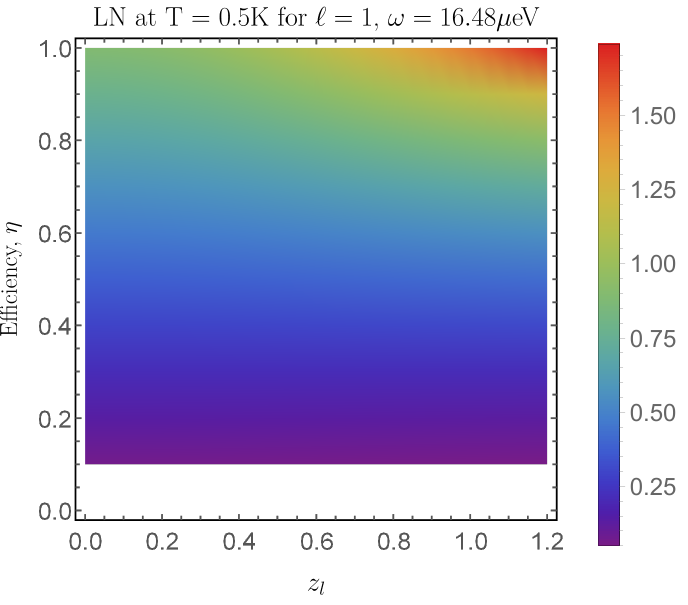}
  \caption{Variation of entanglement (quantified by  LN) between the two OUT modes with the efficiency $\eta$ and the squeezing intensity in the initial state $z_l$. White points have LN$=0$.}
  \label{fig:LNetaz}
\end{figure}

\section{Experimental Proposal for detection of entanglement}\label{sec:ExpProp}
Thermal noise is a roadblock between any experiment aiming at observing amplification by rotational superradiance and the resulting generation of entanglement. Specifically, polaritons couple to the vibrations of the crystalline structure of the semiconductor quantum wells. This heat transfer mechanism results in the generation of ``heat phonons'' (collective excitations) in typical experimental conditions even at vanishing temperatures, although there is a threshold below which the generation of collective excitations is dominated by the photonic vacuum~\cite{frerot_bogoliubov_2023}.  However, the data of Fig.~\ref{fig:ColorLogNegTvsSq} indicates that entanglement can be measured at the output even in the presence of thermal noise, provided that a single-mode squeezed state is made to scatter at the ergosurface and that the sample is cooled to low-enough temperatures, which are within the reach of dilution cryostats ($T<\SI{0.3}{\kelvin}$).

The strategy to observe entanglement thus is to send a mode of orbital angular momentum within the superradiance acceptance window, with one squeezed quadrature (either phase or intensity).
Orbital angular momentum can be readily obtained by shaping the Gaussian mode of a laser with a spatial light modulator~\cite{parigi_storage_2015}.
This can then be used to pump an optical parametric oscillator (OPO).
Type-I OPOs regularly yield above 3dB of squeezing below the shot-noise~\cite{Wu_opo_1986} with records at near-infrared wavelengths (as the resonance of the cavity used in our experiment in section~\ref{subsec:polexp}) of about 9dB with improved detection methods~\cite{Takeno_opo_07}.
These correspond to squeezing intensities of 0.35 and 1.04 respectively.

At the output, intensity correlations and phase anti-correlations can be measured with homodyne detection.
This consists of collecting the photons exiting the cavity at a given angle from a given spot, corresponding to excitations spatially located on either side of the ergosurface.\footnote{The angles at which photons come out of the cavity can be determined thanks to coherent probe spectroscopy, where a continuous-wave, coherent state excites collective excitations of the fluid on either side of the ergosurface, yielding their Bogoliubov spectrum~\cite{claude_2022}.}
Each of the two beams (which share the same frequency) is then interfered with a (spatially matched) reference beam (local oscillator, LO, provided by the same laser that generated the squeezed state at the input) on a beam-splitter.
The resulting interference pattern is measured with photodiodes in both output ports of the beam-splitter, while the relative phase of the signal beams and the LO can be scanned by modulating the path-length of the LO to the beam splitter.
This allows for the resolution of both quadratures of the output state and to thus reconstruct its density matrix by optical state tomography~\cite{lvovsky_tomo_2009}.
From there, the LN~\eqref{eq:defLN} can be evaluated using the analytical tools presented in this article.

\section{Outlook}
Superradiant amplification is a universal  phenomenon in which  radiation  scattered off time-independent potentials exhibits greater amplitude than the incoming radiation. Because of the type of input states commonly used in theoretical and experimental investigations (see the review~\cite{brito_superradiance_2020}), superradiance is commonly believed to yield only classical statistics, i.e., classically correlated radiation resulting from the amplification and redistribution of incoming field quanta.

In this paper, we have revisited the basics of superradiance and described  its field theoretic origin as the scattering of field modes having opposite symplectic norm $Q_{\rm KG}$ (which is the conserved quantity in relativistic bosonic field theories).  From there, we have used  tools from the theory of  Gaussian quantum bosonic systems to identify  that the symplectic transformation describing superradiant scattering is a two-mode squeezer.

From this perspective, superradiance shares similarities with the Hawking effect, since the latter is also described by a two-mode squeezer \cite{Agullo:2021vwj,Brady:2022ffk,Agullo:2022oye}. A key difference is that the pair production from superradiance does not follow a black body spectrum.

As in any two-mode squeezing process, entanglement is not exclusive to vacuum input; rather, it extends to a broad range of input states, including coherent states. However, in the case of coherent states, quantum correlations tend to be drowned by classical correlations. This overshadowing is the origin of the common belief that superradiance only yields classical statistics. We have shown that entanglement generation can be modulated at will, by appropriately choosing the input state: while thermal fluctuations at the input inhibit the generation of entanglement, the use of (unentangled) single-mode squeezed states amplifies the non-separability at the output. 
These findings establish that superradiance is an entanglement-generating phenomenon. 

In the context of rotational superradiance in ergoregions, the superradiant amplification of vacuum fluctuations yields a finite population of negative-norm field excitations in the ergoregion.
In the absence of dissipation, repeated superradiant amplification of this negative-norm population would result in a build-up of its amplitude, resulting in a dynamical instability of the ergoregion~\cite{Friedman:1978}. In this paper, we have proposed a new configuration to quench the intrinsic dynamical instability of ergoregions. In numerical simulations we have used a polaritonic quantum fluid of light, which has a driven-dissipative dynamics, to show that superradiance can occur in irrotational vortices. We explain the stability of the ergoregion observed in numerical simulations as a result of the losses in the system (mostly due to the photonic decay of the cavity). 

This configuration with a horizonless ergoregion has allowed us to apply our theoretical methods to the study of quantum emission by rotational superradiance. 

We have taken noise and losses into account in our calculations and shown that, although entanglement is generically degraded by thermal noise, it can be recovered using  single-mode squeezed states at the input. This strategy can be used in future experiments to facilitate the observation of entanglement generation by superradiance.

In conclusion, our methods permit the study of rotational superradiance with control over the input state, a feat not attainable in the presence of horizons because of the Hawking effect. This presents a new avenue for both theoretical and experimental exploration of this phenomenon. Nevertheless, our analysis also has implications for rotating configurations where dissipation is provided by a horizon, as for spinning black holes or analogue gravity experiments~\cite{torres_rotational_2017,jacquet_polariton_2020,braidotti_measurement_2022,svancara_exploring_2023}. Indeed, in scenarios featuring a horizon, the dynamics of entanglement will be ruled by the interplay between the Hawking effect and rotational superradiance~\cite{Agullo:2023pgp}. An understanding of this interplay requires knowledge of each of these individual mechanisms. Having the capability to investigate quantum emission originating from each effect independently, becomes valuable in this regard.

\begin{acknowledgements}
We have benefited from discussions with Julio Arrechea, Beatriz Elizaga Navascu\'es, Alessandro Fabbri, Elisabeth Giacobino, Luca Giacomelli, Mathieu Isoard, Dimitrios Kranas, Jorge Pullin and Justin Wilson.
I.A., A.D., and P.C. are supported by the NSF grants PHY-2110273, PHY-1903799, and PHY-2206557, by the RCS program of  Louisiana Boards of Regents through the grant LEQSF(2023-25)-RD-A-04, and by the Hearne Institute for Theoretical Physics. A.J.B. is supported by the DARPA Young Faculty Award (YFA) Grant No. N660012014029. 
M.J.J. and A.B. acknowledge funding from Ile de France DIM Sirteq via project ``FOLIAGE''. A.B. is a member of the Institut Universitaire de France.
\end{acknowledgements}

\bibliography{Bibliography.bib}

\providecommand{\noopsort}[1]{}\providecommand{\singleletter}[1]{#1}%
\begin{thebibliography}{55}%
\makeatletter
\providecommand \@ifxundefined [1]{%
 \@ifx{#1\undefined}
}%
\providecommand \@ifnum [1]{%
 \ifnum #1\expandafter \@firstoftwo
 \else \expandafter \@secondoftwo
 \fi
}%
\providecommand \@ifx [1]{%
 \ifx #1\expandafter \@firstoftwo
 \else \expandafter \@secondoftwo
 \fi
}%
\providecommand \natexlab [1]{#1}%
\providecommand \enquote  [1]{``#1''}%
\providecommand \bibnamefont  [1]{#1}%
\providecommand \bibfnamefont [1]{#1}%
\providecommand \citenamefont [1]{#1}%
\providecommand \href@noop [0]{\@secondoftwo}%
\providecommand \href [0]{\begingroup \@sanitize@url \@href}%
\providecommand \@href[1]{\@@startlink{#1}\@@href}%
\providecommand \@@href[1]{\endgroup#1\@@endlink}%
\providecommand \@sanitize@url [0]{\catcode `\\12\catcode `\$12\catcode
  `\&12\catcode `\#12\catcode `\^12\catcode `\_12\catcode `\%12\relax}%
\providecommand \@@startlink[1]{}%
\providecommand \@@endlink[0]{}%
\providecommand \url  [0]{\begingroup\@sanitize@url \@url }%
\providecommand \@url [1]{\endgroup\@href {#1}{\urlprefix }}%
\providecommand \urlprefix  [0]{URL }%
\providecommand \Eprint [0]{\href }%
\providecommand \doibase [0]{https://doi.org/}%
\providecommand \selectlanguage [0]{\@gobble}%
\providecommand \bibinfo  [0]{\@secondoftwo}%
\providecommand \bibfield  [0]{\@secondoftwo}%
\providecommand \translation [1]{[#1]}%
\providecommand \BibitemOpen [0]{}%
\providecommand \bibitemStop [0]{}%
\providecommand \bibitemNoStop [0]{.\EOS\space}%
\providecommand \EOS [0]{\spacefactor3000\relax}%
\providecommand \BibitemShut  [1]{\csname bibitem#1\endcsname}%
\let\auto@bib@innerbib\@empty
\bibitem [{\citenamefont {Dicke}(1954)}]{Dicke:1954zz}%
  \BibitemOpen
  \bibfield  {author} {\bibinfo {author} {\bibfnamefont {R.~H.}\ \bibnamefont
  {Dicke}},\ }\bibfield  {title} {\bibinfo {title} {{Coherence in Spontaneous
  Radiation Processes}},\ }\href {https://doi.org/10.1103/PhysRev.93.99}
  {\bibfield  {journal} {\bibinfo  {journal} {Phys. Rev.}\ }\textbf {\bibinfo
  {volume} {93}},\ \bibinfo {pages} {99} (\bibinfo {year} {1954})}\BibitemShut
  {NoStop}%
\bibitem [{\citenamefont {Zel'dovich}(1971)}]{Zeldovich1}%
  \BibitemOpen
  \bibfield  {author} {\bibinfo {author} {\bibfnamefont {Y.~B.}\ \bibnamefont
  {Zel'dovich}},\ }\href@noop {} {\bibfield  {journal} {\bibinfo  {journal}
  {Zh.Eksp.Teor.Fiz.}\ }\textbf {\bibinfo {volume} {14}},\ \bibinfo {pages}
  {270} (\bibinfo {year} {1971})}\BibitemShut {NoStop}%
\bibitem [{\citenamefont {Klein}(1929)}]{Klein:1929zz}%
  \BibitemOpen
  \bibfield  {author} {\bibinfo {author} {\bibfnamefont {O.}~\bibnamefont
  {Klein}},\ }\bibfield  {title} {\bibinfo {title} {{Die Reflexion von
  Elektronen an einem Potentialsprung nach der relativistischen Dynamik von
  Dirac}},\ }\href {https://doi.org/10.1007/BF01339716} {\bibfield  {journal}
  {\bibinfo  {journal} {Z. Phys.}\ }\textbf {\bibinfo {volume} {53}},\ \bibinfo
  {pages} {157} (\bibinfo {year} {1929})}\BibitemShut {NoStop}%
\bibitem [{\citenamefont {Sauter}(1931)}]{Sauter:1931zz}%
  \BibitemOpen
  \bibfield  {author} {\bibinfo {author} {\bibfnamefont {F.}~\bibnamefont
  {Sauter}},\ }\bibfield  {title} {\bibinfo {title} {{Uber das Verhalten eines
  Elektrons im homogenen elektrischen Feld nach der relativistischen Theorie
  Diracs}},\ }\href {https://doi.org/10.1007/BF01339461} {\bibfield  {journal}
  {\bibinfo  {journal} {Z. Phys.}\ }\textbf {\bibinfo {volume} {69}},\ \bibinfo
  {pages} {742} (\bibinfo {year} {1931})}\BibitemShut {NoStop}%
\bibitem [{\citenamefont {Schwinger}(1951)}]{Schwinger:1951nm}%
  \BibitemOpen
  \bibfield  {author} {\bibinfo {author} {\bibfnamefont {J.~S.}\ \bibnamefont
  {Schwinger}},\ }\bibfield  {title} {\bibinfo {title} {{On gauge invariance
  and vacuum polarization}},\ }\href {https://doi.org/10.1103/PhysRev.82.664}
  {\bibfield  {journal} {\bibinfo  {journal} {Phys. Rev.}\ }\textbf {\bibinfo
  {volume} {82}},\ \bibinfo {pages} {664} (\bibinfo {year} {1951})}\BibitemShut
  {NoStop}%
\bibitem [{\citenamefont {Arvanitaki}\ \emph {et~al.}(2010)\citenamefont
  {Arvanitaki}, \citenamefont {Dimopoulos}, \citenamefont {Dubovsky},
  \citenamefont {Kaloper},\ and\ \citenamefont
  {March-Russell}}]{Arvanitaki:2009fg}%
  \BibitemOpen
  \bibfield  {author} {\bibinfo {author} {\bibfnamefont {A.}~\bibnamefont
  {Arvanitaki}}, \bibinfo {author} {\bibfnamefont {S.}~\bibnamefont
  {Dimopoulos}}, \bibinfo {author} {\bibfnamefont {S.}~\bibnamefont
  {Dubovsky}}, \bibinfo {author} {\bibfnamefont {N.}~\bibnamefont {Kaloper}},\
  and\ \bibinfo {author} {\bibfnamefont {J.}~\bibnamefont {March-Russell}},\
  }\bibfield  {title} {\bibinfo {title} {{String Axiverse}},\ }\href
  {https://doi.org/10.1103/PhysRevD.81.123530} {\bibfield  {journal} {\bibinfo
  {journal} {Phys. Rev. D}\ }\textbf {\bibinfo {volume} {81}},\ \bibinfo
  {pages} {123530} (\bibinfo {year} {2010})},\ \Eprint
  {https://arxiv.org/abs/0905.4720} {arXiv:0905.4720 [hep-th]} \BibitemShut
  {NoStop}%
\bibitem [{\citenamefont {Arvanitaki}\ and\ \citenamefont
  {Dubovsky}(2011)}]{Arvanitaki:2010sy}%
  \BibitemOpen
  \bibfield  {author} {\bibinfo {author} {\bibfnamefont {A.}~\bibnamefont
  {Arvanitaki}}\ and\ \bibinfo {author} {\bibfnamefont {S.}~\bibnamefont
  {Dubovsky}},\ }\bibfield  {title} {\bibinfo {title} {{Exploring the String
  Axiverse with Precision Black Hole Physics}},\ }\href
  {https://doi.org/10.1103/PhysRevD.83.044026} {\bibfield  {journal} {\bibinfo
  {journal} {Phys. Rev. D}\ }\textbf {\bibinfo {volume} {83}},\ \bibinfo
  {pages} {044026} (\bibinfo {year} {2011})},\ \Eprint
  {https://arxiv.org/abs/1004.3558} {arXiv:1004.3558 [hep-th]} \BibitemShut
  {NoStop}%
\bibitem [{\citenamefont {Penrose}(1969)}]{Penrose:1969pc}%
  \BibitemOpen
  \bibfield  {author} {\bibinfo {author} {\bibfnamefont {R.}~\bibnamefont
  {Penrose}},\ }\bibfield  {title} {\bibinfo {title} {{Gravitational collapse:
  The role of general relativity}},\ }\href
  {https://doi.org/10.1023/A:1016578408204} {\bibfield  {journal} {\bibinfo
  {journal} {Riv. Nuovo Cim.}\ }\textbf {\bibinfo {volume} {1}},\ \bibinfo
  {pages} {252} (\bibinfo {year} {1969})}\BibitemShut {NoStop}%
\bibitem [{\citenamefont {Penrose}\ and\ \citenamefont
  {Floyd}(1971)}]{Penrose:1971uk}%
  \BibitemOpen
  \bibfield  {author} {\bibinfo {author} {\bibfnamefont {R.}~\bibnamefont
  {Penrose}}\ and\ \bibinfo {author} {\bibfnamefont {R.~M.}\ \bibnamefont
  {Floyd}},\ }\bibfield  {title} {\bibinfo {title} {{Extraction of rotational
  energy from a black hole}},\ }\href {https://doi.org/10.1038/physci229177a0}
  {\bibfield  {journal} {\bibinfo  {journal} {Nature}\ }\textbf {\bibinfo
  {volume} {229}},\ \bibinfo {pages} {177} (\bibinfo {year}
  {1971})}\BibitemShut {NoStop}%
\bibitem [{\citenamefont {Press}\ and\ \citenamefont
  {Teukolsky}(1972)}]{Press:1972zz}%
  \BibitemOpen
  \bibfield  {author} {\bibinfo {author} {\bibfnamefont {W.~H.}\ \bibnamefont
  {Press}}\ and\ \bibinfo {author} {\bibfnamefont {S.~A.}\ \bibnamefont
  {Teukolsky}},\ }\bibfield  {title} {\bibinfo {title} {{Floating Orbits,
  Superradiant Scattering and the Black-hole Bomb}},\ }\href
  {https://doi.org/10.1038/238211a0} {\bibfield  {journal} {\bibinfo  {journal}
  {Nature}\ }\textbf {\bibinfo {volume} {238}},\ \bibinfo {pages} {211}
  (\bibinfo {year} {1972})}\BibitemShut {NoStop}%
\bibitem [{\citenamefont {Teukolsky}\ and\ \citenamefont
  {Press}(1974)}]{Teukolsky:1974yv}%
  \BibitemOpen
  \bibfield  {author} {\bibinfo {author} {\bibfnamefont {S.~A.}\ \bibnamefont
  {Teukolsky}}\ and\ \bibinfo {author} {\bibfnamefont {W.~H.}\ \bibnamefont
  {Press}},\ }\bibfield  {title} {\bibinfo {title} {{Perturbations of a
  rotating black hole. III - Interaction of the hole with gravitational and
  electromagnet ic radiation}},\ }\href {https://doi.org/10.1086/153180}
  {\bibfield  {journal} {\bibinfo  {journal} {Astrophys. J.}\ }\textbf
  {\bibinfo {volume} {193}},\ \bibinfo {pages} {443} (\bibinfo {year}
  {1974})}\BibitemShut {NoStop}%
\bibitem [{\citenamefont {Blandford}\ and\ \citenamefont
  {Znajek}(1977)}]{Blandford:1977ds}%
  \BibitemOpen
  \bibfield  {author} {\bibinfo {author} {\bibfnamefont {R.~D.}\ \bibnamefont
  {Blandford}}\ and\ \bibinfo {author} {\bibfnamefont {R.~L.}\ \bibnamefont
  {Znajek}},\ }\bibfield  {title} {\bibinfo {title} {{Electromagnetic
  extractions of energy from Kerr black holes}},\ }\href
  {https://doi.org/10.1093/mnras/179.3.433} {\bibfield  {journal} {\bibinfo
  {journal} {Mon. Not. Roy. Astron. Soc.}\ }\textbf {\bibinfo {volume} {179}},\
  \bibinfo {pages} {433} (\bibinfo {year} {1977})}\BibitemShut {NoStop}%
\bibitem [{\citenamefont {Zel'dovich}(1972)}]{Zeldovich2}%
  \BibitemOpen
  \bibfield  {author} {\bibinfo {author} {\bibfnamefont {Y.~B.}\ \bibnamefont
  {Zel'dovich}},\ }\href@noop {} {\bibfield  {journal} {\bibinfo  {journal}
  {Zh.Eksp.Teor.Fiz.}\ }\textbf {\bibinfo {volume} {62}},\ \bibinfo {pages}
  {2076} (\bibinfo {year} {1972})}\BibitemShut {NoStop}%
\bibitem [{\citenamefont {Brito}\ \emph {et~al.}(2020)\citenamefont {Brito},
  \citenamefont {Cardoso},\ and\ \citenamefont
  {Pani}}]{brito_superradiance_2020}%
  \BibitemOpen
  \bibfield  {author} {\bibinfo {author} {\bibfnamefont {R.}~\bibnamefont
  {Brito}}, \bibinfo {author} {\bibfnamefont {V.}~\bibnamefont {Cardoso}},\
  and\ \bibinfo {author} {\bibfnamefont {P.}~\bibnamefont {Pani}},\ }\bibfield
  {title} {\bibinfo {title} {Superradiance -- the 2020 {Edition}},\ }\href@noop
  {} {\bibfield  {journal} {\bibinfo  {journal} {arXiv:1501.06570 [astro-ph,
  physics:gr-qc, physics:hep-ph, physics:hep-th, physics:physics]}\ } (\bibinfo
  {year} {2020})}\BibitemShut {NoStop}%
\bibitem [{\citenamefont {Starobinskii}(1973)}]{Starobinskii:1973hgd}%
  \BibitemOpen
  \bibfield  {author} {\bibinfo {author} {\bibfnamefont {A.~A.}\ \bibnamefont
  {Starobinskii}},\ }\bibfield  {title} {\bibinfo {title} {{Amplification of
  waves during reflection from a rotating ''black hole''}},\ }\href@noop {}
  {\bibfield  {journal} {\bibinfo  {journal} {Sov. Phys. JETP}\ }\textbf
  {\bibinfo {volume} {64}},\ \bibinfo {pages} {48} (\bibinfo {year}
  {1973})}\BibitemShut {NoStop}%
\bibitem [{\citenamefont {Unruh}(1974)}]{Unruh:1974bw}%
  \BibitemOpen
  \bibfield  {author} {\bibinfo {author} {\bibfnamefont {W.~G.}\ \bibnamefont
  {Unruh}},\ }\bibfield  {title} {\bibinfo {title} {{Second quantization in the
  Kerr metric}},\ }\href {https://doi.org/10.1103/PhysRevD.10.3194} {\bibfield
  {journal} {\bibinfo  {journal} {Phys. Rev. D}\ }\textbf {\bibinfo {volume}
  {10}},\ \bibinfo {pages} {3194} (\bibinfo {year} {1974})}\BibitemShut
  {NoStop}%
\bibitem [{\citenamefont {Friedman}(1978{\natexlab{a}})}]{Friedman:1978wla}%
  \BibitemOpen
  \bibfield  {author} {\bibinfo {author} {\bibfnamefont {J.~L.}\ \bibnamefont
  {Friedman}},\ }\bibfield  {title} {\bibinfo {title} {{Generic instability of
  rotating relativistic stars}},\ }\href {https://doi.org/10.1007/BF01202527}
  {\bibfield  {journal} {\bibinfo  {journal} {Commun. Math. Phys.}\ }\textbf
  {\bibinfo {volume} {62}},\ \bibinfo {pages} {247} (\bibinfo {year}
  {1978}{\natexlab{a}})}\BibitemShut {NoStop}%
\bibitem [{\citenamefont {Torres}\ \emph {et~al.}(2017)\citenamefont {Torres},
  \citenamefont {Patrick}, \citenamefont {Coutant}, \citenamefont {Richartz},
  \citenamefont {Tedford},\ and\ \citenamefont
  {Weinfurtner}}]{torres_rotational_2017}%
  \BibitemOpen
  \bibfield  {author} {\bibinfo {author} {\bibfnamefont {T.}~\bibnamefont
  {Torres}}, \bibinfo {author} {\bibfnamefont {S.}~\bibnamefont {Patrick}},
  \bibinfo {author} {\bibfnamefont {A.}~\bibnamefont {Coutant}}, \bibinfo
  {author} {\bibfnamefont {M.}~\bibnamefont {Richartz}}, \bibinfo {author}
  {\bibfnamefont {E.~W.}\ \bibnamefont {Tedford}},\ and\ \bibinfo {author}
  {\bibfnamefont {S.}~\bibnamefont {Weinfurtner}},\ }\bibfield  {title}
  {\bibinfo {title} {Rotational superradiant scattering in a vortex flow},\
  }\href {https://doi.org/10.1038/nphys4151} {\bibfield  {journal} {\bibinfo
  {journal} {Nature Physics}\ }\textbf {\bibinfo {volume} {13}},\ \bibinfo
  {pages} {833} (\bibinfo {year} {2017})}\BibitemShut {NoStop}%
\bibitem [{\citenamefont {Braidotti}\ \emph {et~al.}(2022)\citenamefont
  {Braidotti}, \citenamefont {Prizia}, \citenamefont {Maitland}, \citenamefont
  {Marino}, \citenamefont {Prain}, \citenamefont {Starshynov}, \citenamefont
  {Westerberg}, \citenamefont {Wright},\ and\ \citenamefont
  {Faccio}}]{braidotti_measurement_2022}%
  \BibitemOpen
  \bibfield  {author} {\bibinfo {author} {\bibfnamefont {M.-C.}\ \bibnamefont
  {Braidotti}}, \bibinfo {author} {\bibfnamefont {R.}~\bibnamefont {Prizia}},
  \bibinfo {author} {\bibfnamefont {C.}~\bibnamefont {Maitland}}, \bibinfo
  {author} {\bibfnamefont {F.}~\bibnamefont {Marino}}, \bibinfo {author}
  {\bibfnamefont {A.}~\bibnamefont {Prain}}, \bibinfo {author} {\bibfnamefont
  {I.}~\bibnamefont {Starshynov}}, \bibinfo {author} {\bibfnamefont
  {N.}~\bibnamefont {Westerberg}}, \bibinfo {author} {\bibfnamefont {E.~M.}\
  \bibnamefont {Wright}},\ and\ \bibinfo {author} {\bibfnamefont
  {D.}~\bibnamefont {Faccio}},\ }\bibfield  {title} {\bibinfo {title}
  {Measurement of {Penrose} {Superradiance} in a {Photon} {Superfluid}},\
  }\href {https://doi.org/10.1103/PhysRevLett.128.013901} {\bibfield  {journal}
  {\bibinfo  {journal} {Physical Review Letters}\ }\textbf {\bibinfo {volume}
  {128}},\ \bibinfo {pages} {013901} (\bibinfo {year} {2022})}\BibitemShut
  {NoStop}%
\bibitem [{\citenamefont {Matacz}\ \emph {et~al.}(1993)\citenamefont {Matacz},
  \citenamefont {Davies},\ and\ \citenamefont
  {Ottewill}}]{davies1993QuantumSuperradiance}%
  \BibitemOpen
  \bibfield  {author} {\bibinfo {author} {\bibfnamefont {A.~L.}\ \bibnamefont
  {Matacz}}, \bibinfo {author} {\bibfnamefont {P.~C.~W.}\ \bibnamefont
  {Davies}},\ and\ \bibinfo {author} {\bibfnamefont {A.~C.}\ \bibnamefont
  {Ottewill}},\ }\bibfield  {title} {\bibinfo {title} {Quantum vacuum
  instability near rotating stars},\ }\href
  {https://doi.org/10.1103/PhysRevD.47.1557} {\bibfield  {journal} {\bibinfo
  {journal} {Phys. Rev. D}\ }\textbf {\bibinfo {volume} {47}},\ \bibinfo
  {pages} {1557} (\bibinfo {year} {1993})}\BibitemShut {NoStop}%
\bibitem [{\citenamefont {Agullo}\ \emph {et~al.}(2023)\citenamefont {Agullo},
  \citenamefont {Brady}, \citenamefont {Delhom},\ and\ \citenamefont
  {Kranas}}]{Agullo:2023pgp}%
  \BibitemOpen
  \bibfield  {author} {\bibinfo {author} {\bibfnamefont {I.}~\bibnamefont
  {Agullo}}, \bibinfo {author} {\bibfnamefont {A.~J.}\ \bibnamefont {Brady}},
  \bibinfo {author} {\bibfnamefont {A.}~\bibnamefont {Delhom}},\ and\ \bibinfo
  {author} {\bibfnamefont {D.}~\bibnamefont {Kranas}},\ }\bibfield  {title}
  {\bibinfo {title} {Entanglement from rotating black holes in thermal baths},\
  }\href@noop {} {\  (\bibinfo {year} {2023})},\ \Eprint
  {https://arxiv.org/abs/2307.06215} {arXiv:2307.06215 [gr-qc]} \BibitemShut
  {NoStop}%
\bibitem [{\citenamefont {Friedman}(1978{\natexlab{b}})}]{Friedman:1978}%
  \BibitemOpen
  \bibfield  {author} {\bibinfo {author} {\bibfnamefont {J.~L.}\ \bibnamefont
  {Friedman}},\ }\bibfield  {title} {\bibinfo {title} {{Ergosphere
  instability}},\ }\href {https://doi.org/10.1086/152444} {\bibfield  {journal}
  {\bibinfo  {journal} {Commun. Math. Phys..}\ }\textbf {\bibinfo {volume}
  {63}},\ \bibinfo {pages} {243} (\bibinfo {year}
  {1978}{\natexlab{b}})}\BibitemShut {NoStop}%
\bibitem [{\citenamefont {Comins}\ and\ \citenamefont
  {Schutz}(1978)}]{comins_ergoregion_1978}%
  \BibitemOpen
  \bibfield  {author} {\bibinfo {author} {\bibfnamefont {N.}~\bibnamefont
  {Comins}}\ and\ \bibinfo {author} {\bibfnamefont {B.}~\bibnamefont
  {Schutz}},\ }\bibfield  {title} {\bibinfo {title} {On the ergoregion
  instability},\ }\href {https://doi.org/10.1098/rspa.1978.0196} {\bibfield
  {journal} {\bibinfo  {journal} {Proceedings of the Royal Society of London.
  A. Mathematical and Physical Sciences}\ }\textbf {\bibinfo {volume} {364}},\
  \bibinfo {pages} {211} (\bibinfo {year} {1978})}\BibitemShut {NoStop}%
\bibitem [{\citenamefont {Giacomelli}\ and\ \citenamefont
  {Carusotto}(2021)}]{Giacomelli:2020evu}%
  \BibitemOpen
  \bibfield  {author} {\bibinfo {author} {\bibfnamefont {L.}~\bibnamefont
  {Giacomelli}}\ and\ \bibinfo {author} {\bibfnamefont {I.}~\bibnamefont
  {Carusotto}},\ }\bibfield  {title} {\bibinfo {title} {{Understanding
  superradiant phenomena with synthetic vector potentials in atomic
  Bose-Einstein condensates}},\ }\href
  {https://doi.org/10.1103/PhysRevA.103.043309} {\bibfield  {journal} {\bibinfo
   {journal} {Phys. Rev. A}\ }\textbf {\bibinfo {volume} {103}},\ \bibinfo
  {pages} {043309} (\bibinfo {year} {2021})},\ \Eprint
  {https://arxiv.org/abs/2011.01736} {arXiv:2011.01736 [cond-mat.quant-gas]}
  \BibitemShut {NoStop}%
\bibitem [{\citenamefont {Carusotto}\ and\ \citenamefont
  {Ciuti}(2013)}]{carusotto_quantum_2013}%
  \BibitemOpen
  \bibfield  {author} {\bibinfo {author} {\bibfnamefont {I.}~\bibnamefont
  {Carusotto}}\ and\ \bibinfo {author} {\bibfnamefont {C.}~\bibnamefont
  {Ciuti}},\ }\bibfield  {title} {\bibinfo {title} {Quantum fluids of light},\
  }\bibfield  {journal} {\bibinfo  {journal} {Reviews of Modern Physics}\
  }\textbf {\bibinfo {volume} {85}},\ \href
  {https://doi.org/10.1103/RevModPhys.85.299} {10.1103/RevModPhys.85.299}
  (\bibinfo {year} {2013})\BibitemShut {NoStop}%
\bibitem [{\citenamefont {Kasprzak}\ \emph {et~al.}(2006)\citenamefont
  {Kasprzak}, \citenamefont {Richard}, \citenamefont {Kundermann},
  \citenamefont {Baas}, \citenamefont {Jeambrun}, \citenamefont {Keeling},
  \citenamefont {Marchetti}, \citenamefont {Szymańska}, \citenamefont
  {André}, \citenamefont {Staehli}, \citenamefont {Savona}, \citenamefont
  {Littlewood}, \citenamefont {Deveaud},\ and\ \citenamefont
  {Dang}}]{kasprzak_boseeinstein_2006}%
  \BibitemOpen
  \bibfield  {author} {\bibinfo {author} {\bibfnamefont {J.}~\bibnamefont
  {Kasprzak}}, \bibinfo {author} {\bibfnamefont {M.}~\bibnamefont {Richard}},
  \bibinfo {author} {\bibfnamefont {S.}~\bibnamefont {Kundermann}}, \bibinfo
  {author} {\bibfnamefont {A.}~\bibnamefont {Baas}}, \bibinfo {author}
  {\bibfnamefont {P.}~\bibnamefont {Jeambrun}}, \bibinfo {author}
  {\bibfnamefont {J.~M.~J.}\ \bibnamefont {Keeling}}, \bibinfo {author}
  {\bibfnamefont {F.~M.}\ \bibnamefont {Marchetti}}, \bibinfo {author}
  {\bibfnamefont {M.~H.}\ \bibnamefont {Szymańska}}, \bibinfo {author}
  {\bibfnamefont {R.}~\bibnamefont {André}}, \bibinfo {author} {\bibfnamefont
  {J.~L.}\ \bibnamefont {Staehli}}, \bibinfo {author} {\bibfnamefont
  {V.}~\bibnamefont {Savona}}, \bibinfo {author} {\bibfnamefont {P.~B.}\
  \bibnamefont {Littlewood}}, \bibinfo {author} {\bibfnamefont
  {B.}~\bibnamefont {Deveaud}},\ and\ \bibinfo {author} {\bibfnamefont {L.~S.}\
  \bibnamefont {Dang}},\ }\bibfield  {title} {\bibinfo {title}
  {Bose–{Einstein} condensation of exciton polaritons},\ }\bibfield
  {journal} {\bibinfo  {journal} {Nature}\ }\textbf {\bibinfo {volume} {443}},\
  \href {https://doi.org/10.1038/nature05131} {10.1038/nature05131} (\bibinfo
  {year} {2006})\BibitemShut {NoStop}%
\bibitem [{\citenamefont {Balili}\ \emph {et~al.}(2007)\citenamefont {Balili},
  \citenamefont {Hartwell}, \citenamefont {Snoke}, \citenamefont {Pfeiffer},\
  and\ \citenamefont {West}}]{balili_bose-einstein_2007}%
  \BibitemOpen
  \bibfield  {author} {\bibinfo {author} {\bibfnamefont {R.}~\bibnamefont
  {Balili}}, \bibinfo {author} {\bibfnamefont {V.}~\bibnamefont {Hartwell}},
  \bibinfo {author} {\bibfnamefont {D.}~\bibnamefont {Snoke}}, \bibinfo
  {author} {\bibfnamefont {L.}~\bibnamefont {Pfeiffer}},\ and\ \bibinfo
  {author} {\bibfnamefont {K.}~\bibnamefont {West}},\ }\bibfield  {title}
  {\bibinfo {title} {Bose-{Einstein} {Condensation} of {Microcavity}
  {Polaritons} in a {Trap}},\ }\bibfield  {journal} {\bibinfo  {journal}
  {Science}\ }\textbf {\bibinfo {volume} {316}},\ \href
  {https://doi.org/10.1126/science.1140990} {10.1126/science.1140990} (\bibinfo
  {year} {2007})\BibitemShut {NoStop}%
\bibitem [{\citenamefont {Lagoudakis}\ \emph {et~al.}(2008)\citenamefont
  {Lagoudakis}, \citenamefont {Wouters}, \citenamefont {Richard}, \citenamefont
  {Baas}, \citenamefont {Carusotto}, \citenamefont {André}, \citenamefont
  {Dang},\ and\ \citenamefont {Deveaud-Plédran}}]{lagoudakis_quantized_2008}%
  \BibitemOpen
  \bibfield  {author} {\bibinfo {author} {\bibfnamefont {K.~G.}\ \bibnamefont
  {Lagoudakis}}, \bibinfo {author} {\bibfnamefont {M.}~\bibnamefont {Wouters}},
  \bibinfo {author} {\bibfnamefont {M.}~\bibnamefont {Richard}}, \bibinfo
  {author} {\bibfnamefont {A.}~\bibnamefont {Baas}}, \bibinfo {author}
  {\bibfnamefont {I.}~\bibnamefont {Carusotto}}, \bibinfo {author}
  {\bibfnamefont {R.}~\bibnamefont {André}}, \bibinfo {author} {\bibfnamefont
  {L.~S.}\ \bibnamefont {Dang}},\ and\ \bibinfo {author} {\bibfnamefont
  {B.}~\bibnamefont {Deveaud-Plédran}},\ }\bibfield  {title} {\bibinfo {title}
  {Quantized vortices in an exciton–polariton condensate},\ }\bibfield
  {journal} {\bibinfo  {journal} {Nature Physics}\ }\textbf {\bibinfo {volume}
  {4}},\ \href {https://doi.org/10.1038/nphys1051} {10.1038/nphys1051}
  (\bibinfo {year} {2008})\BibitemShut {NoStop}%
\bibitem [{\citenamefont {Utsunomiya}\ \emph {et~al.}(2008)\citenamefont
  {Utsunomiya}, \citenamefont {Tian}, \citenamefont {Roumpos}, \citenamefont
  {Lai}, \citenamefont {Kumada}, \citenamefont {Fujisawa}, \citenamefont
  {Kuwata-Gonokami}, \citenamefont {Löffler}, \citenamefont {Höfling},
  \citenamefont {Forchel},\ and\ \citenamefont
  {Yamamoto}}]{utsunomiya_observation_2008}%
  \BibitemOpen
  \bibfield  {author} {\bibinfo {author} {\bibfnamefont {S.}~\bibnamefont
  {Utsunomiya}}, \bibinfo {author} {\bibfnamefont {L.}~\bibnamefont {Tian}},
  \bibinfo {author} {\bibfnamefont {G.}~\bibnamefont {Roumpos}}, \bibinfo
  {author} {\bibfnamefont {C.~W.}\ \bibnamefont {Lai}}, \bibinfo {author}
  {\bibfnamefont {N.}~\bibnamefont {Kumada}}, \bibinfo {author} {\bibfnamefont
  {T.}~\bibnamefont {Fujisawa}}, \bibinfo {author} {\bibfnamefont
  {M.}~\bibnamefont {Kuwata-Gonokami}}, \bibinfo {author} {\bibfnamefont
  {A.}~\bibnamefont {Löffler}}, \bibinfo {author} {\bibfnamefont
  {S.}~\bibnamefont {Höfling}}, \bibinfo {author} {\bibfnamefont
  {A.}~\bibnamefont {Forchel}},\ and\ \bibinfo {author} {\bibfnamefont
  {Y.}~\bibnamefont {Yamamoto}},\ }\bibfield  {title} {\bibinfo {title}
  {Observation of {Bogoliubov} excitations in exciton-polariton condensates},\
  }\href {https://doi.org/10.1038/nphys1034} {\bibfield  {journal} {\bibinfo
  {journal} {Nature Physics}\ }\textbf {\bibinfo {volume} {4}},\ \bibinfo
  {pages} {700} (\bibinfo {year} {2008})}\BibitemShut {NoStop}%
\bibitem [{\citenamefont {Amo}\ \emph {et~al.}(2009)\citenamefont {Amo},
  \citenamefont {Lefrère}, \citenamefont {Pigeon}, \citenamefont {Adrados},
  \citenamefont {Ciuti}, \citenamefont {Carusotto}, \citenamefont {Houdré},
  \citenamefont {Giacobino},\ and\ \citenamefont
  {Bramati}}]{amo_superfluidity_2009}%
  \BibitemOpen
  \bibfield  {author} {\bibinfo {author} {\bibfnamefont {A.}~\bibnamefont
  {Amo}}, \bibinfo {author} {\bibfnamefont {J.}~\bibnamefont {Lefrère}},
  \bibinfo {author} {\bibfnamefont {S.}~\bibnamefont {Pigeon}}, \bibinfo
  {author} {\bibfnamefont {C.}~\bibnamefont {Adrados}}, \bibinfo {author}
  {\bibfnamefont {C.}~\bibnamefont {Ciuti}}, \bibinfo {author} {\bibfnamefont
  {I.}~\bibnamefont {Carusotto}}, \bibinfo {author} {\bibfnamefont
  {R.}~\bibnamefont {Houdré}}, \bibinfo {author} {\bibfnamefont
  {E.}~\bibnamefont {Giacobino}},\ and\ \bibinfo {author} {\bibfnamefont
  {A.}~\bibnamefont {Bramati}},\ }\bibfield  {title} {\bibinfo {title}
  {Superfluidity of polaritons in semiconductor microcavities},\ }\href
  {https://doi.org/10.1038/nphys1364} {\bibfield  {journal} {\bibinfo
  {journal} {Nature Physics}\ }\textbf {\bibinfo {volume} {5}},\ \bibinfo
  {pages} {805} (\bibinfo {year} {2009})}\BibitemShut {NoStop}%
\bibitem [{\citenamefont {Baas}\ \emph {et~al.}(2004)\citenamefont {Baas},
  \citenamefont {Karr}, \citenamefont {Eleuch},\ and\ \citenamefont
  {Giacobino}}]{baas_bista_2004}%
  \BibitemOpen
  \bibfield  {author} {\bibinfo {author} {\bibfnamefont {A.}~\bibnamefont
  {Baas}}, \bibinfo {author} {\bibfnamefont {J.~P.}\ \bibnamefont {Karr}},
  \bibinfo {author} {\bibfnamefont {H.}~\bibnamefont {Eleuch}},\ and\ \bibinfo
  {author} {\bibfnamefont {E.}~\bibnamefont {Giacobino}},\ }\bibfield  {title}
  {\bibinfo {title} {Optical bistability in semiconductor microcavities},\
  }\href {https://doi.org/10.1103/PhysRevA.69.023809} {\bibfield  {journal}
  {\bibinfo  {journal} {Phys. Rev. A}\ }\textbf {\bibinfo {volume} {69}},\
  \bibinfo {pages} {023809} (\bibinfo {year} {2004})}\BibitemShut {NoStop}%
\bibitem [{\citenamefont {Claude}\ \emph {et~al.}(2022)\citenamefont {Claude},
  \citenamefont {Jacquet}, \citenamefont {Usciati}, \citenamefont {Carusotto},
  \citenamefont {Giacobino}, \citenamefont {Bramati},\ and\ \citenamefont
  {Glorieux}}]{claude_2022}%
  \BibitemOpen
  \bibfield  {author} {\bibinfo {author} {\bibfnamefont {F.}~\bibnamefont
  {Claude}}, \bibinfo {author} {\bibfnamefont {M.~J.}\ \bibnamefont {Jacquet}},
  \bibinfo {author} {\bibfnamefont {R.}~\bibnamefont {Usciati}}, \bibinfo
  {author} {\bibfnamefont {I.}~\bibnamefont {Carusotto}}, \bibinfo {author}
  {\bibfnamefont {E.}~\bibnamefont {Giacobino}}, \bibinfo {author}
  {\bibfnamefont {A.}~\bibnamefont {Bramati}},\ and\ \bibinfo {author}
  {\bibfnamefont {Q.}~\bibnamefont {Glorieux}},\ }\bibfield  {title} {\bibinfo
  {title} {High-{Resolution} {Coherent} {Probe} {Spectroscopy} of a {Polariton}
  {Quantum} {Fluid}},\ }\href {https://doi.org/10.1103/PhysRevLett.129.103601}
  {\bibfield  {journal} {\bibinfo  {journal} {Physical Review Letters}\
  }\textbf {\bibinfo {volume} {129}},\ \bibinfo {pages} {103601} (\bibinfo
  {year} {2022})}\BibitemShut {NoStop}%
\bibitem [{\citenamefont {Macher}\ and\ \citenamefont
  {Parentani}(2009)}]{macher_blackwhite_2009}%
  \BibitemOpen
  \bibfield  {author} {\bibinfo {author} {\bibfnamefont {J.}~\bibnamefont
  {Macher}}\ and\ \bibinfo {author} {\bibfnamefont {R.}~\bibnamefont
  {Parentani}},\ }\bibfield  {title} {\bibinfo {title} {Black/white hole
  radiation from dispersive theories},\ }\href
  {https://doi.org/10.1103/PhysRevD.79.124008} {\bibfield  {journal} {\bibinfo
  {journal} {Physical Review D}\ }\textbf {\bibinfo {volume} {79}},\ \bibinfo
  {pages} {124008} (\bibinfo {year} {2009})}\BibitemShut {NoStop}%
\bibitem [{\citenamefont {Visser}(1998)}]{visser_acoustic_1998}%
  \BibitemOpen
  \bibfield  {author} {\bibinfo {author} {\bibfnamefont {M.}~\bibnamefont
  {Visser}},\ }\bibfield  {title} {\bibinfo {title} {Acoustic black holes:
  horizons, ergospheres and {Hawking} radiation},\ }\href@noop {} {\bibfield
  {journal} {\bibinfo  {journal} {Classical and Quantum Gravity}\ }\textbf
  {\bibinfo {volume} {15}},\ \bibinfo {pages} {1767} (\bibinfo {year}
  {1998})}\BibitemShut {NoStop}%
\bibitem [{\citenamefont {Jacquet}\ \emph {et~al.}(2022)\citenamefont
  {Jacquet}, \citenamefont {Joly}, \citenamefont {Claude}, \citenamefont
  {Giacomelli}, \citenamefont {Glorieux}, \citenamefont {Bramati},
  \citenamefont {Carusotto},\ and\ \citenamefont
  {Giacobino}}]{jacquet_analogue_2022}%
  \BibitemOpen
  \bibfield  {author} {\bibinfo {author} {\bibfnamefont {M.~J.}\ \bibnamefont
  {Jacquet}}, \bibinfo {author} {\bibfnamefont {M.}~\bibnamefont {Joly}},
  \bibinfo {author} {\bibfnamefont {F.}~\bibnamefont {Claude}}, \bibinfo
  {author} {\bibfnamefont {L.}~\bibnamefont {Giacomelli}}, \bibinfo {author}
  {\bibfnamefont {Q.}~\bibnamefont {Glorieux}}, \bibinfo {author}
  {\bibfnamefont {A.}~\bibnamefont {Bramati}}, \bibinfo {author} {\bibfnamefont
  {I.}~\bibnamefont {Carusotto}},\ and\ \bibinfo {author} {\bibfnamefont
  {E.}~\bibnamefont {Giacobino}},\ }\bibfield  {title} {\bibinfo {title}
  {Analogue quantum simulation of the {Hawking} effect in a polariton
  superfluid},\ }\href {https://doi.org/10.1140/epjd/s10053-022-00477-5}
  {\bibfield  {journal} {\bibinfo  {journal} {The European Physical Journal D}\
  }\textbf {\bibinfo {volume} {76}},\ \bibinfo {pages} {152} (\bibinfo {year}
  {2022})}\BibitemShut {NoStop}%
\bibitem [{\citenamefont {Alperin}\ and\ \citenamefont
  {Berloff}(2021)}]{alperin_multiply_2021}%
  \BibitemOpen
  \bibfield  {author} {\bibinfo {author} {\bibfnamefont {S.~N.}\ \bibnamefont
  {Alperin}}\ and\ \bibinfo {author} {\bibfnamefont {N.~G.}\ \bibnamefont
  {Berloff}},\ }\bibfield  {title} {\bibinfo {title} {Multiply charged vortex
  states of polariton condensates},\ }\href
  {https://doi.org/10.1364/OPTICA.418377} {\bibfield  {journal} {\bibinfo
  {journal} {Optica}\ }\textbf {\bibinfo {volume} {8}},\ \bibinfo {pages} {301}
  (\bibinfo {year} {2021})}\BibitemShut {NoStop}%
\bibitem [{\citenamefont {Jacquet}\ \emph {et~al.}(2020)\citenamefont
  {Jacquet}, \citenamefont {Boulier}, \citenamefont {Claude}, \citenamefont
  {Ma\'itre}, \citenamefont {Cancellieri}, \citenamefont {Adrados},
  \citenamefont {Amo}, \citenamefont {Pigeon}, \citenamefont {Glorieux},
  \citenamefont {Bramati},\ and\ \citenamefont
  {Giacobino}}]{jacquet_polariton_2020}%
  \BibitemOpen
  \bibfield  {author} {\bibinfo {author} {\bibfnamefont {M.~J.}\ \bibnamefont
  {Jacquet}}, \bibinfo {author} {\bibfnamefont {T.}~\bibnamefont {Boulier}},
  \bibinfo {author} {\bibfnamefont {F.}~\bibnamefont {Claude}}, \bibinfo
  {author} {\bibfnamefont {A.}~\bibnamefont {Ma\'itre}}, \bibinfo {author}
  {\bibfnamefont {E.}~\bibnamefont {Cancellieri}}, \bibinfo {author}
  {\bibfnamefont {C.}~\bibnamefont {Adrados}}, \bibinfo {author} {\bibfnamefont
  {A.}~\bibnamefont {Amo}}, \bibinfo {author} {\bibfnamefont {S.}~\bibnamefont
  {Pigeon}}, \bibinfo {author} {\bibfnamefont {Q.}~\bibnamefont {Glorieux}},
  \bibinfo {author} {\bibfnamefont {A.}~\bibnamefont {Bramati}},\ and\ \bibinfo
  {author} {\bibfnamefont {E.}~\bibnamefont {Giacobino}},\ }\bibfield  {title}
  {\bibinfo {title} {Polariton fluids for analogue gravity physics},\ }\href
  {https://doi.org/10.1098/rsta.2019.0225} {\bibfield  {journal} {\bibinfo
  {journal} {Philosophical Transactions of the Royal Society A}\ }\textbf
  {\bibinfo {volume} {378}},\ \bibinfo {pages} {20190225} (\bibinfo {year}
  {2020})}\BibitemShut {NoStop}%
\bibitem [{\citenamefont {Peres}(1996)}]{peres96}%
  \BibitemOpen
  \bibfield  {author} {\bibinfo {author} {\bibfnamefont {A.}~\bibnamefont
  {Peres}},\ }\bibfield  {title} {\bibinfo {title} {Separability criterion for
  density matrices},\ }\href {https://doi.org/10.1103/PhysRevLett.77.1413}
  {\bibfield  {journal} {\bibinfo  {journal} {Physical Review Letters}\
  }\textbf {\bibinfo {volume} {77}},\ \bibinfo {pages} {1413} (\bibinfo {year}
  {1996})}\BibitemShut {NoStop}%
\bibitem [{\citenamefont {Vidal}\ and\ \citenamefont {Werner}(2002)}]{vidal02}%
  \BibitemOpen
  \bibfield  {author} {\bibinfo {author} {\bibfnamefont {G.}~\bibnamefont
  {Vidal}}\ and\ \bibinfo {author} {\bibfnamefont {R.~F.}\ \bibnamefont
  {Werner}},\ }\bibfield  {title} {\bibinfo {title} {Computable measure of
  entanglement},\ }\href {https://doi.org/10.1103/PhysRevA.65.032314}
  {\bibfield  {journal} {\bibinfo  {journal} {Phys. Rev. A}\ }\textbf {\bibinfo
  {volume} {65}},\ \bibinfo {pages} {032314} (\bibinfo {year}
  {2002})}\BibitemShut {NoStop}%
\bibitem [{\citenamefont {Plenio}(2005)}]{plenio05}%
  \BibitemOpen
  \bibfield  {author} {\bibinfo {author} {\bibfnamefont {M.~B.}\ \bibnamefont
  {Plenio}},\ }\bibfield  {title} {\bibinfo {title} {Logarithmic negativity: A
  full entanglement monotone that is not convex},\ }\href
  {https://doi.org/10.1103/PhysRevLett.95.090503} {\bibfield  {journal}
  {\bibinfo  {journal} {Physical Review Letters}\ }\textbf {\bibinfo {volume}
  {95}},\ \bibinfo {pages} {090503} (\bibinfo {year} {2005})}\BibitemShut
  {NoStop}%
\bibitem [{\citenamefont {Weedbrook}\ \emph {et~al.}(2012)\citenamefont
  {Weedbrook}, \citenamefont {Pirandola}, \citenamefont
  {Garc{\'\i}a-Patr{\'o}n}, \citenamefont {Cerf}, \citenamefont {Ralph},
  \citenamefont {Shapiro},\ and\ \citenamefont {Lloyd}}]{weedbrook2012}%
  \BibitemOpen
  \bibfield  {author} {\bibinfo {author} {\bibfnamefont {C.}~\bibnamefont
  {Weedbrook}}, \bibinfo {author} {\bibfnamefont {S.}~\bibnamefont
  {Pirandola}}, \bibinfo {author} {\bibfnamefont {R.}~\bibnamefont
  {Garc{\'\i}a-Patr{\'o}n}}, \bibinfo {author} {\bibfnamefont {N.~J.}\
  \bibnamefont {Cerf}}, \bibinfo {author} {\bibfnamefont {T.~C.}\ \bibnamefont
  {Ralph}}, \bibinfo {author} {\bibfnamefont {J.~H.}\ \bibnamefont {Shapiro}},\
  and\ \bibinfo {author} {\bibfnamefont {S.}~\bibnamefont {Lloyd}},\ }\bibfield
   {title} {\bibinfo {title} {Gaussian quantum information},\ }\href
  {https://doi.org/10.1103/RevModPhys.84.621} {\bibfield  {journal} {\bibinfo
  {journal} {Reviews of Modern Physics}\ }\textbf {\bibinfo {volume} {84}},\
  \bibinfo {pages} {621} (\bibinfo {year} {2012})}\BibitemShut {NoStop}%
\bibitem [{\citenamefont {Serafini}(2017)}]{serafini17QCV}%
  \BibitemOpen
  \bibfield  {author} {\bibinfo {author} {\bibfnamefont {A.}~\bibnamefont
  {Serafini}},\ }\href@noop {} {\emph {\bibinfo {title} {Quantum continuous
  variables: a primer of theoretical methods}}}\ (\bibinfo  {publisher} {CRC
  press},\ \bibinfo {year} {2017})\BibitemShut {NoStop}%
\bibitem [{\citenamefont {Braunstein}(2005)}]{braunstein2005squeezing}%
  \BibitemOpen
  \bibfield  {author} {\bibinfo {author} {\bibfnamefont {S.~L.}\ \bibnamefont
  {Braunstein}},\ }\bibfield  {title} {\bibinfo {title} {Squeezing as an
  irreducible resource},\ }\href {https://doi.org/10.1103/PhysRevA.71.055801}
  {\bibfield  {journal} {\bibinfo  {journal} {Phys. Rev. A}\ }\textbf {\bibinfo
  {volume} {71}},\ \bibinfo {pages} {055801} (\bibinfo {year}
  {2005})}\BibitemShut {NoStop}%
\bibitem [{\citenamefont {Asb\'oth}\ \emph {et~al.}(2005)\citenamefont
  {Asb\'oth}, \citenamefont {Calsamiglia},\ and\ \citenamefont
  {Ritsch}}]{asboth2005EntanglementPotential}%
  \BibitemOpen
  \bibfield  {author} {\bibinfo {author} {\bibfnamefont {J.~K.}\ \bibnamefont
  {Asb\'oth}}, \bibinfo {author} {\bibfnamefont {J.}~\bibnamefont
  {Calsamiglia}},\ and\ \bibinfo {author} {\bibfnamefont {H.}~\bibnamefont
  {Ritsch}},\ }\bibfield  {title} {\bibinfo {title} {{Computable Measure of
  Nonclassicality for Light}},\ }\href
  {https://doi.org/10.1103/PhysRevLett.94.173602} {\bibfield  {journal}
  {\bibinfo  {journal} {Phys. Rev. Lett.}\ }\textbf {\bibinfo {volume} {94}},\
  \bibinfo {pages} {173602} (\bibinfo {year} {2005})}\BibitemShut {NoStop}%
\bibitem [{\citenamefont {Fr\'erot}\ \emph {et~al.}(2023)\citenamefont
  {Fr\'erot}, \citenamefont {Vashisht}, \citenamefont {Morassi}, \citenamefont
  {Lema\^{\i}tre}, \citenamefont {Ravets}, \citenamefont {Bloch}, \citenamefont
  {Minguzzi},\ and\ \citenamefont {Richard}}]{frerot_bogoliubov_2023}%
  \BibitemOpen
  \bibfield  {author} {\bibinfo {author} {\bibfnamefont {I.}~\bibnamefont
  {Fr\'erot}}, \bibinfo {author} {\bibfnamefont {A.}~\bibnamefont {Vashisht}},
  \bibinfo {author} {\bibfnamefont {M.}~\bibnamefont {Morassi}}, \bibinfo
  {author} {\bibfnamefont {A.}~\bibnamefont {Lema\^{\i}tre}}, \bibinfo {author}
  {\bibfnamefont {S.}~\bibnamefont {Ravets}}, \bibinfo {author} {\bibfnamefont
  {J.}~\bibnamefont {Bloch}}, \bibinfo {author} {\bibfnamefont
  {A.}~\bibnamefont {Minguzzi}},\ and\ \bibinfo {author} {\bibfnamefont
  {M.}~\bibnamefont {Richard}},\ }\bibfield  {title} {\bibinfo {title}
  {{Bogoliubov Excitations Driven by Thermal Lattice Phonons in a Quantum Fluid
  of Light}},\ }\href {https://doi.org/10.1103/PhysRevX.13.041058} {\bibfield
  {journal} {\bibinfo  {journal} {Phys. Rev. X}\ }\textbf {\bibinfo {volume}
  {13}},\ \bibinfo {pages} {041058} (\bibinfo {year} {2023})}\BibitemShut
  {NoStop}%
\bibitem [{\citenamefont {Parigi}\ \emph {et~al.}(2015)\citenamefont {Parigi},
  \citenamefont {D’Ambrosio}, \citenamefont {Arnold}, \citenamefont
  {Marrucci}, \citenamefont {Sciarrino},\ and\ \citenamefont
  {Laurat}}]{parigi_storage_2015}%
  \BibitemOpen
  \bibfield  {author} {\bibinfo {author} {\bibfnamefont {V.}~\bibnamefont
  {Parigi}}, \bibinfo {author} {\bibfnamefont {V.}~\bibnamefont
  {D’Ambrosio}}, \bibinfo {author} {\bibfnamefont {C.}~\bibnamefont
  {Arnold}}, \bibinfo {author} {\bibfnamefont {L.}~\bibnamefont {Marrucci}},
  \bibinfo {author} {\bibfnamefont {F.}~\bibnamefont {Sciarrino}},\ and\
  \bibinfo {author} {\bibfnamefont {J.}~\bibnamefont {Laurat}},\ }\bibfield
  {title} {\bibinfo {title} {Storage and retrieval of vector beams of light in
  a multiple-degree-of-freedom quantum memory},\ }\href
  {https://doi.org/10.1038/ncomms8706} {\bibfield  {journal} {\bibinfo
  {journal} {Nature Communications}\ }\textbf {\bibinfo {volume} {6}},\
  \bibinfo {pages} {7706} (\bibinfo {year} {2015})}\BibitemShut {NoStop}%
\bibitem [{\citenamefont {Wu}\ \emph {et~al.}(1986)\citenamefont {Wu},
  \citenamefont {Kimble}, \citenamefont {Hall},\ and\ \citenamefont
  {Wu}}]{Wu_opo_1986}%
  \BibitemOpen
  \bibfield  {author} {\bibinfo {author} {\bibfnamefont {L.-A.}\ \bibnamefont
  {Wu}}, \bibinfo {author} {\bibfnamefont {H.~J.}\ \bibnamefont {Kimble}},
  \bibinfo {author} {\bibfnamefont {J.~L.}\ \bibnamefont {Hall}},\ and\
  \bibinfo {author} {\bibfnamefont {H.}~\bibnamefont {Wu}},\ }\bibfield
  {title} {\bibinfo {title} {Generation of squeezed states by parametric down
  conversion},\ }\href {https://doi.org/10.1103/PhysRevLett.57.2520} {\bibfield
   {journal} {\bibinfo  {journal} {Phys. Rev. Lett.}\ }\textbf {\bibinfo
  {volume} {57}},\ \bibinfo {pages} {2520} (\bibinfo {year}
  {1986})}\BibitemShut {NoStop}%
\bibitem [{\citenamefont {Takeno}\ \emph {et~al.}(2007)\citenamefont {Takeno},
  \citenamefont {Yukawa}, \citenamefont {Yonezawa},\ and\ \citenamefont
  {Furusawa}}]{Takeno_opo_07}%
  \BibitemOpen
  \bibfield  {author} {\bibinfo {author} {\bibfnamefont {Y.}~\bibnamefont
  {Takeno}}, \bibinfo {author} {\bibfnamefont {M.}~\bibnamefont {Yukawa}},
  \bibinfo {author} {\bibfnamefont {H.}~\bibnamefont {Yonezawa}},\ and\
  \bibinfo {author} {\bibfnamefont {A.}~\bibnamefont {Furusawa}},\ }\bibfield
  {title} {\bibinfo {title} {Observation of -9 db quadrature squeezing with
  improvement of phase stability in homodyne measurement},\ }\href
  {https://doi.org/10.1364/OE.15.004321} {\bibfield  {journal} {\bibinfo
  {journal} {Opt. Express}\ }\textbf {\bibinfo {volume} {15}},\ \bibinfo
  {pages} {4321} (\bibinfo {year} {2007})}\BibitemShut {NoStop}%
\bibitem [{\citenamefont {Lvovsky}\ and\ \citenamefont
  {Raymer}(2009)}]{lvovsky_tomo_2009}%
  \BibitemOpen
  \bibfield  {author} {\bibinfo {author} {\bibfnamefont {A.~I.}\ \bibnamefont
  {Lvovsky}}\ and\ \bibinfo {author} {\bibfnamefont {M.~G.}\ \bibnamefont
  {Raymer}},\ }\bibfield  {title} {\bibinfo {title} {Continuous-variable
  optical quantum-state tomography},\ }\href
  {https://doi.org/10.1103/RevModPhys.81.299} {\bibfield  {journal} {\bibinfo
  {journal} {Rev. Mod. Phys.}\ }\textbf {\bibinfo {volume} {81}},\ \bibinfo
  {pages} {299} (\bibinfo {year} {2009})}\BibitemShut {NoStop}%
\bibitem [{\citenamefont {Agullo}\ \emph
  {et~al.}(2022{\natexlab{a}})\citenamefont {Agullo}, \citenamefont {Brady},\
  and\ \citenamefont {Kranas}}]{Agullo:2021vwj}%
  \BibitemOpen
  \bibfield  {author} {\bibinfo {author} {\bibfnamefont {I.}~\bibnamefont
  {Agullo}}, \bibinfo {author} {\bibfnamefont {A.~J.}\ \bibnamefont {Brady}},\
  and\ \bibinfo {author} {\bibfnamefont {D.}~\bibnamefont {Kranas}},\
  }\bibfield  {title} {\bibinfo {title} {{Quantum Aspects of Stimulated Hawking
  Radiation in an Optical Analog White-Black Hole Pair}},\ }\href
  {https://doi.org/10.1103/PhysRevLett.128.091301} {\bibfield  {journal}
  {\bibinfo  {journal} {Phys. Rev. Lett.}\ }\textbf {\bibinfo {volume} {128}},\
  \bibinfo {pages} {091301} (\bibinfo {year} {2022}{\natexlab{a}})},\ \Eprint
  {https://arxiv.org/abs/2107.10217} {arXiv:2107.10217 [gr-qc]} \BibitemShut
  {NoStop}%
\bibitem [{\citenamefont {Brady}\ \emph {et~al.}(2022)\citenamefont {Brady},
  \citenamefont {Agullo},\ and\ \citenamefont {Kranas}}]{Brady:2022ffk}%
  \BibitemOpen
  \bibfield  {author} {\bibinfo {author} {\bibfnamefont {A.~J.}\ \bibnamefont
  {Brady}}, \bibinfo {author} {\bibfnamefont {I.}~\bibnamefont {Agullo}},\ and\
  \bibinfo {author} {\bibfnamefont {D.}~\bibnamefont {Kranas}},\ }\bibfield
  {title} {\bibinfo {title} {{Symplectic circuits, entanglement, and stimulated
  Hawking radiation in analogue gravity}},\ }\href
  {https://doi.org/10.1103/PhysRevD.106.105021} {\bibfield  {journal} {\bibinfo
   {journal} {Phys. Rev. D}\ }\textbf {\bibinfo {volume} {106}},\ \bibinfo
  {pages} {105021} (\bibinfo {year} {2022})},\ \Eprint
  {https://arxiv.org/abs/2209.11317} {arXiv:2209.11317 [gr-qc]} \BibitemShut
  {NoStop}%
\bibitem [{\citenamefont {Agullo}\ \emph
  {et~al.}(2022{\natexlab{b}})\citenamefont {Agullo}, \citenamefont {Brady},\
  and\ \citenamefont {Kranas}}]{Agullo:2022oye}%
  \BibitemOpen
  \bibfield  {author} {\bibinfo {author} {\bibfnamefont {I.}~\bibnamefont
  {Agullo}}, \bibinfo {author} {\bibfnamefont {A.~J.}\ \bibnamefont {Brady}},\
  and\ \bibinfo {author} {\bibfnamefont {D.}~\bibnamefont {Kranas}},\
  }\bibfield  {title} {\bibinfo {title} {{Event horizons are tunable factories
  of quantum entanglement}},\ }\href
  {https://doi.org/10.1142/S0218271822420081} {\bibfield  {journal} {\bibinfo
  {journal} {Int. J. Mod. Phys. D}\ }\textbf {\bibinfo {volume} {31}},\
  \bibinfo {pages} {2242008} (\bibinfo {year} {2022}{\natexlab{b}})},\ \Eprint
  {https://arxiv.org/abs/2209.09980} {arXiv:2209.09980 [gr-qc]} \BibitemShut
  {NoStop}%
\bibitem [{\citenamefont {Švančara}\ \emph {et~al.}(2023)\citenamefont
  {Švančara}, \citenamefont {Smaniotto}, \citenamefont {Solidoro},
  \citenamefont {MacDonald}, \citenamefont {Patrick}, \citenamefont {Gregory},
  \citenamefont {Barenghi},\ and\ \citenamefont
  {Weinfurtner}}]{svancara_exploring_2023}%
  \BibitemOpen
  \bibfield  {author} {\bibinfo {author} {\bibfnamefont {P.}~\bibnamefont
  {Švančara}}, \bibinfo {author} {\bibfnamefont {P.}~\bibnamefont
  {Smaniotto}}, \bibinfo {author} {\bibfnamefont {L.}~\bibnamefont {Solidoro}},
  \bibinfo {author} {\bibfnamefont {J.~F.}\ \bibnamefont {MacDonald}}, \bibinfo
  {author} {\bibfnamefont {S.}~\bibnamefont {Patrick}}, \bibinfo {author}
  {\bibfnamefont {R.}~\bibnamefont {Gregory}}, \bibinfo {author} {\bibfnamefont
  {C.~F.}\ \bibnamefont {Barenghi}},\ and\ \bibinfo {author} {\bibfnamefont
  {S.}~\bibnamefont {Weinfurtner}},\ }\href
  {https://doi.org/10.48550/arXiv.2308.10773} {\bibinfo {title} {Exploring the
  {Quantum}-to-{Classical} {Vortex} {Flow}: {Quantum} {Field} {Theory}
  {Dynamics} on {Rotating} {Curved} {Spacetimes}}} (\bibinfo {year}
  {2023})\BibitemShut {NoStop}%
\bibitem [{\citenamefont {Wald}(1995)}]{Wald:1995yp}%
  \BibitemOpen
  \bibfield  {author} {\bibinfo {author} {\bibfnamefont {R.~M.}\ \bibnamefont
  {Wald}},\ }\href@noop {} {\emph {\bibinfo {title} {{Quantum Field Theory in
  Curved Space-Time and Black Hole Thermodynamics}}}},\ Chicago Lectures in
  Physics\ (\bibinfo  {publisher} {University of Chicago Press},\ \bibinfo
  {address} {Chicago, IL},\ \bibinfo {year} {1995})\BibitemShut {NoStop}%
\bibitem [{\citenamefont {Patrick}(2020)}]{patrick_rotational_2020}%
  \BibitemOpen
  \bibfield  {author} {\bibinfo {author} {\bibfnamefont {S.}~\bibnamefont
  {Patrick}},\ }\bibfield  {title} {\bibinfo {title} {Rotational superradiance
  with {Bogoliubov} dispersion},\ }\href@noop {} {\bibfield  {journal}
  {\bibinfo  {journal} {arXiv:2011.04620 [cond-mat, physics:gr-qc,
  physics:physics]}\ } (\bibinfo {year} {2020})}\BibitemShut {NoStop}%
\end{thebibliography}%

\appendix
\clearpage
\onecolumngrid

\section{Proof of the theorem characterizing superradiance}\label{app:proof}

{Here we provide a proof for {\bf Theorem 1}, which states: \textit{The matrix $\bs{B}$ describing the scattering from IN to OUT modes off a time-independent potential corresponds to superradiant scattering if and only if it is not a unitary matrix.}}

{To prove this theorem, we will actually prove its negation: the scattering is not superradiant if and only if $\bs{B}$ is unitary. Firstly, we recall that $\bs{B}$ is the matrix of complex coefficients
\beq
\bs{B}=\left(\begin{array}{cc} T & r\cr R & t\end{array}\right)\, ,
\eeq
that relate the amplitude of IN and OUT wave packet basis elements with central frequency $\omega$ in the considered scattering process (see section \ref{ClassicalSR}). We find that}
\beq 
\bs{B}^{\dagger}\cdot \bs{B}=
\left(\begin{array}{cc} |T|^2 +|R|^2& T^* r +R^* t\cr T r^* +R t^* & |t|^2 +|r|^2\end{array}\right)\,.
\eeq
Unitarity of $\bs{B}$ requires the following constraints 
\begin{equation}
    |T|^2 +|R|^2=1\, ,\qquad\qquad|t|^2 +|r|^2=1 \qquad\text{and}\qquad T r^* +R t^*=0\,,
\end{equation} 
implying that $|T|,|R|,|t|$ and $|r|$ are all smaller than unity and the scattering is non-superradiant. 

To prove that  non-superradiant scattering implies unitarity of $\bs{B}$, let us use the conserved quantity $Q_{\rm KG}$ to define a (non-positive definite) Hermitian inner product in the space of complex solutions to the Klein-Gordon equation
\beq \langle W_1, W_2\rangle \equiv  Q_{\rm KG}(W_1,W_2)=i/\hbar \int (W_1^*\,  {\Pi_{W_2}}-{\Pi_{W_1}^*} \, W_2)\,  dx\,, \label{eq:SympProdAp}\eeq
{ where $\Pi_{W_i}$ denotes the conjugate momentum of the solution $W_i$.}  This product is commonly referred to as the Klein-Gordon or symplectic product (see, e.g., \cite{Wald:1995yp}). Without loss of generality, we can work with  wave packets normalized to have $Q_{\rm KG}(W_i,W_i)=\pm 1$. Furthermore, it is easy to check that left- and right-moving wave packets are orthogonal. With this, conservation of   $Q_{\rm KG}$ implies
\begin{equation}
\begin{split}
    &\langle W^{\rm in}_{r},W^{\rm in}_{r}\rangle=|T|^2 \langle W^{\rm out}_{r},W^{\rm out}_{r} \rangle + |R|^2 \langle W^{\rm out}_{l},W^{\rm out}_{l}\rangle\\
    &\langle W^{\rm in}_{l},W^{\rm in}_{l}\rangle=|r|^2 \langle W^{\rm out}_{r},W^{\rm out}_{r} \rangle + |t|^2 \langle W^{\rm out}_{l},W^{\rm out}_{l}\rangle\\
    &\langle W^{\rm in}_{r},W^{\rm in}_{l}\rangle=T^* r \langle W^{\rm out}_{r},W^{\rm out}_{r} \rangle + R^* t\langle W^{\rm out}_{l},W^{\rm out}_{l}\rangle \, .
\end{split}
\label{eq:ConsNomrConstraints}
\end{equation}
For a non-superradiant scattering,   $|T|,|R|,|t|$ and $|r|$ are all smaller than unity. With this, relations 
\eqref{eq:ConsNomrConstraints} imply that the KG norm of each of the four wave packets must have the same sign, either $+1$ or $-1$. Furthermore, \eqref{eq:ConsNomrConstraints} also implies 
\begin{equation} \label{eq:NSupConstraints}
    |T|^2 +|R|^2=1\qquad,\qquad|t|^2 +|r|^2=1 \qquad\text{and}\qquad T r^* +R t^*=0\,,
\end{equation} 
implying that $\bs{B}$ is a unitary matrix. This proves the theorem.\\

We conclude this Appendix by presenting several expressions employed in the main text. It has been established that, in the context of non-superradiant scattering, all four modes participating in the scattering process exhibit the same sign of $Q_{\rm KG}$. In contrast, in superradiant scattering scenarios, the two IN modes bear different signs of $Q_{\rm KG}$, and the same applies to the two OUT modes.

{When, for superradiant scattering, there is one mode with norm of each sign at the same side of the interaction region, $\langle W^{in}_{r},W^{in}_{r}\rangle=\langle W^{out}_{r},W^{out}_{r}\rangle$, and} Eq.~\eqref{eq:ConsNomrConstraints} lead to the following constraints 
 \begin{equation}
    |T|^2 - |R|^2 =1\qquad,\qquad |t|^2 - |r|^2=1 \qquad\text{and}\qquad T r^* - R t^*=0\,.
    \label{eq:SupConstraints}
\end{equation} 
{On the other hand, when both modes at each side of the interaction region have norm of the same sign, $\langle W^{in}_{r},W^{in}_{r}\rangle=-\langle W^{out}_{r},W^{out}_{r}\rangle$, } Eq.~\eqref{eq:ConsNomrConstraints} produce
 \begin{equation}
    |R|^2 - |T|^2 =1\qquad,\qquad |r|^2 - |t|^2=1 \qquad\text{and}\qquad T r^* - R t^*=0\,. \label{eq:SupConstraints2}
\end{equation} 
These two sets of constraints indicate that either reflection or transmission coefficients (or both) have an absolute value greater than one, indicating superradiant scattering.  

\section{Gaussian states, entanglement quantifiers, and entanglement witnesses}\label{app:Gaussian}

This appendix contains a brief summary of techniques of Gaussian states for bosonic quantum systems on which we have based the calculations in the main body of the paper. These are well-known tools, and detailed descriptions of them can be found, for instance, in \cite{weedbrook2012,serafini17QCV}. In comparing the content of this appendix with other references, including \cite{weedbrook2012,serafini17QCV}, the reader should bear in mind that we work here with creation and annihilation operators, rather than Hermitian combinations of them. This introduces some differences in the expressions we write below. 

A bosonic quantum system with $N$ degrees of freedom can be described in terms of $2N$ creation and anihilation operators $\hat{\boldsymbol{A}}=(\hat{a}_1,...,\hat{a}_N,\hat{a}^\dagger_1,...,\hat{a}^\dagger_N)$ which satisfy standard conmmutation relations
\begin{equation}
    \lrsq{\hat{A}^I,\hat{A}^J}=\Omega^{IJ}\,,\qquad\text{where}\qquad \bs{\Omega}\coloneqq
    \begin{pmatrix}
        0_N & \mathbb{I}_N\\
        -\mathbb{I}_N & 0_N 
    \end{pmatrix}
\end{equation}
is the symplectic form and $I,J\in\{1,...,2N\}$. A general state for such system will be characterized by a density matrix $\hat{\rho}$, which encodes the information of the infinitely many moments of the creation and annihilation operators 
\begin{equation}\label{creanncmr}
\expval{\hat{A}^{I_1}...\,\hat{{A}}^{I_n}}\coloneqq\Tr[\hat{\rho}\hat{A}^{I_1}...\,\hat{{A}}^{I_n}]\, ,
\end{equation}
with $n\in\mathbb{N}$. Of special relevance for us are the first and the (symmetrized and centered) second moments, known as mean vector and covariance matrix, whose components are respectively
\begin{equation} \label{eq:moments}
    \mu^I=\expval{\hat{A}^I}\qquad\text\qquad \sigma^{IJ}=\expval{\{\hat{A}^I-\mu^I,\hat{A}^J-\mu^J\}}\, ,
\end{equation}
where curly brackets denote the anti-commutator. Among all possible quantum states, there is a subset called Gaussian states, which have the remarkable property that all their moments can be written in terms of their first and second moments, in analogy to what occurs for Gaussian probability distributions. Hence, these states are fully characterized by their mean and covariance matrix, which contain complete information on the physical properties of the state. Furthermore, Gaussian states are ubiquitous, with prominent examples being the ground or thermal states of any quadratic Hamiltonian, and any state that can be prepared by applying symplectic transformations to such states. As example, the vacuum is characterized by 
\begin{equation}
    \bs{\mu}_{\rm vac}={\bs 0} \qquad\text{and}\qquad \bs{\sigma}_{\rm vac}
    =
    \begin{pmatrix}
        0_N & \mathbb{I}_N\\
        \mathbb{I}_N & 0_N
    \end{pmatrix}\,,\label{eq:Vacuum}
\end{equation}
a coherent state is characterized by having the same covariance matrix as the vacuum but a ``displaced'' mean, namely $\bs{\mu}_{\rm coh}\neq {\bs 0}$ and $\bs{\sigma}_{\rm coh}=\bs{\sigma}_{\rm vac}$. A thermal state is of the form $\boldsymbol{\mu}={\bs 0}$ and $\bs{\sigma}=\oplus_j (1+2n_j)\, \bs{\sigma}_{\rm vac}$, where $n_j$ is the mean number of quanta in the mode $j$, with $j\in\{1,...,N\}$, and the covariance matrices in the direct sum are those corresponding to a vacuum 1-mode system. 

Furthermore, if an $N$-mode system is in a Gaussian state, the reduced state of any subset of the $N$ modes is also Gaussian.

The evolution of linear quantum systems ---i.e., systems described by a Hamiltonian at most quadratic in the creation and annihilation operators--- is completely determined by the classical evolution. Hence, evolution can be encoded in a linear symplectic transformation $\bs{S}$. These are $2N\times2N$ complex matrices  of the form 
\beq \label{formS} {\bs S}=\left(\begin{array}{cc} {\bs \alpha} & {\bs \beta}\cr {\bs \beta}^*& {\bs \alpha}^*\end{array}\right)\, ,\eeq 
with ${\bs \alpha}$ and ${\bs \beta}$ complex $N\times N$ matrices, and satisfying that they leave invariant the symplectic form, i.e.,  $\bs{S}\cdot \bs{\Omega} \cdot \bs{S}^{\top}=\bs{\Omega}$.
This condition implies 
\beq  {\bs \alpha}\cdot {\bs \alpha}^{\dagger}-{\bs \beta}\cdot {\bs \beta}^{\dagger}=\mathbb{I}_2\, , \ \ {\bs \alpha}\cdot {\bs \beta}^{\top}-{\bs \beta} \cdot {\bs \alpha}^{\top}=0_2 \, .\eeq
Matrices ${\bs S}$ of this form constitute a group isomorphic to the symplectic group Sp(2N,$\mathbb{R})$.

Moreover, linear evolution preserves Gaussianity. If the initial state at time  $t_0$ is Gaussian and characterized by $\mu^i_{t_0}$ and $\sigma^{ij}_{t_0}$, the state of the system will be Gaussian at all times, and given by $\mu^i_t=S{}^{i}{}_k\mu^k_0$ and $\sigma_t{}^{ij}=S{}^{i}{}_k\sigma_{t_0}{}^{km}S_{}^{j}{}_m$, or, in matrix notation
\begin{equation}
    \bs{\mu}_t=\bs{S}\cdot\bs{\mu}_{t_0}\qquad\text{and}\qquad\bs{\sigma}_t=\bs{S}\cdot\bs{\sigma}_{t_0}\cdot\bs{S}^\top.
    \label{eq:EvolutionGaussian}
\end{equation}
In the Heisenberg picture, this evolution is translated to the creation and annihilation operators as 
\begin{equation}
    \bs{A}_{t}=\bs{S}\cdot\bs{A}_{t_0}\,.
\end{equation}
Because ${\bs S}$ is symplectic, this transformation preserves the standard commutation relations written in \eqref{creanncmr}.  

We can express the mean number of quanta in a given mode $i$ as
\begin{equation}
    \langle \hat{N}_i\rangle=\frac{1}{2}\sigma^{i,i+N}+\mu^i\mu^{i+N}-\frac{1}{2},
\label{eq:NquantaGaussian}
\end{equation}
We use this expression to compute the number of quanta for the cases considered in Eqs.~\eqref{eq:NquantVac} to \eqref{eq:NquantlSq}. Starting with the vacuum case, we have an initial state characterized by \eqref{eq:Vacuum} for $N=2$. As in the text, we consider a scenario where modes at the left of the interacting region, $W_r^{\rm in}$ and $W_l^{\rm out}$, have negative norm. The matrix $\bs{S}$ describing superradiant scattering (Eq.~\eqref{eq:GeneralSuperradiantMatrix} in the main text) is
\begin{equation}
 \bS_{\rm SR}=
 \begin{pmatrix}
     0 & r & T & 0 \\
     R^*& 0 & 0  &t^* \\
     T^* &0 & 0 &r^*  \\
     0 & t &R & 0  
 \end{pmatrix}\,.
\end{equation}
The  state after the scattering is 
\begin{equation}
    \begin{split}
    &\bs{\mu}_{\rm out}=\bS\cdot\bs{\mu}_{\rm vac}={\bs 0}
    \\
    &\bs{\sigma}_{\rm out}
    =\bs{S}\cdot\bs{\sigma}_{\rm vac}\cdot\bS^\top=
    \begin{pmatrix}
        0 & 2TR^* & 1+2|T|^2 & 0 \\
        2TR^* & 0 & 0 & 1+2|T|^2\\
        1+2|T|^2 & 0 & 0 & 2RT^* \\
        0 & 1+2|T|^2 & 2RT^* & 0
    \end{pmatrix}\,,    
    \end{split}
    \label{eq:Vacuum}
\end{equation}
where we have used $|R|^2=1+|T|^2$, $|r|=|R|$, $|t|=|T|$ as well as $tr^*=RT^*$, which hold for superradiant scattering. We see that the reduced covariance matrices correspond to a thermal state with $|T|^2$ number of quanta for each of the modes, as expected for a two-mode squeezed vacuum. Indeed, using Eq.~\eqref{eq:NquantaGaussian}, we find that mean numbers of quanta for each mode are
\begin{equation}
    \langle \hat{N}_r^{\rm out}\rangle=\frac{1}{2}(\sigma^{13}-1)=|T|^2\qquad\text{and}\qquad \langle \hat{N}_\ell^{\rm out}\rangle=\frac{1}{2}(\sigma^{24}-1)=|T|^2\, 
\end{equation}
As a remark, note that the fact that each pair of modes is in a thermal state does not mean that the number of quanta per frequency -- i.e. the dependence of $|T|$ on $\omega$ -- follows a black body distribution with a common temperature for all modes. 
 
Starting now with a coherent input state $\ket{\gamma_r,\gamma_\ell}$, with  mean $\bs{\mu}_{\rm coh}=(\gamma_r,\gamma_\ell,\gamma_r^*,\gamma_\ell^*)^{\top}$ and same covariance matrix as the vacuum, we find that
\begin{equation}
    \bs{\mu}_{\rm out}=\big(T\gamma_r^*+r \gamma_\ell\;,\;R^*\gamma_r+t^*\gamma_\ell^*\;,\;T^*\gamma_r + r^* \gamma_\ell^* \;,\;R\gamma_r^* + t\gamma_\ell\big)^\top\, ,
\end{equation}
and the  $\bs{\sigma}_{\rm out}$ given in \eqref{eq:Vacuum}. Thus, we find
\begin{align}
    &\langle \hat{N}_r^{\rm out}\rangle=\frac{1}{2}(\sigma_{\rm out}^{13}-1)+{\mu}_{\rm out}^{1}{\mu}_{\rm out}^{3}={|\gamma_l|^2}+{|T|^2}+{|T|^2\lr{|\gamma_r|^2+|\gamma_\ell|^2}+2\text{Re}\lrsq{rT^*\gamma_r\gamma_\ell}},\\
    &\langle \hat{N}_\ell^{\rm out}\rangle=\frac{1}{2}(\sigma_{\rm out}^{24}-1)+{\mu}_{\rm out}^{2}{\mu}_{\rm out}^{4}={|\gamma_r|^2}+{|T|^2}+{|T|^2\lr{|\gamma_\ell|^2+|\gamma_r|^2}+2\text{Re}\lrsq{tR^*\gamma_\ell\gamma_r}}.
\end{align}
These are the expressions given in expressions \eqref{eq:NquantrCoh} and \eqref{eq:NquantlCoh} in the main text. 

Finally,  a single-mode squeezed state is a Gaussian state 
input characterized by $\bs{\mu}_{\rm sq}=0$ and 
\begin{equation}
    \bs{\sigma}_{\rm sq}=
    \begin{pmatrix}
        0 & 0 & 1 & 0\\
        0 & -\ee^{\ci\phi_l}\sinh{2z_l} & 0 & \cosh{2z_l}\\
        1 & 0 & 0 & 0\\
        0 & \cosh{2z_l} & 0 & -\ee^{-\ci\phi_l}\sinh{2z_l}
    \end{pmatrix}\,,
\end{equation}
where we have squeezed only the second mode $W_l^{\rm in}$. $z_l\in \mathbb{R}$ is called the squeezing intensity, and $\phi_l\in [0,2\pi)$ is the squeezing angle. For this single-mode squeezed state as an input, the state after superradiant scattering is characterized by $\bs{\mu}_{\rm out}=0$ and
\begin{equation}
    \bs{\sigma}_{\rm out}
    =
    \begin{pmatrix}
        -\ee^{\ci\phi_l} r^2 \sinh{2z_l} & 2TR^*\cosh^2z_l & (1+2|T|^2)\cosh^2z_l+\sinh^2z_l & -\ee^{\ci\phi_l} rt \sinh{2z_l} \\
        2TR^*\cosh^2z_l & -\ee^{-\ci\phi_l} t^*{}^2 \sinh{2z_l} & -\ee^{-\ci\phi_l} r^*t^* \sinh{2z_l} & 1+2|T|^2\cosh^2 z_l\\
        (1+2|T|^2)\cosh^2z_l+\sinh^2z_l & -\ee^{-\ci\phi_l} r^*t^* \sinh{2z_l} & -\ee^{-\ci\phi_l} r^*{}^2 \sinh{2z_l} & 2RT^*\cosh^2z_l\\
        -\ee^{\ci\phi_l} rt \sinh{2z_l} & 1+2|T|^2\cosh^2 z_l & 2RT^*\cosh^2z_l & -\ee^{\ci\phi_l} t^2 \sinh{2z_l}
    \end{pmatrix}\,.
\end{equation}
where we have used identities between hyperbolic trigonometric functions and the relations satisfied the scattering coefficients.

Again, using formula \eqref{eq:NquantaGaussian} to compute the number of particles in each mode, we find
\begin{align}
    &\langle \hat{N}_r^{\rm out}\rangle=\frac{1}{2}(\sigma_{\rm out}^{13}-1)= {\sinh^2{z_l}}+{|T|^2}+{|T|^2\sinh^2{z_l}}\;,\\
    &\langle \hat{N}_\ell^{\rm out}\rangle=\frac{1}{2}(\sigma_{\rm out}^{24}-1)={|T|^2} + {|T|^2 \sinh^2{z_l}}
\end{align}
which correspond to formulas \eqref{eq:NquantrSq} and \eqref{eq:NquantlSq} in the main text.

Many properties of the state can be extracted from $\bs{\sigma}$ alone. For instance, the absolute value of the eigenvalues of $\bs{\sigma}\cdot\bs{\Omega}^{-1}$ is always greater or equal to 1, and the state is pure if and only if all of these eigenvalues are equal to $\pm 1$. For Gaussian states, the information in $\bs{\sigma}$ is also enough to compute the von Neumann entropy of the state and the Logarithmic Negativity (LN) associated with any bi-partition of the system. These are quantities that we have used to quantify the entanglement generated by superradiance.

The  computation of these quantities from the covariance matrix of a given Gaussian state proceeds as follows. The von Neumann entropy of a Gaussian state is given by (see, e.g. \cite{serafini17QCV})
\begin{equation}
S[\bs{\sigma}]=\sum_k^N\left[\left(\frac{\nu_k+1}{2}\right) {\log_2} \left(\frac{\nu_k+1}{2}\right)-\left(\frac{\nu_k-1}{2}\right) {\log_2} \left(\frac{\nu_k-1}{2}\right)\right]\, ,
\label{eq:VNEntropy}
\end{equation}
where $\nu_k$ are the so-called symplectic eigenvalues of the covariance matrix $\bs{\sigma}$, namely the modulus of the eigenvalues of $\bs{\sigma}\cdot\bs{\Omega}^{-1}$. There are only $N$ such eigenvalues, even though these matrices are $2N$-dimensional, because eigenvalues come in pairs, with the members of each pair differing by a sign. This sign is irrelevant since $\nu_k$ are defined as the modulus of the eigenvalues.

Similarly, given the Gaussian state $(\bs{\mu}_{AB},\bs{\sigma}_{AB})$ (with $A$ and $B$ two subsystems), the Logarithmic Negativity can be computed from the symplectic eigenvalues of the so-called partially transposed state~\cite{vidal02,plenio05} --- where the transpose is taken with respect to either of the two subsystems. If subsystem $A$ consists of $p$ modes, labeled by $\{k_1,...,k_p\}$, with $k_j\in\{1,...,N\}$ and $1\leq p\leq N-1$, the partially transposed state with respect to $A$ is the Gaussian state with the same mean $\bs{\mu}_{AB}$ but covariance matrix $\tilde{\bs{\sigma}}_{AB}$ obtained from $\bs{\sigma}_{AB}$ just by swapping the rows $k_j$ and $k_{j+N}$ and the columns $k_j$ and $k_{j+N}$ for all the modes of the subsystem $A$. The Logarithmic Negativity of the bi-partition $AB$ in state $\bs{\sigma}_{AB}$ is thus given by the formula
\begin{equation}
    {\rm LogNeg}[{\bm\sigma}_{AB}]=\sum_i {\rm Max}[0, -\log_2 \tilde \nu_i]
    \label{eq:LN}
\end{equation}
where $\tilde{\nu}_i$ are the symplectic eigenvalues of $\tilde{\bm\sigma}_{AB}$. 

\subsection*{Modeling loss}

In this subsection, we provide some additional details of the loss model used in section \ref{sec:ProductionOfEntanglement}.

A simple theory to model loss of a bosonic mode $\hat{a}$ is given by the simple input-output relation $\hat{a}\rightarrow \sqrt{\eta}\, \hat{a}+\sqrt{1-\eta}\, \hat{e}$, where $0\leq\eta\leq1$ is the transmittance and $\hat{e}$ is the annihilation operator of an environment mode following thermal Gaussian statistics---i.e., with mean zero, 
$\expval{\hat{e}}=0$, and variance determined by the thermal population, $\expval{\hat{e}^\dagger\hat{e}}=n_{\rm env}$. To illustrate the model with a short example, consider a strong classical wave, i.e., a coherent state, with intensity (mean number of quanta) $I_{\rm in}\gg n_{\rm env}$. The output intensity is then $I_{\rm in}\rightarrow I_{\rm out}=\eta I_{\rm in}+ (1-\eta)n_{\rm env}\approx \eta I_{\rm in}$. In other words, the input intensity is attenuated by an amount $\eta$. This model is often employed to represent detector losses, where the efficiency of the detector, expressed as a percentage, is $100\times\eta$. 

For simplicity, we consider ``pure loss'', such that $n_{\rm env}=0$, i.e. the environment is close to the vacuum. On Gaussian states,  the pure loss channel transforms the mean and the covariance matrix via

\bea {\bs \mu}  &\rightarrow &  \sqrt{\eta}\, {\bs \mu} \nonumber \\ {\bs \sigma}  & \rightarrow &  \eta\, {\bs \sigma} + (1-\eta)\,  \bs{\sigma}_{\rm vac}\, , \eea 
where $\bs{\sigma}_{\rm vac}$ is the covariance matrix of the vacuum. This transformation reflects the fact that quanta make it to the detector with probability $\eta$: for $\eta=0$, all quanta are lost and only the vacuum remains, whilst $\eta=1$ corresponds to perfect collection of the outgoing radiation.

\section{Kinematics of collective excitations in rotating polariton fluids}\label{app:dispbogo}

\begin{figure}[ht]
    \centering
    \includegraphics[width=.95\textwidth]{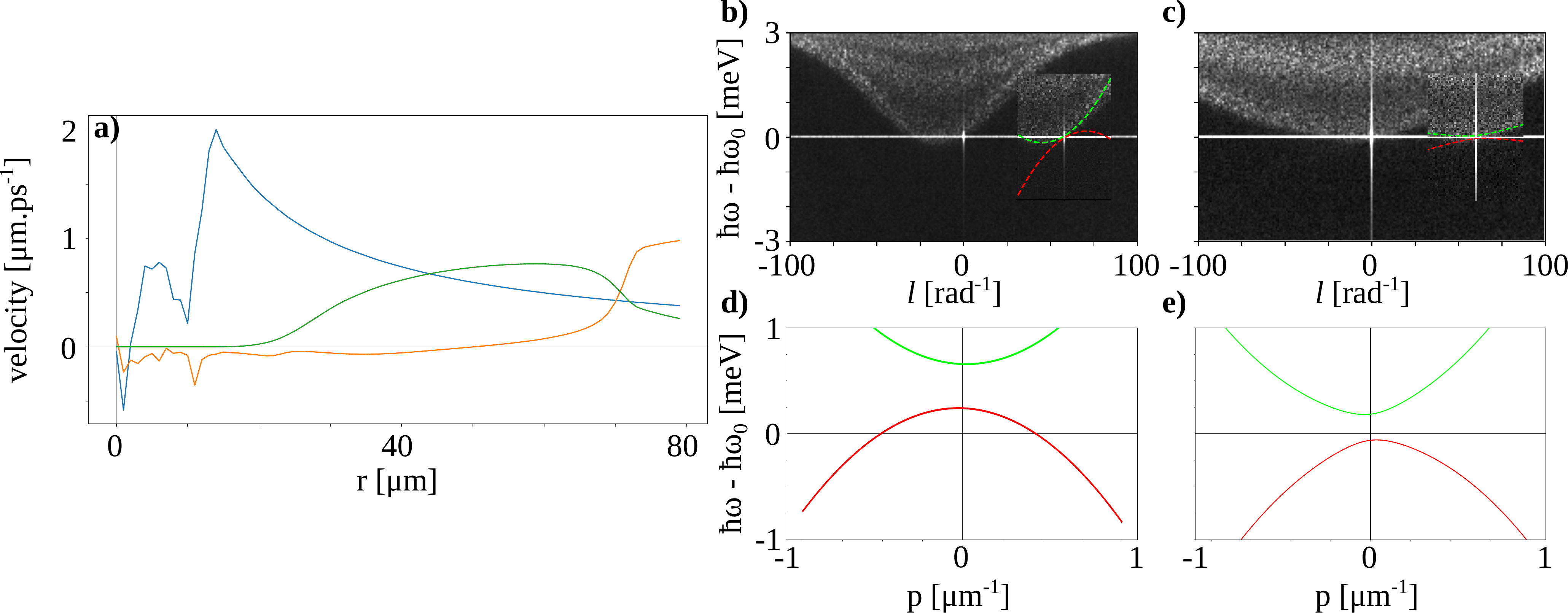}
    \caption{Collective excitation in the inhomogeneous fluid.
    \textbf{a)} Velocities along a radial cut in Fig.~\ref{fig:fig_num_vort_pol} \textbf{f)}.
    Green, $c_s$; orange, $v_r$; blue, $v_\theta$.
    \textbf{b)} $\ell-\omega$ dispersion in the supersonic region.
    \textbf{c)} $\ell-\omega$ dispersion in the subsonic region.
    Insets: zoom near $\ell=\ell_{vortex}$ and $\omega=0$ and fit of WKB dispersion.
    Green, positive-norm branch; red, negative-norm branch.
    \textbf{d)} $p-\omega$ WKB dispersion in the subsonic region.
    \textbf{e)} $p-\omega$ WKB dispersion in the supersonic region.}
    \label{fig:num_disp}
\end{figure}

In this appendix, we study the kinematics of collective excitations ---which are described by a Klein-Gordon field--- on the effective curved geometry created by the mean-field of the fluid.

The equation or linear perturbations $\delta\psi$ around the background fluid described by $\psi^0$ satisfying the stationary equation~\eqref{eq:bistab} is
\begin{equation}
    \ci (\partial_t+\bs{\tv}^0\cdot\bs{\nabla})\delta\psi=\lrsq{-\frac{\hbar}{2m_{LP}}\nabla^2-\frac{\ci \hbar}{2m_{LP}}\lr{\bs{\nabla} \cdot\bs{v}^0}-\delta+2gn^0-\ci\frac{\gamma}{2}}\delta\psi+g n^0 \delta\psi^*,
    \label{eq:appDefPertGPE}
\end{equation}
where  $n^0$ and $\bs{v}^0$ are the mean-field density and flow velocity, respectively.
$\delta$ is the effective frequency detuning between the laser and the lower polariton branch at $k$. 
Eqn~\eqref{eq:appDefPertGPE} can be rewritten as a system of linear equations for the fields $\delta\psi$ and $\delta\psi^*$ as
\begin{equation}
    \ci (\partial_t+\bs{\tv}^0\cdot\bs{\nabla})
    \begin{pmatrix}
    \delta\psi\\
    \delta\psi^*
    \end{pmatrix}
    =
    \mathcal{L}\begin{pmatrix}
    \delta\psi\\
    \delta\psi^*
    \end{pmatrix}
    =
    \begin{pmatrix}
    A& gn^0\\
    -gn^0& -A^*
    \end{pmatrix}
    \begin{pmatrix}
    \delta\psi\\
    \delta\psi^*
    \end{pmatrix}
    \label{eq:appDefPertGPEMatrix}
\end{equation}
where $A=-\frac{\hbar}{2m_{LP}}\nabla^2-\frac{\ci \hbar}{2m_{LP}}\lr{\bs{\nabla} \cdot\bs{v}^0}-\delta+2gn^0-\ci\frac{\gamma}{2}$.

In time-independent backgrounds, modes with different frequencies decouple, and Eqn~\eqref{eq:appDefPertGPEMatrix} turns into an eigenvalue problem for the differential operator $\mathcal{L}$.
If the background is also homogeneous, the modes can be expanded in plane waves with fluid-rest-frame wavevector $\bs{k}$, yielding the Bogoliubov dispersion relation~\cite{carusotto_quantum_2013}
\begin{equation}
    \omega-\bs{\tv}^0\cdot\bs{k}=-\frac{\ci\gamma}{2}
    \pm\sqrt{\frac{\hbar^2k^4}{4m_{LP}^2}+\frac{\hbar k^2}{m_{LP}}\lr{2 g n^0-\delta}+\lr{g n^0 - \delta}\lr{3 g n^0 - \delta}}.
    \label{eq:appDispersion}
\end{equation}
In this driven-dissipative system, the frequency detuning $\delta$ controls a mass gap in the dispersion (which vanishes only if $\delta=g n^0,\,3gn^0$).

While Eqn~\eqref{eq:appDispersion} describes well the spectrum of collective excitations in configurations without rotation in which plane waves are the eigenfunctions of the system~\cite{claude_2022}, this is not the case when working with configurations with rotational symmetry around an axis.
In that case, the basis is best obtained by decomposing in the eigenfunctions of the angular momentum operator $L_\theta=-\ci\partial/\partial\theta$, which have the form $\ee^{i\ell\theta}$ with eigenvalue $\ell$, and which will not mix during the evolution.

\begin{figure}[ht]
    \centering
    \includegraphics[width=.4\textwidth]{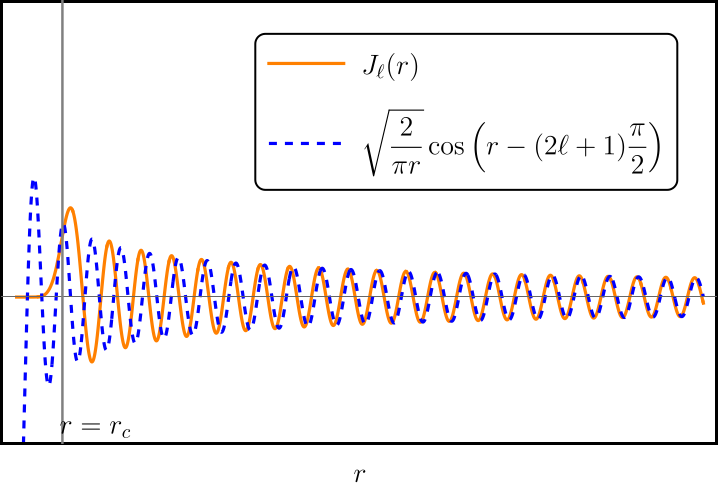}
    \includegraphics[width=.4\textwidth]{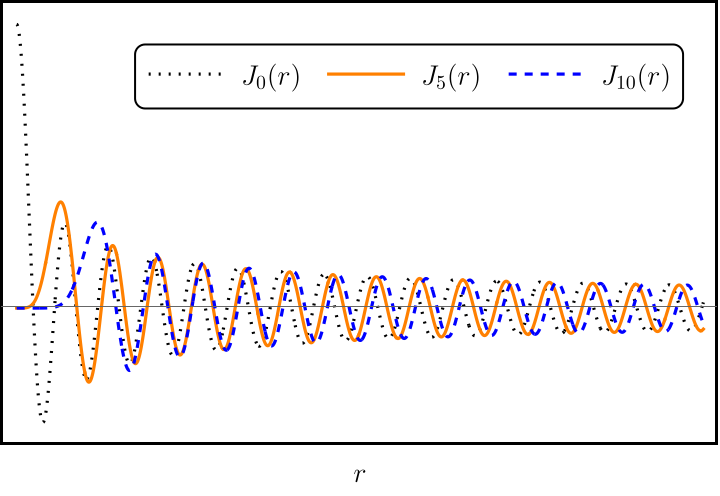}
    \includegraphics[width=.4\textwidth]{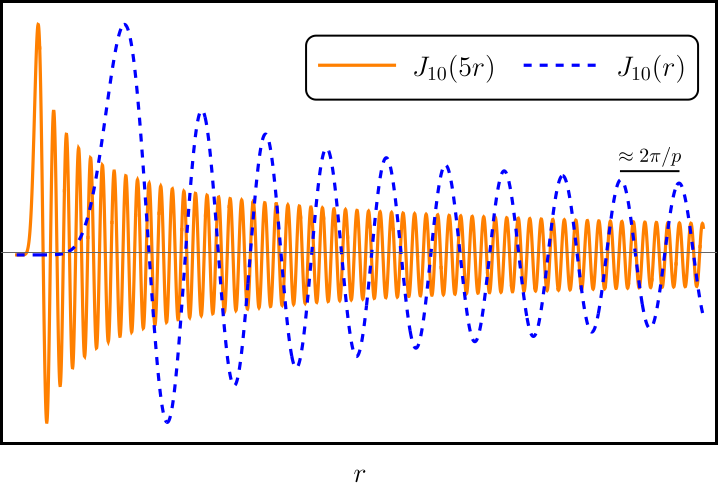}
    \includegraphics[width=.4\textwidth]{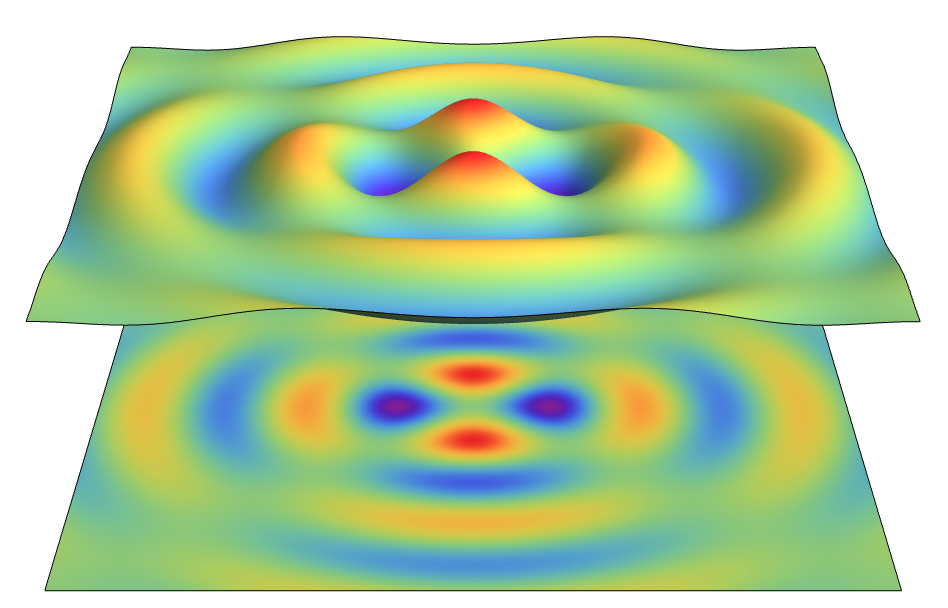}
    \caption{Bessel functions of the first kind for different values of $\ell$ and p as functions of the radius. In the top left graphic, we plot $J_\ell(r)$ and its asymptotic plane wave expansion for $\ell=10$. We see how there is a radius $r_c$ below which the oscillatory behavior turns into an exponentially decaying tail that vanishes at the origin. In the top right graphic we plot $J_\ell(r)$ for $\ell\in\{0,5,10\}$, which shows how $r_0$ monotonically increases with $\ell$, and vanishes for $\ell=0$. In the bottom left graphic we plot $J_{10}(pr)$ for $p\in\{1,5\}$. This plot shows how $r_c$ decreases with $p$, and how the value of $p$ can can be seen as a local radial wave number that controls how fast the oscillations occur in the radial direction. $p$ locally defines a wavelength as $\lambda=2\pi/p$, which is the wavelength that appears in the asymptotic plane wave expansion. The bottom right graphic is the real part of $\ee^{\ci\ell\theta}J_\ell(pr)$ for $\ell=2$ and $p=1$. We can see how $\ell$ is the number of oscillations of the corresponding eigenfunction as $\theta$ goes from 0 to $2\pi$ at a fixed radius.}
    \label{fig:Bessel}
\end{figure}

In the rest frame of the fluid, the problem is solved by finding eigenfunctions of the operator $\mathcal{L}$.
For a given $\ell$, these are the eigenfunctions of both the Laplacian operator $\nabla^2$ and the angular momentum operator $L_\theta$ and are of the form $\ee^{\ci\ell\theta}J_\ell(p r), \ee^{\ci\ell\theta}Y_\ell(p r)$, {with eigenvalues $-p^2$. $J_\ell$ and $Y_\ell$ are Bessel functions of the first and second kind respectively. They asymptotically approach radial plane waves far from the origin, where their amplitude has an oscillatory wave-like profile.
However, below an $\ell$- and $p$-dependent radius $r_c$, $Y_\ell$ diverges while the amplitude of $J_{\ell}$ exponentially decays (much like an evanescent wave). $r_c$ decreases with $p$ and increases with $\ell$, vanishing for $\ell=0$ (see figure \ref{fig:Bessel}). Physically, this can be thought of as a consequence of the effective centrifugal potential barrier, which is not felt by $\ell=0$ waves, but diverges as $r\to 0$ for $\ell\neq0$ waves.}

{Thus, while the eigenvalue $\ell$ is an angular wave number that counts the number of oscillations that occur in a circulation at a fixed radius, the eigenvalue $p$ is not exactly a wave number --- it can be thought of as a local wave number that controls how fast oscillations occur in the radial direction, defining a local wavelength as $2\pi/p$. Indeed, it is possible to arrive at a local dispersion relation through the WKB approximation, which for a profile with vanishing radial velocity reads}
\begin{equation}
    \omega-\frac{v_\theta\ell}{r}=-\frac{\ci\gamma}{2}
    \pm\sqrt{\frac{\hbar^2}{4m_{LP}^2}\lr{p^2+\frac{\ell^2}{r^2}}^2+\frac{\hbar \lr{2 g n^0-\delta}}{m_{LP}}\lr{p^2+\frac{\ell^2}{r^2}}+\lr{g n^0 - \delta}\lr{3 g n^0 - \delta}}
    \label{eq:appDispersionp}
\end{equation}
{where the $\pm$ branches are related to modes of positive/negative norm.} Using the WKB approximation, Eqn~\eqref{eq:appDispersion} can be adapted, with $\bs{k}=(p,\ell/r)$, to recover an expression in which the labels can be physically interpreted as usual~\cite{patrick_rotational_2020}.
The resulting $\ell$-dependent terms that do not vanish in the $p\rightarrow0$ limit yield a position- and $\ell$-dependent contribution to the mass gap in the spectrum of cylindrical waves, which effectively accounts for the effect of the centrifugal barrier near the origin (locally evanescent amplitudes).
In the hydrodynamical approximation, neglecting cavity losses and with $v_r=0$, the WKB dispersion is
\begin{equation}
    \omega-\frac{v_\theta\ell}{r}=
    \pm c_s\sqrt{p^2+\frac{\ell^2}{r^2}},
    \label{eq:appWKBdisp}
\end{equation}
{where $c_s=\sqrt{\hbar gn^0/m}$ (assuming that $\delta=gn^0$). From Eqn~\eqref{eq:appWKBdisp}, we find the condition for positive-(negative-)norm modes to be ``locally propagating'' at a given radius $r_0$, namely, to have real-valued $p$ at positive laboratory-frame frequency at a given radius $r_0$, is given by}
\begin{eqnarray}
    \omega-\frac{v_\theta(r_0)\ell}{r_0}>\frac{c_s \ell}{r_0}\hspace{2cm}\text{and}\hspace{2cm} \omega-\frac{v_\theta(r_0)\ell}{r_0}<-\frac{c_s \ell}{r_0},
\end{eqnarray}
{respectively. Laboratory frequencies outside this range have imaginary solutions for $p$, so that they describe a mode that is ``locally evanescent'' --- energy in such a mode cannot propagate in that spatial region.}

We now study the spectrum of collective excitations.
We use the Truncated Wigner approximation to add small amplitude noise at all spatial and temporal frequencies to the numerical simulation of the GPE~\eqref{eq:GPEpol}.
This noise creates collective excitations at resonant-frequencies in the fluid.

Fig.~\ref{fig:num_disp} shows the dispersion of collective excitations on either side of the ergosurface.
In numerical simulations, after having reached the steady-state of the GPE~\eqref{eq:GPEpol}, we subtract the mean-field background and perform angular and temporal Fourier transforms inside $r-$windows to obtain the local noise spectrum in the $\ell-\omega$ plane.
We plot spectra at radii \textbf{b)} inside ($r=\SI{25}{\micro\meter}$) and \textbf{c)} outside ($r=\SI{67}{\micro\meter}$) the ergoregion.
The vortex mean-field is still visible as a vertical spike at $\ell_{vortex}=0$ ($\ell$ and $\bs{p}$ are defined in the rest frame of the fluid). We observe that the positive-norm branch is more visible than the negative-norm branch.

Insets show zooms on the azimuthal numbers near $\ell_{vortex}$, to which the dispersion obtained by the WKB approximation is superimposed (green, positive-norm branch; red, negative-norm branch).
The WKB dispersion describes well the noise spectrum over regions where the density is high enough for nonlinear self-interactions. As explained above, modes with $\ell>0$ have an $\ell/r$-dependent mass gap induced by the fluid rotation. This has also been evidenced in experiments in superfluid helium~\cite{svancara_exploring_2023}. 

We also calculate the dispersion~\eqref{eq:appDispersion} in the laboratory frame for modes with $\ell=10$ at the same two radii with the WKB approximation ($\delta\neq gn^0$ as in the experiment).
We remark that, due to the finite gap inside and outside the ergosurface, there are intervals in which we find no real roots at positive laboratory-frame frequencies on either side of the interface.
As explained above, this signals the presence of evanescent waves in these regions~\cite{patrick_rotational_2020}.

In Fig.\ref{fig:num_disp} \textbf{d)}, we see that the increase in $v_\theta$ across the ergosurface pulls the negative-norm branch to positive frequency, signaling the possibility to excite negative-energy modes and to have superradiant scattering.
And indeed, there is a narrow positive-frequency interval for which there are positive-norm mode solutions outside, and negative-norm mode solutions inside the ergosurface.
Note that, because of the finite lifetime of polaritons, the group velocity of these rightward and leftward propagating modes has to be taken into account when characterizing the scattering coefficients across the ergosurface --- this defines their mean-free path as well as amplitude damping before and after the scattering process has occurred.

\section{Calculation of the \textit{\textbf{S}} matrix {and wave packet picture}}\label{sec:SmatrixObtention}

Our goal is to obtain, for each mode $(\omega,\ell)$, the coefficients $T_{\omega\ell},R_{\omega\ell},t_{\omega\ell}$ and $r_{\omega\ell}$ defined from these two   scattering processes of wave packets (see Fig.~\ref{fig:scattering} for an illustration)
\bea W^{\rm in}_r &\to& T_{\omega\ell}\, W^{\rm out}_r +R_{\omega\ell}\, W^{\rm out}_l\, , \nonumber \\ W^{\rm in}_l&\to& t_{\omega\ell}\, W^{\rm out}_l +r_{\omega\ell}\, W^{\rm out}_r   \, . \eea
Because of the time-independence and rotational symmetry of the problem, the calculation reduces to solve an ordinary differential equation in the radial direction (Eq.~\eqref{eq:SpatialODE}). We proceed as follows. 

Eq.~\eqref{assymp} contains the (asymptotic) form of our basis of IN and OUT modes near $r_{\rm min}$ and $r_{\rm max}$
\begin{equation}
\begin{split}
\varphi^{\rm in}_r(r_{\rm min})&=N_{\omega\ell}\, {\rm exp}\left\{\ci\lr{\frac{\omega r_{\rm min}}{c}-\frac{\ell v_\theta(r_{\rm min})}{c_s}\log{\frac{r}{r_{\rm min}}}}\right\}\, , \\ 
\varphi^{\rm in}_l(r_{\rm max})&= \tilde N_{\omega\ell}\, {\rm exp}\left\{-\ci\lr{\frac{\omega r_{\rm max}}{c}-\frac{\ell v_\theta(r_{\rm max})}{c_s}\log{\frac{r}{r_{\rm max}}}}\right\}\, ,  \\ 
\varphi^{\rm out}_r(r_{\rm max})&=\tilde N_{\omega\ell}\, {\rm exp}\left\{\ci\lr{\frac{\omega r_{\rm max}}{c}-\frac{\ell v_\theta(r_{\rm max})}{c_s}\log{\frac{r}{r_{\rm max}}}}\right\}\, ,  \\
\varphi^{\rm out}_l(r_{\rm min})&=  N_{\omega\ell}\, {\rm exp}\left\{-\ci\lr{\frac{\omega r_{\rm min}}{c}-\frac{\ell v_\theta(r_{\rm min})}{c_s}\log{\frac{r}{r_{\rm min}}}}\right\}\,.
\end{split}
\label{eq:AsymptoticModes}
\end{equation}
The names in/out and $r/l$ are motivated from the direction in which these modes propagate once the time-dependence $\ee^{\ci\, \omega t}$ is added to them.  $N_{\omega\ell}$ and $\tilde N_{\omega\ell}$ are two normalization constants. When $v_{\theta}(r_{\rm max})\approx 0$, as in the case under consideration, $\tilde N_{\omega\ell}$ can be obtained by noticing that Bessel functions are exact solutions when $r\sim r_{\rm max}$, resulting in $\tilde N_{\omega\ell}=1/\sqrt{4\pi\omega}$. This strategy does not work to obtain $N_{\omega\ell}$, because $v_{\theta}(r_{\rm min})\neq0$. We explain below how to compute this normalization constant.

For each mode $(\omega,\ell)$, we solve the radial differential equation \eqref{eq:SpatialODE} twice. On the one hand, we solve it using boundary data corresponding to the function $\varphi^{\rm in}_l(r)$ and its first derivative at $r_{\rm max}$. We propagate the solution until $r_{\rm min}$ where it becomes a linear combination of the two independent solutions $\varphi^{\rm in}_r(r_{\rm min})$ and $\varphi^{\rm out}_l(r_{\rm min})$. In other words, the asymptotic behavior of this solution is
\begin{equation}
    \varphi^{(1)}_{\omega\ell}(r)=
    \begin{cases}
        a_{\omega\ell} \, \varphi^{\rm out}_{l} + b_{\omega\ell} \, \varphi^{\rm in}_{r} &r\rightarrow r_{\rm min}\\[10pt]
         \varphi^{\rm in}_{l} &r\rightarrow r_{\rm max}
    \end{cases}
\end{equation}
By fitting the numerical solution near $r_{\rm min}$ to a linear combination of $\varphi^{\rm out}_l$ and $\varphi^{\rm in}_r$ we extract the coefficients $a_{\omega\ell}$ and $b_{\omega\ell}$.
On the other hand, we repeat the calculation, this time with boundary data specified by $\varphi^{\rm out}_r(r)$ at $r_{\rm max}$. The asymptotic form of the resulting solution is
\beq 
\varphi^{(2)}_{\omega\ell}(r)=
    \begin{cases}
        c_{\omega\ell}\, \varphi^{\rm in}_{r} + d_{\omega\ell}\, \varphi^{\rm out}_{l} &r\rightarrow r_{\rm min}\\[10pt]
        \varphi^{\rm out}_{r} &r\rightarrow r_{\rm max}
    \end{cases}\,,
\end{equation}
from which we extract $c_{\omega\ell}$ and $d_{\omega\ell}$.

If the time and angular dependence $e^{-i \omega t}e^{i \ell \theta}$ is added to these radial functions, we obtain two ``stationary'' solutions to the wave equation. They are stationary in the sense that they do not describe propagating waves. Instead, they oscillate perpetually in time by virtue of their harmonic time dependence $e^{-i \omega t}$. 

The trick to transform these stationary solutions into a dynamical scattering of waves is to integrate these solutions in a frequency range as follows:
\begin{equation}
\label{eq:WavepacketConstruction}
{\frac{1}{\sqrt{\epsilon}}}\int_{j\epsilon}^{(j+1)\epsilon} \dd\omega \, \ee^{\ci\omega \frac{n}{\epsilon}}\, \ee^{-\ci\omega\, t}\ee^{\ci\ell\theta}\varphi^{(i)}_{\omega\ell}(r)\,,
\end{equation}
for $i=1,2$. In this equation, $j>0$ and $n$ are integers, and $\epsilon$ is a real number with dimensions of frequency. For $i=1$, this function describes  the following dynamical  process 
\beq b_{\omega\ell}\, W^{\rm in}_r + W^{\rm in}_l \xrightarrow{\text{time}}  a_{\omega\ell}\,  W^{\rm out}_l \, ,\eeq
where $W^{\rm in}_r, W^{\rm in}_l$ and $W^{\rm out}_l$ are wave packets with mean frequency $\omega=(j+1/2)\epsilon$ and bandwidth $\Delta \omega=\epsilon$. The integer $n$ controls the region in spacetime in which these packets are supported. 
One can select $\epsilon$ to be small enough for the coefficients $a_{\omega\ell}$ and $b_{\omega\ell}$  not to change significantly with $\omega$ within the interval $\Delta \omega$. Similarly, the solution for $i=2$ describes the scattering 
\beq c_{\omega\ell}\, W^{\rm in}_r \xrightarrow{\text{time}}  d_{\omega\ell}\, W^{\rm out}_l  +  W^{\rm out}_l \, .\eeq
The linearity of the dynamics allows us to take linear combinations of these two scattering processes to obtain information about another scattering process of interest. In particular, two appropriate linear combinations produce 
\begin{equation}
    W_r^{\rm in} \xrightarrow{\text{time}} \frac{d_{\omega\ell}}{c_{\omega\ell}}\, W_l^{\rm out}+ \frac{1}{c_{\omega\ell}}\, \Phi_r^{\rm out}\,,
\end{equation}
and 
\begin{equation}
W_l^{\rm in}\xrightarrow{\text{time}} \lr{a_{\omega\ell}-\frac{b_{\omega\ell}d_{\omega\ell}}{c_{\omega\ell}}}W_l^{\rm out} + \frac{b_{\omega\ell}}{c_{\omega\ell}}W_r^{\rm out},
\end{equation}
from which we can read  the coefficients we are interested in
\begin{equation}
    R_{\omega\ell}=\frac{d_{\omega\ell}}{c_{\omega\ell}} \, , \hspace{0.5cm} T_{\omega\ell}=\frac{1}{c_{\omega\ell}}\, , \hspace{0.5cm} t_{\omega\ell}=a_{\omega\ell}-\frac{b_{\omega\ell}d_{\omega\ell}}{c_{\omega\ell}} \hspace{0.5cm}\text{and}\hspace{0.5cm}  r_{\omega\ell}=\frac{b_{\omega\ell}}{c_{\omega\ell}}\,.
\end{equation} 
This strategy allows us to determine the scattering coefficients $T_{\omega\ell}, R_{\omega\ell}, t_{\omega\ell}$ and $r_{\omega\ell}$ for each mode $(\omega,\ell)$, up to the normalization constant $N_{\omega\ell}$. This constant 
can be determined by demanding  the scattering matrix $\bs{S}_{SR}$ to be symplectic.  Notice that this is a non-trivial requirement, since symplecticity of $\bs{S}_{SR}$ imposes four constraints on  $T_{\omega\ell}, R_{\omega\ell}, t_{\omega\ell}$ and $r_{\omega\ell}$ (Eq.~\eqref{eq:SupConstraints}, which correspond to four real conditions). We have checked that there is a unique value of $N_{\omega\ell}$ for which all four constraints are satisfied. Therefore, besides obtaining this normalization constant, our calculation actually provides a non-trivial test of symplecticity. In table \ref{tab:TableData} we collect the result of our calculations, from which the plots shown in sections \ref{sec:ProductionOfEntanglement} of this article can be reproduced. 

{As a useful remark let us note that, in general, given any solution $\Phi(t,\bs{x})$ of our field equations, we can define an associated ``wave packet'' $W$ as above
\begin{equation}
W=\frac{1}{\sqrt{\epsilon}}\int_{j\epsilon}^{(j+1)\epsilon} \dd\omega \, \ee^{\ci\omega \frac{n}{\epsilon}}\,\Phi(t,\bs{x})\,.
\end{equation}
In a stationary problem, where solutions are labeled by frequency $\omega$, we can build an orthonormal basis of solutions with wave packets associated to the Fourier modes. Each wave packet can be understood as encoding one degree of freedom of the field and, as such, each has a canonically conjugated momentum $\Pi_W$ defined as
\begin{equation}
\Pi_{W}=\frac{c_W}{\sqrt{\epsilon}}\int_{j\epsilon}^{(j+1)\epsilon} \dd\omega \, \ee^{\ci\omega \frac{n}{\epsilon}}\,\Pi(t,\bs{x})\,
\end{equation}
where $\Pi$ is the momentum canonically conjugate to $\Phi$. If the wavepacket basis is orthonormal with respect to the symplectic product \eqref{eq:SympProdAp} ---see also equation \eqref{eq:KGNorm}--- the constant $c_W$ can be chosen so that the canonical commutation relations between independent degrees of freedom are satisfied for all the basis elements.
}

\begin{table}[!htb]
    \centering
    \begin{minipage}[t]{.46\linewidth}
      \centering
        \setlength{\tabcolsep}{8pt} 
        \renewcommand{\arraystretch}{1.2} 
        \resizebox{\textwidth}{!} {
        \begin{tabular}{|c| c c c|}
            \hline
            
            \multicolumn{4}{|c|}{$\ell = 1$} \\
            \hline
            $\omega$ [$\mu$eV] & T & r $=-$ R & t \\
            \hline
            
            6.540 & $1.3410+0.2178 i$ & $0.0809+0.9161 i$ & $1.2821-0.4495 i$ \\
            
            7.063 & $1.3839+0.2507 i$ & $0.1415+0.9788 i$ & $1.2563-0.6324 i$ \\
            
            7.586 & $1.426+0.285 i$ & $0.2104+1.0351 i$ & $1.202-0.819 i$ \\
            
            8.110 & $1.468+0.320 i$ & $0.2870+1.0836 i$ & $1.117-1.004 i$ \\
            
            8.633 & $1.507+0.356 i$ & $0.3704+1.1228 i$ & $1.00-1.183 i$ \\
            
            9.156 & $1.544+0.392 i$ & $0.4594+1.1514 i$ & $0.850-1.347 i$ \\
            
            9.679 & $1.577+0.428 i$ & $0.5522+1.1682 i$ & $0.670-1.490 i$ \\
            
            10.20 & $1.605+0.465 i$ & $0.6470+1.1724 i$ & $0.462-1.606 i$ \\
            
            10.73 & $1.628+0.502 i$ & $0.7417+1.1636 i$ & $0.233-1.688 i$ \\
            
            11.25 & $1.646+0.539 i$ & $0.8338+1.1420 i$ & $-0.012-1.732 i$ \\
            
            11.77 & $1.657+0.577 i$ & $0.921+1.108 i$ & $-0.265-1.734 i$ \\
            
            12.30 & $1.660+0.615 i$ & $1.002+1.063 i$ & $-0.516-1.694 i$ \\
            
            12.82 & $1.657+0.653 i$ & $1.075+1.009 i$ & $-0.757-1.612 i$ \\
            
            13.34 & $1.646+0.693 i$ & $1.138+0.946 i$ & $-0.982-1.492 i$ \\
            
            13.86 & $1.628+0.733 i$ & $1.191+0.878 i$ & $-1.182-1.338 i$ \\
            
            14.39 & $1.603+0.775 i$ & $1.233+0.806 i$ & $-1.353-1.157 i$ \\
            
            14.91 & $1.570+0.817 i$ & $1.264+0.731 i$ & $-1.491-0.954 i$ \\
            
            15.43 & $1.531+0.860 i$ & $1.285+0.657 i$ & $-1.594-0.737 i$ \\
            
            15.96 & $1.485+0.903 i$ & $1.297+0.583 i$ & $-1.662-0.512 i$ \\
            
            16.48 & $1.434+0.947 i$ & $1.3003+0.5120 i$ & $-1.695-0.285 i$ \\
            
            17.00 & $1.377+0.990 i$ & $1.2959+0.4439 i$ & $-1.695-0.062 i$ \\
            
            17.53 & $1.315+1.032 i$ & $1.2849+0.3795 i$ & $-1.665+0.152 i$ \\
            
            18.05 & $1.249+1.073 i$ & $1.2683+0.3193 i$ & $-1.608+0.354 i$ \\
            \hline
        \end{tabular}
        }
    \end{minipage}%
   \hspace{.5cm}
    \begin{minipage}[t]{.46\linewidth}
      \centering
            \setlength{\tabcolsep}{8pt} 
            \renewcommand{\arraystretch}{1.29} 
            \resizebox{\textwidth}{!} {
            \begin{tabular}{|c| c c c|}
            \hline
            
            \multicolumn{4}{|c|}{$\ell = 2$} \\
            \hline
            $\omega$ [$\mu$eV] & T & r $=-$ R & t \\
            \hline
            
            13.34 & $1.1841+0.0743 i$ & $-0.1037+0.6300 i$ & $1.1455+0.3090 i$ \\
            
            14.39 & $1.2189+0.1010 i$ & $-0.0403+0.7030 i$ & $1.2224+0.0389 i$ \\
            
            15.43 & $1.2520+0.1346 i$ & $0.0362+0.7645 i$ & $1.2337-0.2525 i$ \\
            
            16.48 & $1.2816+0.1760 i$ & $0.1229+0.8114 i$ & $1.1720-0.5476 i$ \\
            
            17.53 & $1.3053+0.2260 i$ & $0.2158+0.8416 i$ & $1.0355-0.8262 i$ \\
            
            18.57 & $1.3211+0.2850 i$ & $0.3107+0.8544 i$ & $0.8295-1.0671 i$ \\
            
            19.62 & $1.3271+0.3533 i$ & $0.4034+0.8504 i$ & $0.5659-1.2513 i$ \\
            
            20.67 & $1.3215+0.4304 i$ & $0.4904+0.8313 i$ & $0.2625-1.3648 i$ \\
            
            21.71 & $1.3026+0.5156 i$ & $0.5687+0.7995 i$ & $-0.0597-1.3997 i$ \\
            
            22.76 & $1.2693+0.6074 i$ & $0.6366+0.7580 i$ & $-0.3789-1.3551 i$ \\
            
            23.81 & $1.2204+0.7039 i$ & $0.6934+0.7099 i$ & $-0.6751-1.2365 i$ \\
            
            24.85 & $1.1552+0.8026 i$ & $0.7390+0.6577 i$ & $-0.9312-1.0543 i$ \\
            
            25.90 & $1.0735+0.9009 i$ & $0.7741+0.6040 i$ & $-1.1349-0.8222 i$ \\
            
            26.94 & $0.9752+0.9957 i$ & $0.7996+0.5504 i$ & $-1.2782-0.5555 i$ \\
            
            27.99 & $0.8607+1.0838 i$ & $0.8167+0.4985 i$ & $-1.3574-0.2700 i$ \\
            
            29.04 & $0.7312+1.1621 i$ & $0.8267+0.4490 i$ & $-1.3728+0.0193 i$ \\
            
            30.08 & $0.5880+1.2274 i$ & $0.8307+0.4026 i$ & $-1.3278+0.2989 i$ \\
            
            31.13 & $0.4331+1.2769 i$ & $0.8299+0.3597 i$ & $-1.2279+0.5571 i$ \\
            
            32.18 & $0.2689+1.3080 i$ & $0.8251+0.3202 i$ & $-1.0809+0.7843 i$ \\
            
            33.22 & $0.0984+1.3187 i$ & $0.8172+0.2841 i$ & $-0.8953+0.9731 i$ \\
            
            34.27 & $-0.0752+1.3071 i$ & $0.8068+0.2514 i$ & $-0.6807+1.1184 i$ \\
            
            35.32 & $-0.2483+1.2724 i$ & $0.7946+0.2218 i$ & $-0.4467+1.2170 i$ \\
            
            36.36 & $-0.4173+1.2140 i$ & $0.7809+0.1952 i$ & $-0.2029+1.2676 i$ \\
            \hline
            \end{tabular}
            }
    \end{minipage} 
    
    \begin{minipage}[t]{.46\linewidth}
      \centering
            \setlength{\tabcolsep}{8pt} 
            \renewcommand{\arraystretch}{1.2} 
            \resizebox{\textwidth}{!} {
            \begin{tabular}{|c| c c c|}
            \hline
            
            \multicolumn{4}{|c|}{$\ell = 3$} \\
            \hline
            $\omega$ [$\mu$eV] & T & r $=-$ R & t \\
            \hline
            
            19.62 & $1.1085+0.0208 i$ & $-0.2272+0.4214 i$ & $0.6264+0.9148 i$ \\
            
            21.19 & $1.1390+0.0352 i$ & $-0.1721+0.5185 i$ & $0.9341+0.6527 i$ \\
            
            22.76 & $1.1686+0.0585 i$ & $-0.0910+0.6007 i$ & $1.1335+0.2905 i$ \\
            
            24.33 & $1.1943+0.0937 i$ & $0.0092+0.6596 i$ & $1.1912-0.1271 i$ \\
            
            25.90 & $1.2125+0.1440 i$ & $0.1193+0.6905 i$ & $1.0939-0.5426 i$ \\
            
            27.47 & $1.2198+0.2117 i$ & $0.2290+0.6929 i$ & $0.8534-0.8969 i$ \\
            
            29.04 & $1.2124+0.2974 i$ & $0.3296+0.6706 i$ & $0.5051-1.1416 i$ \\
            
            30.61 & $1.1868+0.4003 i$ & $0.4153+0.6296 i$ & $0.0992-1.2486 i$ \\
            
            32.18 & $1.1396+0.5173 i$ & $0.4834+0.5767 i$ & $-0.3103-1.2124 i$ \\
            
            33.75 & $1.0674+0.6436 i$ & $0.5340+0.5181 i$ & $-0.6755-1.0475 i$ \\
            
            35.32 & $0.9677+0.7729 i$ & $0.5688+0.4587 i$ & $-0.9604-0.7820 i$ \\
            
            36.89 & $0.8392+0.8974 i$ & $0.5903+0.4016 i$ & $-1.1429-0.4510 i$ \\
            
            38.46 & $0.6820+1.0090 i$ & $0.6013+0.3488 i$ & $-1.2144-0.0911 i$ \\
            
            40.02 & $0.4983+1.0989 i$ & $0.6042+0.3012 i$ & $-1.1775+0.2633 i$ \\
            
            41.59 & $0.2922+1.1590 i$ & $0.6011+0.2593 i$ & $-1.0435+0.5829 i$ \\
            
            43.16 & $0.0701+1.1821 i$ & $0.5939+0.2228 i$ & $-0.8302+0.8445 i$ \\
            
            44.73 & $-0.1596+1.1627 i$ & $0.5838+0.1913 i$ & $-0.5593+1.0318 i$ \\
            
            46.30 & $-0.3871+1.0973 i$ & $0.5718+0.1643 i$ & $-0.2546+1.1354 i$ \\
            
            47.87 & $-0.6014+0.9851 i$ & $0.5586+0.1415 i$ & $0.0601+1.1526 i$ \\
            
            49.44 & $-0.7911+0.8282 i$ & $0.5449+0.1221 i$ & $0.3620+1.0866 i$ \\
            
            51.01 & $-0.9456+0.6317 i$ & $0.5309+0.1058 i$ & $0.6311+0.9460 i$ \\
            
            52.58 & $-1.0550+0.4035 i$ & $0.5170+0.0922 i$ & $0.8505+0.7433 i$ \\
            
            54.15 & $-1.1118+0.1541 i$ & $0.5033+0.0808 i$ & $1.0076+0.4945 i$ \\
            \hline
         \end{tabular}
         }
    \end{minipage} 
   \hspace{.4cm}
    \begin{minipage}[t]{.46\linewidth}
      \centering
        \setlength{\tabcolsep}{8pt} 
        \renewcommand{\arraystretch}{1.2} 
        \resizebox{\textwidth}{!} {
        \begin{tabular}{|c| c c c|}
            \hline
            
            \multicolumn{4}{|c|}{$\ell = 4$} \\
            \hline
            $\omega$ [$\mu$eV] & T & r $=-$ R & t \\
            \hline
            
            26.68 & $1.0795+0.0172 i$ & $-0.3017+0.2730 i$ & $-0.0905+1.0758 i$ \\
            
            28.78 & $1.1089+0.0287 i$ & $-0.2626+0.4018 i$ & $0.4714+1.0041 i$ \\
            
            30.87 & $1.1372+0.0498 i$ & $-0.1757+0.5147 i$ & $0.9302+0.6562 i$ \\
            
            32.96 & $1.1596+0.0865 i$ & $-0.0532+0.5910 i$ & $1.1564+0.1219 i$ \\
            
            35.05 & $1.1706+0.1450 i$ & $0.0841+0.6199 i$ & $1.0897-0.4515 i$ \\
            
            37.15 & $1.1651+0.2295 i$ & $0.2141+0.6036 i$ & $0.7601-0.9123 i$ \\
            
            39.24 & $1.1375+0.3407 i$ & $0.3215+0.5537 i$ & $0.2681-1.1568 i$ \\
            
            41.33 & $1.0817+0.4748 i$ & $0.4001+0.4851 i$ & $-0.2602-1.1523 i$ \\
            
            43.43 & $0.9912+0.6239 i$ & $0.4508+0.4105 i$ & $-0.7138-0.9286 i$ \\
            
            45.52 & $0.8609+0.7761 i$ & $0.4786+0.3385 i$ & $-1.0186-0.5531 i$ \\
            
            47.61 & $0.6881+0.9169 i$ & $0.4891+0.2737 i$ & $-1.1414-0.1067 i$ \\
            
            49.70 & $0.4741+1.0299 i$ & $0.4878+0.2179 i$ & $-1.0835+0.3341 i$ \\
            
            51.80 & $0.2253+1.0991 i$ & $0.4789+0.1714 i$ & $-0.8717+0.7063 i$ \\
            
            53.89 & $-0.0462+1.1100 i$ & $0.4653+0.1334 i$ & $-0.5491+0.9658 i$ \\
            
            55.98 & $-0.3235+1.0525 i$ & $0.4493+0.1028 i$ & $-0.1666+1.0884 i$ \\
            
            58.08 & $-0.5857+0.9219 i$ & $0.4322+0.0785 i$ & $0.2240+1.0690 i$ \\
            
            60.17 & $-0.8099+0.7210 i$ & $0.4150+0.0594 i$ & $0.5752+0.9192 i$ \\
            
            62.26 & $-0.9741+0.4602 i$ & $0.3983+0.0446 i$ & $0.8484+0.6640 i$ \\
            
            64.35 & $-1.0594+0.1578 i$ & $0.3823+0.0332 i$ & $1.0164+0.3379 i$ \\
            
            66.45 & $-1.0533-0.1615 i$ & $0.3672+0.0246 i$ & $1.0654-0.0195 i$ \\
            
            68.54 & $-0.9514-0.4689 i$ & $0.3532+0.0183 i$ & $0.9949-0.3678 i$ \\
            
            70.63 & $-0.7593-0.7344 i$ & $0.3401+0.0139 i$ & $0.8168-0.6699 i$ \\
            
            72.72 & $-0.4922-0.9303 i$ & $0.3279+0.0110 i$ & $0.5535-0.8952 i$ \\
            \hline
        \end{tabular}
        }
    \end{minipage}%
    \caption{Numerical data for the scattering coefficients,  obtained as described above, with the velocity profile \eqref{vtheta}, with values $\alpha_1=2 c_s$, $c_s=0.7\mu$m ps$^{-1}$, $\alpha_2=0.3\mu$m$^{-1}$ and $\alpha_3=45\mu$m.}
    \label{tab:TableData}
\end{table}
\end{document}